%% file: ghostSignals--CAV-paper--extended.tex
\documentclass[runningheads]{llncs}

    \usepackage{xparse}
    \usepackage{xspace}
    \usepackage{amsmath}
    \usepackage{thmtools}
    \usepackage{hyperref}
    \usepackage{mathpartir}
    \usepackage{caption}
    \usepackage{subcaption}
    
    \usepackage{cite}
    \usepackage{doi}

            \usepackage{resources}
            \usepackage{notationFonts}
            \usepackage{misc}
            \usepackage{genMath}
            \usepackage{collections}
            \usepackage{functions}
            \usepackage{graphs}
            \usepackage{notes}
            \usepackage{threadPools}
            \usepackage{levels}
            \usepackage{obligations}
            \usepackage{permissions}
            \usepackage{forkPathIDs}
            \usepackage{lists}
            \usepackage{subfiles}

            \usepackage{sigExt-syntax}
            \usepackage{sigExt-ProofRules}
            \usepackage{sigExt-reduction}
            \usepackage{sigExt-programOrderGraph}
            \usepackage{sigExt-universe}

\spnewtheorem{observation}{Observation}{\bfseries}{\itshape}

        \RenewDocumentCommand{\cmdIfML}{o m m}
                {\cmdIfMLllncs[#1]{#2}{#3}}

        \RenewDocumentCommand{\hoareCondML}{o o m}
                {\hoareCondMLllncs[#1][#2]{#3}}
                
        \RenewDocumentCommand{\htProvesML}{O{\fpidVar} m m o m}
                {\htProvesMLllncs[#1]{#2}{#3}[#4]{#5}}
                
        \RenewDocumentCommand{\progProofML}{o m}
                {\progProofMLllncs[#1]{#2}}
                
        \RenewDocumentCommand{\proofRuleHintML}{m}
                {\proofRuleHintMLllncs{#1}}
        
        \RenewDocumentCommand{\referenceHintML}{m}
                {\referenceHintMLllncs{#1}}
        
        \RenewDocumentCommand{\proofHintML}{m}
                {\proofHintMLllncs{#1}}

\RenewDocumentCommand{\alt}{}{\altMedium}

\title{Ghost Signals}
\subtitle{
	Verifying Termination of Busy Waiting\\
	(Extended Version)
}

\author{
        Tobias Reinhard
                \inst{1}
                \orcidID{0000-0003-1048-8735}
        \and
        Bart Jacobs
                \inst{1}
                \orcidID{0000-0002-3605-249X}
}

\institute{
        imec-DistriNet Research Group, KU Leuven, Belgium
        \email{\{tobias.reinhard,bart.jacobs\}@kuleuven.be}
}

\begin{document}

\maketitle

\begin{abstract}
        Programs for multiprocessor machines commonly perform busy waiting for synchronization.
        We propose the first separation logic for modularly verifying termination of such programs under fair scheduling.
        Our logic requires the proof author to associate a \emph{ghost signal} with each busy-waiting loop and allows such loops to iterate while their corresponding signal \sigVar is not set.
        The proof author further has to define a well-founded order on signals and to prove that if the looping thread holds an obligation to set a signal \nextSigVar, then \nextSigVar is ordered above \sigVar.
        By using conventional shared state invariants to associate the state of ghost signals with the state of data structures, programs busy-waiting for arbitrary conditions over arbitrary data structures can be verified.
\end{abstract}


\NewDocumentCommand{\sep}{m}{\progProofChanged{\text{#1}}}
\NewDocumentCommand{\hyp}{m}{\progProofNew{\text{#1}}}

\NewDocumentCommand{\etAl}{}{et al.\xspace}


\section{Introduction}
\subfile{CAV_paper_sections/extended_version/intro}

\section{A Guide on Verifying Termination of Busy Waiting}\label{sec:ApproachInANutshell}
    \subfile{CAV_paper_sections/extended_version/approachInNutshell}

\section{A Realistic Example}\label{sec:RealisticExample}
    \subfile{CAV_paper_sections/extended_version/realisticExample}

\section{Specifying Busy-Waiting Concurrent Objects}\label{sec:FineGrainedConcurrency}
    \subfile{CAV_paper_sections/extended_version/fineGrainedConcurrency}

\section{Tool Support}\label{sec:ToolSupport}
    \subfile{CAV_paper_sections/extended_version/toolSupport}

\section{Integrating Higher-Order Features}\label{sec:HigherOrder}
    \subfile{CAV_paper_sections/extended_version/higherOrder}

\section{Related \& Future Work}\label{sec:RelatedWork}
    \subfile{CAV_paper_sections/extended_version/relatedAndFutureWork}

\section{Conclusion}\label{sec:Conclusion}
    \subfile{CAV_paper_sections/extended_version/conclusion}

\bibliographystyle{splncs04}
\bibliography{bibliography}

\section*{Appendix}
\appendix
    \subfile{CAV_paper_sections/extended_version/appendix}

\section{Case Studies}\label{appendix:sec:Case Studies}

In this section, we verify two example programs in detail.
In \S~\ref{appendix:sec:VerificationOfRealisticExample} we verify the realistic example program presented in \S~\ref{sec:RealisticExample}.
It involves two threads communicating via a shared bounded FIFO.
In \S~\ref{appendix:sec:caseStudy:unbounded} we present and verify a program similar to the one from \S~\ref{sec:RealisticExample} but with an unbounded number of producer and consumer threads.
In order to lower the visual burden on the reader, we use the following colour coding:

As in the paper body, we present the proof state in \ghostFont{blue}, 
applied proof and view shift rules in \proofRuleHint{\text{purple}},
general hints in \proofHint{\text{red}}
and abbreviations in \proofDef{\text{brown}}.
Since the verification outlines we present in this section span multiple figures, we include hints and explanations concerning other figures in \referenceHint{green} (e.g. a hint pointing to the figure where an invariant was defined).
We occasionally remind the reader of earlier proof steps performed in a previous figure by repeating them in a \lowlightText{light grey} font at the beginning of the current figure.

Further, we highlight how our proof steps effect the proof state as follows:
Consider a proof state of the form \progProof{A \slStar \slPointsTo{\hlocVar}{\valVar}}.
(i)~When a proof step adds a new chunk~$C$, we highlight it in \progProofNew{\text{green}}, i.e.
\progProof{A \slStar \slPointsTo{\hlocVar}{\valVar} \slStar \progProofNew{C}}.
(ii)~When a proof step removes the chunk $A$, we highlight this change by underlying the removed part of the assertion with a \progProofCancel{\text{dark grey}} background, i.e., 
\progProof{\progProofCancel{A\ \slStar}\ \slPointsTo{\hlocVar}{\valVar}}.
Note that in this case, the greyed out \progProofCancel{A\ \slStar} is not a part of the proof state anymore.
(iii)~When a proof step changes only part of a chunk, we highlight this change in \progProofChanged{\text{yellow}}.
For instance,  if the step changes the value of heap location \hlocVar from \valVar to \nextValVar, we highlight it in the resulting state as
\progProof{A \slStar \slPointsTo{\hlocVar}{\progProofChanged{\nextValVar}}}.

\subsection{Verification of Realistic Example}\label{appendix:sec:VerificationOfRealisticExample}

\subfile{CAV_paper_sections/extended_version/realisticExampleTransformation}

\clearpage    
\subsection{Case Study : Statically Unbounded Number of Communicating Parties}\label{appendix:sec:caseStudy:unbounded}
    \subfile{CAV_paper_sections/extended_version/caseStudyUnboundedMutliParty}

\end{document}

%% file: CAV_paper_sections/extended_version/intro.tex
    Programs for multiprocessor machines commonly perform busy waiting for synchronization~\cite{Muehlemann1980MethodFR, MellorCrummey1991AlgorithmsFS}. 
    In this paper, we propose a separation logic~\cite{Reynolds2002SeparationLA, OHearn2001LocalRA} to modularly verify termination of such programs under fair scheduling.
    Specifically, we consider programs where some threads busy-wait for a certain condition $C$ over a shared data structure to hold, e.g., a memory flag being set by other threads.
    By modularly, we mean that we reason about each thread and each function in isolation.
    That is, we do not reason about thread scheduling or interleavings.
    We only consider these issues when proving the soundness of our logic.
    Assuming fair scheduling is necessary since busy-waiting for a condition $C$ only terminates if the thread responsible for establishing the condition is sufficiently often scheduled to establish $C$.
    
    Busy waiting is an example of \emph{blocking} behavior, where a thread's progress \emph{requires interference} from other threads. This is not to be confused with \emph{non-blocking} concurrency, where a thread's progress does not rely on---and may in fact be \emph{impeded} by---interference from other threads. Existing proposed approaches for verifying termination of concurrent programs consider only programs that only involve non-blocking concurrent objects \cite{total-tada}, or \emph{primitive blocking constructs} of the programming language, such as acquiring built-in mutexes, receiving from built-in channels, joining threads, or waiting for built-in monitor condition variables \cite{Leino2010DeadlockFreeCA,finite-blocking,Hamin2018DeadlockFreeM}, or both \cite{Jacobs2018ModularTerminationVerification}. Existing techniques that do support busy-waiting are not Hoare logics; instead, they verify termination-preserving \emph{contextual refinements} between more concrete and more abstract implementations of busy-waiting concurrent objects \cite{Liang2017LiliProgressOC,Kim2017LayerByLayer}. In contrast, we here propose the first conventional program logic for modular verification of termination of programs involving busy-waiting, using Hoare triples as module specifications.

    In order to prove that a busy-waiting loop terminates, we have to prove that it performs only finitely many iterations.
    To do this we introduce a special form of  \emph{ghost resources}~\cite{Jung2016HigherorderGS} which we call \emph{ghost signals}.
    As ghost resources they only exist on the verification level and hence do not affect the program's runtime behaviour.
    Signals are initially unset and come with an obligation to set them.
    Setting a signal does not by definition correspond to any runtime condition.
    So, in order to use a signal \sigVar effectively, anyone using our approach has to prove an invariant stating that \sigVar is set if and only if the condition of interest holds.
    Further, the proof author must prove that every thread discharges all its obligations by performing the corresponding actions, e.g., by setting a signal and establishing the corresponding condition by setting the memory flag.
    
    In our verification approach we tie every busy-waiting loop to a finite set of ghost signals \SigSetVar that correspond to the set of conditions the loop is waiting for.
    Every iteration that does not terminate the loop must be justified by the proof author proving that some signal $\sigVar \in \SigSetVar$ has indeed not been set, yet.
    This way, we reduce proving termination to proving that no signal is waited for infinitely often.
    
    Our approach ensures that no thread directly or indirectly waits for itself by requiring the proof author
    (i)~to choose a well-founded and partially ordered set of levels \LevelSet and
    (ii)~to assign a level to every signal
    and by
    (iii)~only allowing a thread to wait for a signal if the signal's level is lower than the level of each held obligation.
    This guarantees that every signal is waited for only finitely often and hence that every busy-waiting loop terminates.
    We use this to prove that every program that is verified using our approach indeed terminates.

    We start by gradually introducing the intuition behind our verification approach and the concepts we use.
    In \S~\ref{sec:Nutshell:ThreadSafePhysicalSignals} and \S~\ref{sec:Nutshell:NonThreadSafePhysicalSignals} we present the main aspects of using signals to verify termination.
    We start by treating them as physical thread-safe resources and only consider busy waiting for a signal to be set.
    Then, we drop thread-safety and explain how to prove data-race- and deadlock-freedom.
    In \S~\ref{sec:Nutshell:AbitraryDataStructures} and \S~\ref{sec:Nutshell:SignalErasure} we generalize our approach to busy waiting for arbitrary conditions over arbitrary data structures and then lift signals to the verification level by introducing ghost signals. 

    In \S~\ref{sec:RealisticExample} we sketch the verification of a realistic producer-consumer example involving a bounded FIFO to demonstrate our approach's usability and address fine grained concurrency in \S~\ref{sec:FineGrainedConcurrency}.
    Further, we describe the available tool support in \S~\ref{sec:ToolSupport} and discuss integrating higher-order features in \S~\ref{sec:HigherOrder}.
    We conclude by comparing our approach to related work and reflecting on it in \S~\ref{sec:RelatedWork} and \S~\ref{sec:Conclusion}.
    
    We formally define our logic and prove its soundness in the appendix.
    To keep the presentation in this paper simple, we assume busy-waiting loops to have a certain syntactical form. 
    In our technical report~\cite{Reinhard2020GhostSignalsTR} we present a generalised version of our logic and its soundness proof.    
    Further, we verify the realistic example presented in \S~\ref{sec:RealisticExample} in full detail in the appendix and in the technical report, using the respective version of our logic.
    We used our tool support to verify C versions of the bounded FIFO example and the CLH lock.
    The tool we used and the annotated .c files can be found at~\cite{verifast2104,ArtifactJacobs2020VerifastGhostSignalConsumerProducerBoundedFifo,ArtifactJacobs2020VerifastCLHLock}.

%% file: CAV_paper_sections/extended_version/approachInNutshell.tex
    \NewDocumentCommand{\m}{}{\progVar{m}}
    \NewDocumentCommand{\x}{}{\progVar{x}}
    \NewDocumentCommand{\y}{}{\progVar{y}}
            
    \NewDocumentCommand{\locX}{}{\metaVar{\hlocVar_\x}}
    \NewDocumentCommand{\locM}{}{\metaVar{\hlocVar_\m}}
            
    \NewDocumentCommand{\invM}{}{\metaVar{\assLockInvVar}}
    \NewDocumentCommand{\levM}{}{0\xspace}
    \NewDocumentCommand{\meMut}{}{\metaVar{mut}} 
    \NewDocumentCommand{\meM}{}{\progVar{mut}}
            
    \NewDocumentCommand{\valX}{}{\metaVar{\valVar_\x}}
            
    \NewDocumentCommand{\meStartDeg}{}{1}
    \NewDocumentCommand{\meWaitDeg}{}{0}
    \NewDocumentCommand{\meLevS}{}{1}
    \NewDocumentCommand{\meIdS}{}{\progVar{sig}}
    \NewDocumentCommand{\meS}{}{\sigVar}

    When we try to verify termination of busy-waiting programs, multiple challenges arise.
    Throughout this section, we describe these challenges and our approach to overcome them.
    In \S~\ref{sec:Nutshell:ThreadSafePhysicalSignals} we start by discussing the core ideas of our logic.
    In order to simplify the presentation we initially consider a simple language with built-in thread-safe \emph{signals} and a corresponding minimal example where one thread busy-waits for such a signal.
    Signals are heap cells containing boolean values that are specially marked as being solely used for busy waiting.
    Throughout this section, we generalize our setting as well as our example towards one that allows to verify programs with busy waiting for arbitrary conditions over arbitrary shared data structures.
    In \S~\ref{sec:Nutshell:NonThreadSafePhysicalSignals} we present the concepts necessary to verify data-race-, deadlock-freedom and termination in the presence of  built-in signals that are not thread safe.
    In \S~\ref{sec:Nutshell:AbitraryDataStructures} we explain how to use these non-thread-safe signals to verify programs that wait for arbitrary conditions over shared data structures.
    We illustrate this by an example waiting for a shared heap cell to be set.
    In \S~\ref{sec:Nutshell:SignalErasure} we erase the signals from our program and lift them to the verification level in the form of a concept we call \emph{ghost signals}.

    \subsection{Simplest Setting: Thread-Safe Physical Signals}\label{sec:Nutshell:ThreadSafePhysicalSignals}
    We want to verify programs that busy-wait for arbitrary conditions over arbitrary shared data structures. 
    As a first step towards achieving this, we first consider programs that busy-wait for simple boolean flags, specially marked as being used for the purpose of busy-waiting. 
    We call these flags \emph{signals}. 
    For now, we assume that read and write operations on signals are thread-safe.
    Consider a simple programming language with built-in signals and with the following commands:
    (i)~\cmdNewSignal for creating a new unset signal, 
    (ii)~\cmdSetSignal{x} for setting $x$ and 
    (iii)~\cmdAwaitNoMut{\cmdIsSignalSet{\varVar}} for busy-waiting until $x$ is set.
    Fig.~\ref{fig:MinEx:RealThreadSafeSignals} presents a minimal example where two threads communicate via a shared signal \sigProgVar.
    The main thread creates the signal \sigProgVar and forks a new thread that busy-waits for \sigProgVar to be set.
    Then, the main thread sets the signal.
    As we assume signal operations to be thread-safe in this example, we do not have to care about potential data races.
    Notice that like all busy-waiting programs, this program is guaranteed to terminate only under fair thread scheduling: 
    Indeed, it does not terminate if the main thread is never scheduled after it forks the new thread. 
    In this paper we verify termination under fair scheduling.

    \begin{figure}
            \vspace{-0.3cm}
            $$
            \begin{array}{l}
                    \keyword{let}\ \sigProgVar := \cmdNewSignal\ \keyword{in}\\
                    \cmdFork{
                            \cmdAwaitNoMut{\cmdIsSignalSet{\sigProgVar}}
                    };
                    \\
                    \cmdSetSignal{\sigProgVar}
            \end{array}
            $$
            \vspace{-0.3cm}
            \caption{
                    Minimal example with two threads communicating via a physical thread-safe signal.
            }
            \label{fig:MinEx:RealThreadSafeSignals}
            \vspace{-0.5cm}
    \end{figure}

\newpage
\subsubsection{Augmented Semantics}
    \paragraph{Obligations}
	The only construct in our language that can lead to non-termination are busy-waiting loops of the form \cmdAwaitNoMut{\cmdIsSignalSet{\sigProgVar}}.
	In order to prove that programs terminate it is therefore sufficient to prove that all created signals are eventually set.
	We use so-called \emph{obligations}~\cite{Hamin2019TransferringOT, Hamin2018DeadlockFreeM, Leino2010DeadlockFreeCA, Kobayashi2006ANT} to ensure this.
	These are \emph{ghost resources}~\cite{Jung2016HigherorderGS}, i.e., resources that do not exist during runtime and can hence not influence a program's runtime behaviour.
	They carry, however, information relevant to the program's verification.
	Generally, holding an obligation requires a thread to discharge it by performing a certain action.
	For instance, when the main thread in our example creates signal \sigProgVar, it simultaneously creates an obligation to set it.
	The only way to discharge this obligation is to set \sigProgVar.
	
	We denote thread IDs by \tidVar and describe which obligations a thread \tidVar holds by bundling them into an obligations chunk \obsQual{\tidVar}{\obBagVar}, where \obBagVar is a multiset of signals.
	We denote multisets by double braces \multiset{\dots} and multiset union by \msCup.
	Each occurrence of a signal \sigVar in \obBagVar corresponds to an obligation by thread \tidVar to set \sigVar.
	Consequently, \noObsQual{\tidVar} asserts that thread \tidVar does not hold any obligations.

	\paragraph{Augmented Semantics} 
	In the \emph{real} semantics of the programming language we consider here, ghost resources such as obligations do not exist during runtime.
	To prove termination, we consider an \emph{augmented} version of it that keeps track of ghost resources during runtime.
	In this semantics, we maintain the invariant that every thread holds exactly one \obsPred chunk.
	That is, for every running thread \tidVar, our heap contains a unique heap cell \obsQualPred{\tidVar} that stores the thread's bag of obligations.
	Further, we let a thread get stuck if it tries to finish while it still holds undischarged obligations.
	Note that we use the term \emph{finish} to refer to thread-local behaviour while we write \emph{termination} to refer to program-global behaviour, i.e., meaning that every thread finishes.
	For every augmented execution there trivially exists a corresponding execution in the real semantics.

	Fig.~\ref{fig:AugmentedReductionRules} presents some of the reduction rules we use to define the augmented semantics.
	We use \augHeapVar to refer to augmented heaps, i.e., heaps that can contain ghost resources.
	A reduction step has the form 
	\augRedStep
		{\augHeapVar}{\cmdVar}
		[\tidVar]
		{\nextAugHeapVar}{\nextCmdVar}
		[\ftsVar]
	expresses that thread \tidVar reduces heap \augHeapVar (which is shared by all threads) and command \cmdVar to heap \nextAugHeapVar and command \nextCmdVar.
	Further, \ftsVar represents the set of threads forked during this step.
	It is either empty or a singleton containing the new thread's ID and the command it is going to execute, i.e., \setOf{(\tidForkedVar, \cmdForkedVar)}.
	We omit it whenever it is clear from the context that no thread is forked.
	Further, we denote disjoint union of sets by \disjCup.


	Our reduction rules comply with the intuition behind obligations we outlined above.
	\augRedNewSignalName creates a new signal and simultaneously a corresponding obligation.
	The only way to discharge it is by setting the signal using \augRedSetSignalName.

\begin{figure}
    \vspace{-0.5cm}
	\begin{mathpar}
		\augRedNewSignal
		\and
		\augRedSetSignal
		\and
		\augRedFork
		\and
		\augRedAwait
	\end{mathpar}

    \vspace{-0.2cm}
	\caption{Reduction rules for augmented semantics.}
	\label{fig:AugmentedReductionRules}
    \vspace{-0.8cm}
\end{figure}

    \paragraph{Forking}
    Whenever a thread forks a new thread, it can pass some of its obligations to the newly forked thread, cf.\@ \augRedForkName.
    Forking a new thread with ID \tidForkedVar also allocates a new heap cell \obsQualPred{\tidForkedVar} to store its bag of obligations.
    Since this is the only way to allocate a new \obsPred heap cell, we will never run into a heap
    $\augHeapVar
    	\disjCup \setOf{\obsQual{\tidVar}{\obBagVar}}
    	\disjCup \setOf{\nextObsQual{\tidVar}{\nextObBagVar}}
    $
    that contains multiple obligations chunks belonging to the same thread \tidVar.    
    Remember that threads cannot finish while holding obligations.
    This prevents them from dropping obligations via dummy~forks.

	\paragraph{Levels}
	In order to prove that a busy-waiting loop \cmdAwaitNoMut{\cmdIsSignalSet{\sigProgVar}} terminates, we must ensure that the waiting thread does not directly or indirectly wait for itself.
	We could just check that it does not hold an obligation for the signal it is waiting for, but that is not sufficient as the following example demonstrates:
	Consider a program with two signals $\sigProgVar_1, \sigProgVar_2$ and two threads.
	Let one thread hold the obligation for $\sigProgVar_2$ and execute 
	$
	\cmdAwaitNoMut{\cmdIsSignalSet{\sigProgVar_1}};\
	\cmdSetSignal{\sigProgVar_2}
	$.
	Likewise, let the other thread hold the obligation for $\sigProgVar_1$ and let it execute
	$
	\cmdAwaitNoMut{\cmdIsSignalSet{\sigProgVar_2}};\
	\cmdSetSignal{\sigProgVar_1}
	$.

	To prevent such \emph{wait cycles} modularly, we apply the usual approach~\cite{Leino2010DeadlockFreeCA, Boyapati2002OwnershipTF, Flanagan2002ExtendedStaticCheckingJava}.
	For every program that we want to execute in our augmented semantics, we choose a partially ordered set of levels \LevelSet.
	Further, during every reduction step in the augmented semantics that creates a signal \sigVar, we pick a level $\levVar \in \LevelSet$ and associate it with \sigVar.
	Note that much like obligations, levels do not exit during runtime in the real semantics.
	Signal chunks in the augmented semantics have the form \assSignalNoBool{(\idVar, \levVar)} where \idVar is the unique signal identifier returned by \cmdNewSignal.
	The level assigned to any signal can be chosen freely, cf.\@ \augRedNewSignalName.
	In practice, determining levels boils down to solving a set of constraints that reflect the dependencies.
	In our example, however, the choice is trivial as it only involves a single signal.
	We choose $\LevelSet = \setOf{0}$ and 0 as level for \sigProgVar and thereby get \assSignalNoBool{(\sigProgVar, 0)}.
	Generally, we denote signal tuples by $\sigVar = (\idVar, \levVar)$.
	Now we can rule out cyclic wait dependencies by only allowing a thread to busy-wait for a signal \sigVar if its level \sigGetLev{\sigVar} is smaller than the level of each held obligation, cf.\@ \augRedAwaitName~\footnote
        {
        	For simplicity, our augmented semantics assumes that the level order and the level associated with any object remains fixed for the entire execution.
        	However, following the approach presented in \cite{Leino2009MultithreadedPrograms}, it would be sound to add a step rule that allows a thread to change the level of an object it has exclusive access to (cf.\@~\S~\ref{sec:Nutshell:NonThreadSafePhysicalSignals}).
        }.
	Given a bag of obligations \obBagVar, we denote this by $\sigGetLev{\sigVar} \levObsLt \obBagVar$.

\myComment
{
	Below, we sketch one possible execution of our example program in the augmented semantics.
	To simplify the notation, we represent a thread with ID \tidVar that executes command \cmdVar as a tuple $(\tidVar, \cmdVar)$.
	Further, we represent thread pools and heaps as sets of threads and heap cells, respectively, and we denote our augmented thread pool reduction relation by \augTpRedStepSymb[].
	Let $\cmdVar_e$ denote the our example program, let \tidMainVar, \tidForkedVar be the IDs of the main and the forked thread, respectively.
	Initially, our thread pool contains only the main thread and our heap contains only its \obsPred chunk.
	Threads that finish without obligations are removed from the thread pool and their \obsPred chunk is removed from the heap.
	We end terminate our execution with an empty heap and an empty thread pool.
	$$\small
	\begin{array}{l}
		\setOf{\obsQual{\tidMainVar}{\msEmpty}},\ \setOf{(\tidMainVar, \cmdVar_e)}
		\\
	\augTpRedStepSymb[\tidMainVar]\
		\setOf{\obsQual{\tidMainVar}{\multiset{(\sigProgVar, 0)}}},\
		\setOf{(\tidMainVar, (\cmdFork{\cmdAwaitNoMut{\cmdIsSignalSet{\sigProgVar}}}); \cmdSetSignal{\sigProgVar})}
		\\
	\augTpRedStepSymb[\tidMainVar]\
		\setOf{\obsQual{\tidMainVar}{\multiset{(\sigProgVar, 0)}},\ \obsQual{\tidForkedVar}{\msEmpty}},\
		\setOf{
			(\tidMainVar, \cmdSetSignal{\sigProgVar}),\
			(\tidForkedVar, \cmdAwaitNoMut{\cmdIsSignalSet{\sigProgVar}})
		}
		\\
	\augTpRedStepSymb[\tidMainVar]\
		\setOf{\obsQual{\tidMainVar}{\msEmpty},\ \obsQual{\tidForkedVar}{\msEmpty}},\
		\setOf{
			(\tidMainVar, \valUnit),\
			(\tidForkedVar, \cmdAwaitNoMut{\cmdIsSignalSet{\sigProgVar}})
		}
		\\
	\augTpRedStepSymb[\tidMainVar]\
		\setOf{\obsQual{\tidForkedVar}{\msEmpty}},\
		\setOf{
			(\tidForkedVar, \cmdAwaitNoMut{\cmdIsSignalSet{\sigProgVar}})
		}
		\\
	\augTpRedStepSymb[\tidForkedVar]\
		\setOf{\obsQual{\tidForkedVar}{\msEmpty}},\
		\setOf{
			(\tidForkedVar, \valUnit)
		}
	\ \augTpRedStepSymb[\tidForkedVar]\
		\emptyset,\ \emptyset
	\end{array}
	$$
}

	\paragraph{Proving Termination}
	As we will explain below, the augmented semantics has no fair infinite executions.
	We can use this as follows to prove that a program \cmdVar terminates under fair scheduling:
	For every fair infinite execution of \cmdVar, show that we can construct a corresponding augmented execution.
	(This requires that each step's side conditions in the augmented semantics are satisfied.
	Note that we thereby prove certain properties for the real execution, like absence of cyclic wait dependencies.)
	As there are no fair infinite executions in the augmented semantics, we get a contradiction.
	It follows that \cmdVar has no fair infinite executions in the real semantics.

    \paragraph{Soundness}
	In order to prove soundness of our approach, we must prove that there indeed are no fair infinite executions in the augmented semantics.
	This boils down to proving that no signal can be waited for infinitely often.
	Consider any program and any fair augmented execution of it.
	Consider the execution's \emph{program order graph}, 
	(i)~whose nodes are the execution steps and 
	(ii)~which has an edge from a step to the next step of the same thread and to the first step of the forked thread, if it is a fork step. 
	Notice that for each obligation created during the execution, the set of nodes corresponding to a step made by a thread while that thread holds the obligation constitutes a path that ends when the obligation is discharged.
	We say that this path \emph{carries} the obligation.
	
	It is not possible that a signal is waited for infinitely often. 
	Indeed, suppose some signals \InfWaitSigSet are. 
	Take $\minSig \in \InfWaitSigSet$ with minimal level. 
	Since \minSig is never set, the path in the program order graph that carries the obligation must be infinite as well. 
	Indeed, suppose it is finite. 
	The final node $N$ of the path cannot discharge the obligation without setting the signal, so it must pass the obligation on either to the next step of the same thread or to a newly forked thread. 
	By fairness of the scheduler, both of these threads will eventually be scheduled.
	This contradicts $N$ being the final node of the path.    
	
	The path carrying the obligation for \minSig waits only for signals that are waited for finitely often. 
	(Remember that \augRedAwaitName requires the signal waited for to be of a lower level than all held obligations, i.e.,  a lower level than that of \minSig.)
	It is therefore a finite path.
	A contradiction. 
	
	Notice that the above argument relies on the property that every non-empty set of levels has a minimal element. 
	For this reason, for termination verification we require that \LevelSet is not just partially ordered, but also well-founded.

\subsubsection{Program Logic}

	\paragraph{}\!\!\!\!\!\!\!
	Directly using the augmented semantics to prove that our example program terminates is cumbersome.
	In the following, we present a separation logic that simplifies this task.
 
	\paragraph{Safety}
	We call a program \cmdVar \emph{safe} under a (partial) heap \augHeapVar if it provides all the resources necessary such that both \cmdVar and any threads it forks can execute without getting stuck in the augmented semantics. (This depends on the angelic choices.)
	We denote this by \safe{\augHeapVar}{\cmdVar}~\cite{VafeiadisCSLSound}~\footnote{
        For a formal definition see the appendix (\S~\ref{appendix:sec:Soundness:ModelRelation}) and the technical report~\cite{Reinhard2020GhostSignalsTR}.
    }.

	Consider a program \cmdVar that is safe under an augmented heap \augHeapVar.
	Let \pheapVar be the real heap that matches \augHeapVar apart from the ghost resources.
	Then, for every real execution that starts with \pheapVar we can construct a corresponding augmented execution.

    \paragraph{Specifications}
	We use Hoare triples \hoareTriple{\assPreVar}{\cmdVar}[\resVar]{\assPostVar(\resVar)}~\cite{Hoare1968HoareLogic} to specify the behaviour of a program \cmdVar.
%
	Such a triple expresses the following:
	Consider any evaluation context \evalCtxt, such that for every return value \valVar, running \evalCtxt[\valVar] from a state that satisfies $\assPostVar(\valVar)$ is safe.
	Then, running \evalCtxt[\cmdVar] from a state that satisfies \assPreVar is safe.
\myComment
{
	Let \assModels{\augHeapVar}{\assVar} express that assertion \assVar holds under heap \augHeapVar.
	Then we can formalize this interpretation as follows:
	$$
	\begin{array}{c}
		\hoareTriple{\assPreVar}{\cmdVar}[\resVar]{\assPostVar(\resVar)}\\
		\Longleftrightarrow\\
		\begin{array}{l}
			\forall \augHeapFrameVar.\
			\forall \evalCtxt.\
			(\forall \valVar.\
			\forall \augHeapPostVar.\
			\assModelsOR
			{\augHeapPostVar}
			{\assPostVar(\valVar)}
			\ \rightarrow\
			\safe
			{\augHeapPostVar \disjCup \augHeapFrameVar}
			{\,\evalCtxt[\valVar]}
			)
			\\
			\phantom{
				\forall \augHeapFrameVar.\
			}
			\rightarrow\ \
			\forall \augHeapPreVar.\
			\assModelsOR{\augHeapPreVar}{\assPreVar}
			\ \rightarrow\
			\safe
			{\augHeapPreVar \disjCup \augHeapFrameVar}
			{\,\evalCtxt[\cmdVar]}	
		\end{array}
	\end{array}
	$$
}

    \paragraph{Proof System}
    We define a proof relation \htProvesSymb which ensures that whenever we can prove 
    \htProvesShort{\assPreVar}{\cmdVar}[\resVar]{\assPostVar(\resVar)},
    then \cmdVar complies with the specification 
    \hoareTripleShort{\assPreVar}{\cmdVar}[\resVar]{\assPostVar(\resVar)}.
    Fig.~\ref{fig:MinEx:RealThreadSafeSignals:ProofRules} presents some of the proof rules we use to define~\htProvesSymb.
    As we evolve our setting throughout this section, we also adapt our proof rules.
    Rules that will be changed later are marked with a prime in their name.
    The full set of rules is presented in the appendix (cf.\@ Fig.~\ref{appendix:fig:ProofRulesPartOne} and~\ref{appendix:fig:ProofRulesPartTwo}).
	Our proof rules~\prMinSigSetSignalName and~\prMinSigAwaitName are similar to the rules for sending and receiving on a channel presented in~\cite{Leino2010DeadlockFreeCA}.
    
    Notice how the proof rules enforce the side-conditions of the augmented semantics.
    Hence, all we have to do to prove that a program \cmdVar terminates is to prove that every thread eventually discharges all its obligations.
    That is, we have to prove \htProves{\noObs}{\cmdVar}{\noObs}.
    Fig.~\ref{fig:MinEx:RealThreadSafeSignals:ProofSketch} illustrates how we can apply our rules to verify that our minimal example terminates.

\myComment
{
	\paragraph{Proving Termination}
	As explained above, we can prove that a program \cmdVar terminates in two steps:
	(i)~Showing that for every fair real execution of \cmdVar, we can construct an augmented counterpart that does not get stuck.
	(ii)~Using the property that there are no infinite fair executions in the augmented semantics.
	As our proof rules' restrictions reflect those in the augmented semantics, a proof in our logic corresponds to such a construction.
	For instance, during a proof, we show how to choose levels and how to delegate obligations to newly forked threads.
	Hence, all that we have to do is to prove that all threads in the constructed augmented execution eventually discharge their obligations.
	Otherwise they would get stuck when trying to finish.
	That is, we can prove termination by proving \htProves{\noObs}{\cmdVar}{\noObs}.	
}

\myComment
{
	\paragraph{Interpreting Obligations Chunks}
	The assertion \obs{\obBagVar} asserts that the current thread's bag of obligations is exactly \obBagVar. 
	We can link this to regular separation logic as follows:
	Whenever a new thread $t$ is created in our augmented semantics, a new heap cell $t.\obsPred$ is allocated as well.
	\obs{\obBagVar} then means that heap cell $t.\obsPred$ (where $t$ is the current thread) stores the bag of obligations \obBagVar, i.e.,
	\slPointsTo{t.\fixedPredNameFont{obs}}{\obBagVar}
	in standard separation logic syntax (cf.\@ \S~\ref{sec:Nutshell:AbitraryDataStructures}).
}

\myComment
{
	We denote the standard separating conjunction by \slStar.
	Considering the augmented semantics, we see that the assertion
	$\obs{\obBagVar \msCup \nextObBagVar}$
	is not equivalent to
	$\obs{\obBagVar} \slStar \obs{\nextObBagVar}$.
	The assertion
	$\obs{\obBagVar \msCup \nextObBagVar}$ represents a single heap cell $t.\obsPred$ that stores the union $\obBagVar \msCup \nextObBagVar$, i.e., a single heap chunk \slPointsTo{t.\fixedPredNameFont{obs}}{\obBagVar \msCup \nextObBagVar}.
	In contrast, $\obs{\obBagVar} \slStar \obs{\nextObBagVar}$ represents a heap where  \obBagVar and \nextObBagVar are stored in two different heap cells, i.e.,
	\slPointsTo{t.\obsPred}{\obBagVar}
	and
	\slPointsTo{t.\obsPred^\prime}{\nextObBagVar}.
}

\myComment
{
	Further, our proof system and the augmented semantics both maintain the invariant that every thread holds exactly one obligations chunk \obs{\obBagVar}, provided that we start with exactly one in the precondition.
	Consider an arbitrary precondition $\obs{\obBagVar} \slStar \assPreVar$ where \assPreVar does not contain any obligations chunks.
	Then, using our proof rules it is not possible to prove a specification of the form
	\htProves
	{\obs{\obBagVar} \slStar \assPreVar}
	{\cmdVar}
	{\obs{\nextObBagVar} \slStar \obs{\nextNextObBagVar} \slStar{} \dots}.
}

    \begin{figure}
            \begin{subfigure}{\textwidth}
                    $$
                    \begin{array}{l p{0.5cm} l}
                            \progProof{\noObs}
                            \\
                            \keyword{let}\ \sigProgVar := \cmdNewSignal\ \keyword{in}
                                    &&\proofRuleHint{\prMinSigNewSignalName with $\levVar = 0$}
                            \\
                            \progProof{
                                    \obsOf{(\sigProgVar, 0)}
                                    \slStar
                                    \assSignalNoBool{(\sigProgVar, 0)}
                            }
                                    &&\proofDef{\sigVar := (\sigProgVar, 0)}
                            \\
                            \keyword{fork}\ (
                                    \progProof{
                                            \noObs
                                            \slStar
                                            \assSignalNoBool{\sigVar}
                                    }
                            \\
                            \phantom{\keyword{fork}\ (}\
                                    \cmdAwaitNoMut{\cmdIsSignalSet{\sigProgVar}}
                                            &&\proofHint{$\sigGetLev{\sigVar} = 0  \levObsLt \msEmpty$}
                            \\
                            \phantom{\keyword{fork}\ (}
                                    \progProof{
                                            \noObs
                                            \slStar
                                            \assSignalNoBool{\sigVar}
                                    });
                            \\
                            \progProof{
                                    \obsOf{\sigVar}
                            }
                            \\
                            \cmdSetSignal{\sigProgVar}
                            \\
                            \progProof{\noObs}
                    \end{array}
                    $$
                    \caption{
                            Proof outline for program from Fig.~\ref{fig:MinEx:RealThreadSafeSignals}.
                            Applied proof rule marked in \proofRuleHint{purple}.
                            Abbreviation marked in \proofDef{\text{brown}}.
                            General hint marked in \proofHint{red}.
                    }
                    \label{fig:MinEx:RealThreadSafeSignals:ProofSketch}
            \end{subfigure}
            
            \begin{subfigure}{\textwidth}
                    \begin{mathpar}
                            \prMinSigNewSignal
                            \and
                            \prMinSigSetSignal
                            \and
                            \and
                            \prMinSigFork
                            \and
                            \prMinSigAwait
                            \and
                            \prLet
                    \end{mathpar}
                    \caption{Proof rules. Rules only used in this section marked with '.}
                    \label{fig:MinEx:RealThreadSafeSignals:ProofRules}
            \end{subfigure}
            \caption{Verifying termination of minimal example with physical thread-safe signal.}
    \end{figure}

    \subsection{Non-Thread-Safe Physical Signals}\label{sec:Nutshell:NonThreadSafePhysicalSignals}
    As a step towards supporting waiting for arbitrary conditions over shared data structures, including non-thread-safe ones, we now move to non-thread-safe signals. 
    For simplicity, in this paper we consider programs that use mutexes to synchronize concurrent accesses to shared data structures. 
    (Our ideas apply equally to programs that use other constructs, such as atomic machine instructions.) 
    Fig.~\ref{fig:MinEx:RealNonThreadSafeSignals:CodeAndSyntacticSugar} presents our updated example.
    
    \begin{figure}
            \begin{subfigure}[b]{0.4\textwidth}
            $$
            \begin{array}{l}
                    \keyword{let}\ \sigProgVar := \cmdNewSignal\ \keyword{in}\\
                    \keyword{let}\ \mutProgVar := \cmdNewMut\ \keyword{in}\\
                    \cmdFork{
                            \cmdAwait{\mutProgVar}{\cmdIsSignalSet{\sigProgVar}}
                    };
                    \\
                    \cmdAcquireMut{\mutProgVar};
                    \\
                    \cmdSetSignal{\sigProgVar};
                    \\
                    \cmdReleaseMut{\mutProgVar}
            \end{array}
            $$
            \caption{Code.}
            \end{subfigure}
            \hspace{0.65cm}
            \begin{subfigure}[b]{0.45\textwidth}
                    $$
                    \begin{array}{l c l}
                            \cmdAwait{\mutProgVar}{\cmdVar} &\ :=\
                                    &(\keyword{while}\
                                            \cmdAcquireMut{\mutProgVar};
                            \\
                                    &&\phantom{\keyword{while}\ }\
                                            \keyword{let}\ \resVar := \cmdVar\ \keyword{in}
                            \\
                                    &&\phantom{\keyword{while}\ }\
                                            \cmdReleaseMut{\mutProgVar};
                            \\
                                    &&\phantom{\keyword{while}\ }\
                                            \neg \resVar
                            \\
                                    &&\phantom{(}\keyword{do}\ \keyword{skip})
                    \end{array}
                    $$
                    \caption{Syntactic sugar. \resVar not free in \mutProgVar.}
            \end{subfigure}
            
            \caption{Minimal example with two threads communicating via a physical non-thread-safe signal protected by a mutex.}
            \label{fig:MinEx:RealNonThreadSafeSignals:CodeAndSyntacticSugar}
    \end{figure}

    As signal \sigProgVar is no longer thread-safe, the two threads can no longer use it directly to communicate.
    Instead, we have to synchronize accesses to avoid data races.
    Hence, we protect the signal by a mutex \mutProgVar created by the main thread.
    In each iteration, the forked thread acquires the mutex, checks whether \sigProgVar has been set and releases it again.
    After forking, the main thread acquires the mutex, sets the signal and releases it again.

    \paragraph{Exposing Signal Values}
    Signals are specially marked heap cells storing boolean values.
    We make this explicit by extending our signal chunks from \assSignalNoBool{\sigVar} to \assSignal{\sigVar}{\slBoolVar} where \slBoolVar is the current value of \sigVar and by updating our proof rules accordingly.
    Upon creation, signals are unset.
    Hence, creating a signal \sigProgVar now spawns an \emph{unset} signal chunk \assSignal{(\sigProgVar, \levVar)}{\slFalse} for some freely chosen level \levVar and an obligation for $(\sigProgVar, \levVar)$, cf.~\prRealNonSafeSigsNewSignalName.
    We present our new proof rules in Fig.~\ref{fig:MinEx:RealNonThreadSafeSignals:ProofRules} and demonstrate their application in Fig.~\ref{fig:MinEx:RealNonThreadSafeSignals:ProofSketch}.

    \begin{figure}
                    $$
                    \begin{array}{l p{0cm} l}
                            \progProof{\noObs}
                            \\
                            \keyword{let}\ \sigProgVar := \cmdNewSignal\ \keyword{in}
                                    &&\proofRuleHint{\prRealNonSafeSigsNewSignalName with $\levVar = \meLevS$}
                            \\
                            \progProof{
                                    \obsOf{(\sigProgVar, \meLevS)}
                                    \slStar
                                    \assSignal{(\sigProgVar, \meLevS)}{\slFalse}
                            }
                                    &&\proofRuleHint{\prViewShiftName \& \vsSemImpName}
                            \\
                            \progProof{
                                    \obsOf{(\sigProgVar, \meLevS)}
                                    \slStar
                                    \exists \slBoolVar.\ \assSignal{(\sigProgVar, \meLevS)}{\slBoolVar}
                            }
                                    &&\proofDef{\sigVar := (\sigProgVar, \meLevS),\
                                            \invM := \exists \slBoolVar.\ \assSignal{\sigVar}{\slBoolVar}
                                    }
                            \\
                            \keyword{let}\ \mutProgVar := \cmdNewMut\ \keyword{in}
                                    &&\proofRuleHint{\prRealNonSafeSigsNewMutexName with $\levVar = \levM$}
                            \\
                            \progProof{
                                            \obsOf{\sigVar}
                                            \slStar
                                            \assMutex{\mutVar}{\invM}
                                    }
                                            &&\proofRuleHint{\prViewShiftName}
                            \\
                            \progProofML{
                                            \obsOf{\sigVar}
                                            \slStar
                                            \assMutex{\mutVar}{\invM}
                                            \slStar
                                            \assMutex{\mutVar}{\invM}
                                    }
                                            &&\quad\quad\proofRuleHint{\& \vsRealNonSafeSigsCloneMutName}
                            \\
                            \keyword{fork}\ (
                                    \progProof{
                                            \noObs
                                            \slStar
                                            \assMutex{\mutVar}{\invM}
                                    }
                            \\
                            \phantom{\keyword{fork}\ (}\
                                    \keyword{with}\ \mutVar\ \keyword{await}
                                            &&\proofHint{$\mutGetLev{\mutVar}, \sigGetLev{\sigVar} \levObsLt \msEmpty$}
                            \\
                            \phantom{\keyword{fork}\ (}\quad\quad
                                    \progProof{
                                            \obsOf{\mutVar}
                                            \slStar
                                            \invM
                                    }
                                            &&\proofRuleHint{\prExistsName}
                            \\
                            \phantom{\keyword{fork}\ (}\quad\quad
                                    \ghostFont{\forall \slBoolVar.}\
                                    \progProof{
                                            \obsOf{\mutVar}
                                            \slStar
                                            \assSignal{\sigVar}{\slBoolVar}
                                    }
                            \\
                            \phantom{\keyword{fork}\ (\quad\quad \ghostFont{\forall \slBoolVar.}}\
                                    \cmdIsSignalSet{\sigProgVar}
                            \\
                            \phantom{\keyword{fork}\ (\quad\quad \ghostFont{\forall \slBoolVar.}}\
                                    \progProof[\resVar]{
                                            \obsOf{\mutVar}
                                            \slStar
                                            \assSignal{\sigVar}{\slBoolVar}
                                            \wedge
                                            \resVar = \slBoolVar
                                    }
                                            &&\proofRuleHint{\prViewShiftName \& \vsSemImpName}
                            \\
                            \phantom{\keyword{fork}\ (\quad\quad \ghostFont{\forall \slBoolVar.}}\
                                    \progProofML[\resVar]{
                                            \obsOf{\mutVar}
                                            \\
                                            \slStar\,
                                            \slIfElse{\resVar}
                                                    {\invM}
                                                    {\assSignal{\sigVar}{\slFalse}}
                                    }
                            \\
                            \phantom{\keyword{fork}\ (}
                                    \progProof{
                                            \noObs
                                            \slStar
                                            \assMutex{\mutVar}{\invM}
                                    }
                                            &&\proofRuleHint{\prViewShiftName \& \vsSemImpName}
                            \\
                            \phantom{\keyword{fork}\ (}
                                    \progProof{
                                            \noObs
                                    });
                            \\
                            \progProof{
                                    \obsOf{\sigVar}
                                    \slStar
                                    \assMutex{\mutVar}{\invM}
                            }
                            \\
                            \cmdAcquireMut{\mutProgVar};
                                    &&\proofHint{$\mutGetLev{\mutVar} = \levM < \meLevS = \sigGetLev{\sigVar}$}
                            \\
                            \progProof{
                                    \obsOf{\sigVar, \mutVar}
                                    \slStar
                                    \assMutLockedNoFrac{\mutVar}{\invM}
                                    \slStar
                                    \exists \slBoolVar.\ \assSignal{\sigVar}{\slBoolVar}
                            }
                                    &&\proofRuleHint{\prExistsName}
                            \\
                            \ghostFont{\forall \slBoolVar.}\
                            \progProof{
                                    \obsOf{\sigVar, \mutVar}
                                    \slStar
                                    \assMutLockedNoFrac{\mutVar}{\invM}
                                    \slStar
                                    \assSignal{\sigVar}{\slBoolVar}
                            }
                            \\
                            \phantom{\ghostFont{\forall \slBoolVar.}\ }\
                            \cmdSetSignal{\sigProgVar};
                            \\
                            \phantom{\ghostFont{\forall \slBoolVar.}\ }\
                            \progProof{
                                    \obsOf{\mutVar}
                                    \slStar
                                    \assMutLockedNoFrac{\mutVar}{\invM}
                                    \slStar
                                    \assSignal{\sigVar}{\slTrue}
                            }
                                    &&\proofRuleHint{\prViewShiftName \& \vsSemImpName}
                            \\
                            \phantom{\ghostFont{\forall \slBoolVar.}\ }\
                            \progProof{
                                    \obsOf{\mutVar}
                                    \slStar
                                    \assMutLockedNoFrac{\mutVar}{\invM}
                                    \slStar
                                    \invM
                            }
                            \\
                            \phantom{\ghostFont{\forall \slBoolVar.}\ }\
                            \cmdReleaseMut{\mutProgVar}
                            \\
                            \phantom{\ghostFont{\forall \slBoolVar.}\ }\
                            \progProof{
                                    \noObs
                                    \slStar
                                    \assMutex{\mutVar}{\invM}
                            }
                                    &&\proofRuleHint{\prViewShiftName \& \vsSemImpName}
                            \\
                            \phantom{\ghostFont{\forall \slBoolVar.}\ }\
                            \progProof{\noObs}
                    \end{array}
                    $$
                    \vspace{-0.4cm}
                    \caption{
                            Proof outline for program \ref{fig:MinEx:RealNonThreadSafeSignals:CodeAndSyntacticSugar},
                            verifying termination with mutexes \& non-thread safe signals.
                            Applied proof and view shift rules marked in \proofRuleHint{purple}.
                            Abbreviations marked in \proofDef{\text{brown}}.
                            General hints marked in \proofHint{red}.
                    }
            
            \label{fig:MinEx:RealNonThreadSafeSignals:ProofSketch}
    \end{figure}

    \begin{figure}
            \begin{subfigure}{\textwidth}
                    \begin{mathpar}
                            \prRealNonSafeSigsNewSignal
                            \and
                            \prRealNonSafeSigsSet
                            \and
                            \prRealNonSafeSigsIsSet
                            \and
                            \prRealNonSafeSigsAwait
                    \end{mathpar}
                    \caption{Signals \& busy waiting.}
            \end{subfigure}

            \begin{subfigure}{\textwidth}
                    \begin{mathpar}
                            \prRealNonSafeSigsNewMutex
                            \and
                            \prRealNonSafeSigsAcquire
                            \and
                            \prRealNonSafeSigsRelease
                    \end{mathpar}
                    \caption{Mutexes.}
            \end{subfigure}

            \begin{subfigure}{\textwidth}
                    \begin{mathpar}
                            \prFrame
                            \and
                            \prExists
                            \and
                            \prForkPaper
                            \and
                            \prSimpleViewShiftPaper
                    \end{mathpar}
                    \caption{Standard rules.}
            \end{subfigure}

            \begin{subfigure}{\textwidth}
                    \begin{mathpar}
                            \vsSemImp
                            \and
                            \vsTrans
                            \and
                            \vsRealNonSafeSigsCloneMut
                    \end{mathpar}
                    \caption{View shifts.}
            \end{subfigure}
            
            \caption{
                    Proof rules \& view shift rules for mutexes \& non-thread safe signals. 
                    Rules only used in this section marked with ''.
            }
            \label{fig:MinEx:RealNonThreadSafeSignals:ProofRules}
    \end{figure}

    \paragraph{Data Races}
    As read and write operations on signals are no longer thread-safe, our logic has to ensure that two threads never try to access \sigProgVar at the same time.
    Hence, in our logic possession of a signal chunk \assSignal{\sigVar}{\slBoolVar} expresses (temporary) \emph{exclusive ownership} of \sigVar.
    Further, our logic requires threads to own any signal they are trying to access.
    Specifically, when a thread wants to set \sigProgVar, it must hold a chunk of the form \assSignal{(\sigProgVar, \levVar)}{\slBoolVar}, cf.~\prRealNonSafeSigsSetName.
    The same holds for reading a signal's value, cf.~\prRealNonSafeSigsIsSetName.
    Note that signal chunks are not duplicable and only created upon creation of the signal they refer to.
    Therefore, holding a signal chunk for \sigProgVar indeed guarantees that the holding thread has the exclusive right to access \sigProgVar (while holding the signal chunk).
 
    \paragraph{Synchronization \& Lock Invariants}
    After the main thread creates \sigProgVar, it exclusively owns the signal.
    The main thread can transfer ownership of this resource during forking, cf.~\prMinSigForkName, and thereby allow the forked thread to busy-wait for \sigProgVar.
    This would, however, leave the main thread without any permission to set the signal and thereby discharge its obligation.

    We use mutexes to let multiple threads share ownership of a common set of resources in a synchronized fashion.
    Every mutex is associated with a \emph{lock invariant} \assLockInvVar, an assertion chosen by the proof author that specifies which resources the mutex protects.
    In our example, we want both threads to share \sigProgVar.
    To reflect the fact that the signal's value changes over time, we choose a lock invariant that abstracts over its concrete value.
    We choose
    $\assLockInvVar := \exists \slBoolVar.\ \assSignal{(\sigProgVar, \levVar)}{\slBoolVar}$.
    Let us ignore the chosen signal level \levVar for now.
    Creating the mutex \mutProgVar consumes this lock invariant and binds it to \mutProgVar by creating a mutex chunk \assMutex{(\mutProgVar, \dots)} {\invM}, cf.~\prRealNonSafeSigsNewMutexName.
    Thereby, the main thread loses access to \sigProgVar.
    The only way to regain access is by acquiring \mutProgVar, cf.~\prRealNonSafeSigsAcquireName.
    Once the thread releases \mutProgVar, it again loses access to all resources protected by the mutex, cf.~\prRealNonSafeSigsReleaseName.
    
    \paragraph{Deadlocks}
    We have to ensure that any acquired mutex is eventually released, again.
    Hence, acquiring a mutex spawns a release obligation for this mutex and the only way to discharge this obligation is indeed by releasing it, cf.~\prRealNonSafeSigsAcquireName and \prRealNonSafeSigsReleaseName.
    
    Any attempt to acquire a mutex will block until the mutex becomes available.
    In order to prove that our program terminates, we have to prove that it does not get stuck during an acquisition attempt.
    To prevent wait cycles involving mutexes, we require the proof author to associate every mutex as well (just like signals) with a level \levVar.
    This level can be freely chosen during the mutex' creation, cf.~\prRealNonSafeSigsNewMutexName.
    Mutex chunks therefore have the form \assMutex{(\hlocVar, \levVar)}{\assLockInvVar} where \hlocVar is the heap location the mutex is stored at.
    Their only purpose is to record the level and lock invariant a mutex is associated with.
    Hence, these chunks can be freely duplicated as we will see later.
    Generally, we denote mutex tuples by $\mutVar = (\hlocVar, \levVar)$.
    We only allow to acquire a mutex if its level is lower than the level of each held obligation, cf.~\prRealNonSafeSigsAcquireName.
    This also prevents any thread from attempting to acquire mutexes twice, e.g., 
    $\cmdAcquireMut{\mutProgVar}; \cmdAcquireMut{\mutProgVar}$ or \cmdAwait{\mutProgVar}{\cmdAcquireMut{\mutProgVar}}.

    \paragraph{View Shifts}
    When verifying a program, it can be necessary to reformulate the proof state and to draw semantic conclusions.
    To allow this we introduce a so-called \emph{view shift} relation~\viewShiftSymb~\cite{Jung2018IrisGroundUp}.
    By applying proof rule~\prViewShiftName and \vsSemImpName we can strengthen the precondition and weaken the postcondition.
    In our example, we use this to convert the unset signal chunk into the lock invariant which abstracts over the signal's value, i.e.,
    \viewShift
            {\assSignal{\sigVar}{\slFalse}}
            {\exists \slBoolVar.\ \assSignal{\sigVar}{\slBoolVar}}.

    The logic we present in this work is an intuitionistic separation logic that allows us to drop chunks.
    \footnote{
            This allows a thread to drop its obligations chunk \obs{\obBagVar}.
            Note, however, that by dropping this chunk the thread does not drop its obligations, but only its ability to show what its obligations are.
            In particular the thread would be unable to present an empty obligations chunk upon termination.
    }
    This allows us to simplify the postcondition of our fork proof rule's premise from $\noObs \slStar \assPostVar$ to \noObs, cf.~\prForkName, and drop all unneeded chunks via a semantic implication 
    \viewShift{\noObs \slStar \assPostVar}{\noObs}.
    
    We also allow to clone mutex chunks via view shifts, cf.~\vsRealNonSafeSigsCloneMutName.
    In our example, this is necessary to inform both threads which level and lock invariant mutex \mutProgVar is associated with.
    That is, the main thread clones the mutex chunk \assMutex{\mutVar}{\assLockInvVar} and passes one chunk on when it forks the busy-waiting thread.

    In \S~\ref{sec:Nutshell:SignalErasure} we extend our view shift relation and revisit our interpretation of what a view shift expresses. 
    The full set of rules we use to define \viewShiftSymb is presented in the appendix (cf.\@~Fig.~\ref{appendix:fig:ViewShiftRules}).

    \paragraph{Busy Waiting}
    In the approach presented in this paper, for simplicity we only support busy-waiting loops of the form \cmdAwait{\mutProgVar}{\cmdVar}, which is syntactic sugar for
    \cmdWhileSkip
            {
                    \cmdAcquireMut{\mutProgVar};
                    \cmdLet{\resVar}{\cmdVar}
                            {
                                    \cmdReleaseMut{\mutProgVar};
                                    \neg \resVar
                            }
            }
    where \resVar denotes a fresh variable.~\footnote{
            As we discuss in \S~\ref{sec:ToolSupport}, in the technical report accompanying this paper we present a more general logic that imposes no such syntactic restrictions.
     }
    In each iteration, the loop tries to acquire \mutProgVar, executes \cmdVar, releases \mutProgVar again and lets the result returned by \cmdVar determine whether the loop continues.
    Such loops can fail to terminate for two reasons:
    (i)~Acquiring \mutProgVar can get stuck and
    (ii)~the loop could diverge.
    
    We prevent the loop from getting stuck by requiring \mutProgVar's level to be lower than the level of each held obligation, cf.~\prRealNonSafeSigsAwaitName.
    Further, we enforce termination by requiring the loop to wait for a signal.
    That is, when verifying a busy-waiting loop using our approach, the proof author must choose a fixed signal and prove that this signal remains unset at the end of every non-finishing iteration.
    This way, we can prove that the loop terminates by proving that every signal is eventually set, just as in \S~\ref{sec:Nutshell:ThreadSafePhysicalSignals}.
    And just as before, our logic requires the level of the waited-for signal to be lower than the level of each held obligation.
    
    Acquiring the mutex in every iteration makes the lock invariant available during the verification of the loop body \cmdVar.
    This lock invariant has to be restored at the end of the iteration such that it can be consumed during the mutex's release.
    \prRealNonSafeSigsAwaitName allows for an additional view shift to restore the invariant.
    In our example, we end our busy-waiting loop's non-finishing iterations with the assertion \assSignal{\sigVar}{\slFalse}.
    We use a semantic implication view shift to convert the signal chunk into the mutex invariant 
    $\exists \slBoolVar.\ \assSignal{\sigVar}{\slBoolVar}$.

    \paragraph{Choosing Levels}
    In our example, we have to assign levels to the mutex \mutProgVar and to the signal \sigProgVar.
    Our proof rules for mutex acquisition and busy-waiting impose some restrictions on the levels of the involved mutexes and signals.
    By analysing the corresponding rule applications that occur in our proof, we can derive which constraints our level choice must comply with.
    Our example's verification involves one application of \prRealNonSafeSigsAcquireName and one application of \prRealNonSafeSigsAwaitName:
    (i)~Our main thread tries to acquire \mutProgVar while holding an obligation to set \sigProgVar.
    (ii)~The forked thread busy-waits for \sigProgVar while not holding any obligations.    
    Our assignment of levels must therefore satisfy the single constraint $\mutGetLev{\mutVar} \levLt \sigGetLev{\sigVar}$. 
    So, we choose $\LevelSet = \setOf{0, 1}$,
    $\mutGetLev{\mutVar} = \levM$ and
    $\sigGetLev{\sigVar} = \meLevS$.

    \subsection{Arbitrary Data Structures}\label{sec:Nutshell:AbitraryDataStructures}
    
    The proof rules we introduced in \S~\ref{sec:Nutshell:NonThreadSafePhysicalSignals} allow us to verify programs busy-waiting for arbitrary conditions over arbitrary shared data structures as follows:
    For every condition $C$ the program waits for, the proof author inserts a signal \sigVar into the program.
    They ensure that \sigVar is set at the same time the program establishes $C$ and prove an invariant stating that the signal's value expresses whether $C$ holds.
    Then, the waiting thread can use \sigVar to wait for $C$.
    We illustrate this here for the simplest case of setting a single heap cell in Fig.~\ref{fig:MinEx:Heap:Code}.

    \begin{figure}
            \vspace{-0.7cm}
            \begin{subfigure}[t]{0.45\textwidth}
                    $$
                    \begin{array}[t]{l}
                            \keyword{let}\ \x := \cmdAlloc{0}\ \keyword{in}\\
                            \keyword{let}\ \mutProgVar := \cmdNewMut\ \keyword{in}\\
                            \cmdFork{
                                    \cmdAwait{\mutProgVar}{\cmdHeapLocReadEq{\x}{1}}
                            };
                            \\
                            \cmdAcquireMut{\mutProgVar};
                            \\
                            \cmdAssignToHeap{\x}{1};
                            \\
                            \cmdReleaseMut{\mutProgVar}
                    \end{array}
                    $$
                    \vspace{-0.3cm}
                    \caption{Example program with busy waiting for heap cell \x to be set.}
                    \label{fig:MinEx:Heap:Code}
            \end{subfigure}
            \hspace{0.7cm}
            \begin{subfigure}[t]{0.45\textwidth}
                    $$
                    \begin{array}[t]{l}
                            \keyword{let}\ \x := \cmdAlloc{0}\ \keyword{in}\\
                            \!\!\progProofNew{
                                    \keyword{let}\ \sigProgVar := \cmdNewSignal\ \keyword{in}
                            }\\
                            \keyword{let}\ \mutProgVar := \cmdNewMut\ \keyword{in}
        \vspace{-0.072cm}
                            \\
                            \cmdFork{
                                    \cmdAwait{\mutProgVar}{\cmdHeapLocReadEq{\x}{1}}
                            };
        \vspace{-0.072cm}
                            \\
                            \cmdAcquireMut{\mutProgVar};
                            \\
                            \cmdAssignToHeap{\x}{1};
                            \\
                            \!\!\progProofNew{
                                    \cmdSetSignal{\sigProgVar};
                            }
                            \\
                            \cmdReleaseMut{\mutProgVar}
                    \end{array}
                    $$
                    \vspace{-0.3cm}
                    \caption{
                            Example program~\ref{fig:MinEx:Heap:Code} with additional signal \sigProgVar inserted, marked in \progProofNew{\text{green}}.
                            \sigProgVar and \x are kept in sync.
                    }
                    \label{fig:MinEx:HeapSignal:Code}
            \end{subfigure}
            
            \begin{subfigure}{\textwidth}
                    $$
                    \begin{array}{l c l}
                            \cmdHeapLocReadEq{\expVar}{\nextExpVar} &\ := \
                                    &(\cmdLet{\resVar}
                                            {\cmdReadHeapLoc{\expVar}}
                                            {\resVar = \nextExpVar})
                    \end{array}
                    $$
                    \vspace{-0.4cm}
                    \caption{Syntactic sugar. \resVar free in \nextExpVar.}
            \end{subfigure}

            \caption{Minimal example illustrating busy waiting for condition over heap cell.}
            \vspace{-0.3cm}
    \end{figure}

    The program involves three new non-thread-safe commands:
    (i)~\cmdAlloc{\valVar} for allocating a new heap cell and initializing it with value~\valVar,
    (ii)~\cmdAssignToHeap{\hlocVar}{\valVar} for assigning value~\valVar to heap location~\hlocVar,
    (iii)~\cmdReadHeapLoc{\hlocVar} for reading the value stored in heap location~\hlocVar.
    We use \cmdHeapLocReadEq{\hlocVar}{\valVar} as syntactic sugar for 
    \cmdLet{\resVar}
            {\cmdReadHeapLoc{\expVar}}
            {\resVar = \nextExpVar}.
    
    In our example, the main thread allocates \x, initializes it with the value 0 and protects it using mutex \mutProgVar.
    It forks a new thread busy-waiting for \x to be set.
    Afterwards, the main thread sets \x.
    As explained above, we verify the program by inserting a signal \sigProgVar that reflects whether \x has been set, yet.
    Fig.~\ref{fig:MinEx:HeapSignal:Code} presents the resulting code.
    The main thread creates the signal and sets it when it sets \x.

    \begin{figure}
            \vspace{-0.6cm}
            \begin{subfigure}{\textwidth}
                    $$
                    \begin{array}{l p{0.1cm} l}
                            \progProof{\noObs}
                            \\
                            \keyword{let}\ \x := \cmdAlloc{0}\ \keyword{in}
                            \\
                            \progProof{
                                    \noObs
                                    \slStar
                                    \slPointsTo{\x}{0}
                            }
                            \\
                            \keyword{let}\ \sigProgVar := \cmdNewSignal\ \keyword{in}
                                    &&\proofRuleHint{\prRealNonSafeSigsNewSignalName with $\levVar = \meLevS$}
                            \\
                            \keyword{let}\ \mutProgVar := \cmdNewMut\ \keyword{in}
                                    &&\proofRuleHint{\prRealNonSafeSigsNewMutexName with $\levVar = \levM$}
                            \\
                            \proofDef{
                                    \sigVar := (\sigProgVar, \meLevS),\
                                    \mutVar := (\mutProgVar, \levM)
                            }
                            \\
                            \proofDef{
                                    \invM := \exists \valVar.\ 
                                                    \slPointsTo{\x}{\valVar}
                                                    \slStar
                                                    \assSignal{\sigVar}{\valVar = 1}
                            }
                            \\
                            \progProof{
                                    \obsOf{\sigVar}
                                    \slStar
                                    \assMutex{\mutVar}{\invM}
                                    \slStar
                                    \assMutex{\mutVar}{\invM}
                            }
                            \\
                            \keyword{fork}\ (
                                    \progProof{
                                            \noObs
                                            \slStar
                                            \assMutex{\mutVar}{\invM}
                                    }
                            \\
                            \phantom{\keyword{fork}\ (}\
                                    \keyword{with}\ \mutVar\ \keyword{await}
                                            &&\proofHint{$\mutGetLev{\mutVar}, \sigGetLev{\sigVar} \levObsLt \msEmpty$}
                            \\
                            \phantom{\keyword{fork}\ (}\quad\quad
                                    \progProof{
                                            \obsOf{\mutVar}
                                            \slStar
                                            \invM
                                    }
                            \\
                            \phantom{\keyword{fork}\ (}\quad\quad
                                    \ghostFont{\forall \valVar.}\
                                    \progProof{
                                            \obsOf{\mutVar}
                                            \slStar
                                            \slPointsTo{\x}{\valVar}
                                            \slStar
                                            \assSignal{\sigVar}{\valVar = 1}
                                    }
                            \\
                            \phantom{\keyword{fork}\ (\quad\quad\ghostFont{\forall \valVar.}\ }
                                    \cmdHeapLocReadEq{\x}{1}
                            \\
                            \phantom{\keyword{fork}\ (\quad\quad\ghostFont{\forall \valVar.}\ }
                                    \progProofML[\resVar]{
                                            \obsOf{\mutVar}
                                            \\
                                            \slStar\,
                                            \slIfElseFalseBranchNextLine{\resVar}
                                                    {\invM}
                                                    {
                                                            \slPointsTo{\x}{\valVar}
                                                            \wedge
                                                            \valVar \neq 1
                                                            \slStar
                                                            \assSignal{\sigVar}{\slFalse}
                                                    }
                                    }
                                    \hspace{-5cm}
                            \\
                            \phantom{\keyword{fork}\ (}
                                    \progProof{
                                            \noObs
                                    });
                            \\
                            \progProof{
                                    \obsOf{\sigVar}
                                    \slStar
                                    \assMutex{\mutVar}{\invM}
                            }
                            \\
                            \cmdAcquireMut{\mutProgVar};
                                    &&\proofHint{$\mutGetLev{\mutVar} = \levM < \meLevS = \sigGetLev{\sigVar}$}
                            \\
                            \ghostFont{\forall \valVar.}\
                            \progProof{
                                    \obsOf{\sigVar, \mutVar}
                                    \slStar
                                    \assMutLockedNoFrac{\mutVar}{\invM}
                                    \slStar
                                    \slPointsTo{\x}{\valVar}
                                    \slStar
                                    \assSignal{\sigVar}{\valVar = 1}
                            }
                            \hspace{-5cm}
                            \\
                            \phantom{\ghostFont{\forall \valVar.}\ }
                            \cmdAssignToHeap{\x}{1};
                            \\
                            \phantom{\ghostFont{\forall \valVar.}\ }
                            \progProof{
                                    \obsOf{\sigVar, \mutVar}
                                    \slStar
                                    \assMutLockedNoFrac{\mutVar}{\invM}
                                    \slStar
                                    \slPointsTo{\x}{1}
                                    \slStar
                                    \assSignal{\sigVar}{\valVar = 1}
                            }
                            \hspace{-5cm}
                            \\
                            \phantom{\ghostFont{\forall \valVar.}\ }
                            \cmdSetSignal{\sigProgVar};
                            \\
                            \phantom{\ghostFont{\forall \valVar.}\ }
                            \progProof{
                                    \obsOf{\mutVar}
                                    \slStar
                                    \assMutLockedNoFrac{\mutVar}{\invM}
                                    \slStar
                                    \slPointsTo{\x}{1}
                                    \slStar
                                    \assSignal{\sigVar}{\slTrue}
                            }
                            \\
                            \phantom{\ghostFont{\forall \valVar.}\ }
                            \cmdReleaseMut{\mutProgVar}
                            \\
                            \phantom{\ghostFont{\forall \valVar.}\ }
                            \progProof{\noObs}
                    \end{array}
                    $$
                    \vspace{-0.4cm}
                    \caption{
                            Proof outline for program~\ref{fig:MinEx:HeapSignal:Code}.
                            Applied proof rules marked in \proofRuleHint{purple}.
                            Abbreviations marked in \proofDef{\text{brown}}.
                            General hints marked in \proofHint{red}.
                    }
                    \label{fig:MinEx:HeapSignal:ProofSketch}
            \end{subfigure}

            \begin{subfigure}{\textwidth}
                    \begin{mathpar}
                            \prAlloc
                            \and
                            \prAssignToHeapPaper
                            \and
                            \prHeapCellsReadHeapLoc
                            \and
                            \prExp
                    \end{mathpar}
                    \vspace{-0.4cm}
                    \caption{
                            Proof rules.
                            Evaluation function \evalExp{\cdot}.
                            Rules only used in this section marked with~'''.
                    }
                    \label{fig:MinEx:Heap:ProofRules}
            \end{subfigure}
            \caption{Verifying termination of busy waiting for condition over heap cell.}
            \vspace{-0.5cm}
    \end{figure}

    \paragraph{Heap Cells}
    Verifying this example does not conceptually differ from the example we presented in \S~\ref{sec:Nutshell:NonThreadSafePhysicalSignals}.
    Fig.~\ref{fig:MinEx:Heap:ProofRules} presents the new proof rules we need and Fig.~\ref{fig:MinEx:HeapSignal:ProofSketch} sketches our example's verification.
    As with non-thread-safe signals, we have to prevent multiple threads from trying to access \x at the same time in order to prevent data races.
    For this we use so-called \emph{points-to} chunks~\cite{Reynolds2002SeparationLA, OHearn2001LocalRA}.
    They have the form \slPointsTo{\hlocVar}{\valVar} and express that heap location~\hlocVar stores the value~\valVar.
    When a thread holds such a chunk, it exclusively owns the right to access heap location~\hlocVar.
    
    Heap locations are unique and the only way to create a new points-to chunk is to allocate and initialize a new heap cell via \cmdAlloc{\valVar}, cf.~\prAllocName.
    Hence, there will never be two points-to chunks involving the same heap location.
    In order to read or write a heap cell via \cmdReadHeapLoc{\hlocVar} or \cmdAssignToHeap{\hlocVar}{\expVar}, the acting thread must first acquire possession of the corresponding points-to chunk, cf.~\prAssignToHeapName and~\prHeapCellsReadHeapLocName.
    
    \paragraph{Relating Signals to Conditions}
    In our example, the forked thread busy-waits for \x to be set while our proof rules require us to justify each iteration by showing an unset signal.
    That is, we must prove an invariant stating that the value of \x matches \sigProgVar.
    As this invariant must be shared between both threads, we encode it in the lock invariant:
    $\assLockInvVar := \exists \valVar.\ \slPointsTo{\x}{\valVar} \slStar \assSignal{\sigVar}{\valVar = 1}$.
    This does not only allow both threads to share the heap cell and the signal but it also automatically enforces that they maintain the invariant whenever they acquire and release the mutex.

\subsection{Signal Erasure}\label{sec:Nutshell:SignalErasure}
    
    In the program from Fig.~\ref{fig:MinEx:HeapSignal:Code} signal \sigProgVar is never read and does hence not influence the waiting thread's runtime behaviour.
    Therefore, we can verify the original program presented in Fig.~\ref{fig:MinEx:Heap:Code} by erasing the physical signal and treating it as ghost code.

    \paragraph{Ghost Signals}
    Central aspects of the proof sketch we presented in Fig.~\ref{fig:MinEx:HeapSignal:ProofSketch} are that
    (i)~the main thread was obliged to set \sigProgVar and that
    (ii)~the value of \sigProgVar reflected whether \x was already set.
    \emph{Ghost signals} allow us to keep this information but at the same to remove the physical signals from the code.
    Ghost signals are essentially identical to the physical non-thread-safe signals we used so far.
    However, as ghost resources they cannot influence the program's runtime behaviour.
    They merely carry information we can use during the verification process.
    
    \paragraph{View Shifts Revisited}
    We implement ghost signals by extending our view shift relation.
    In particular, we introduce two new view shift rules: \vsNewSignalName and \vsSetSignalName presented in Fig.~\ref{fig:MinExSignalsErased:ViewShiftRules}.
    The former creates a new unset signal and simultaneously spawns an obligation to set it.
    The latter can be used to set a signal and thereby discharge a corresponding obligation.
    We say that these rules change the \emph{ghost state} and therefore call their application a \emph{ghost proof step}.
    With this extension, a view shift \viewShift{\assPreVar}{\assPostVar} expresses that we can reach postcondition \assPostVar from precondition \assPreVar by
    (i)~drawing semantic conclusions or by
    (ii)~manipulating the ghost state.
    In Fig.~\ref{fig:MinExSignalsErased:ProofSketch} we use ghost signals to verify the program from~\ref{fig:MinEx:Heap:Code}.

    Note that lifting signals to the verification level does not affect the soundness of our approach.
    The argument we presented in \S~\ref{sec:Nutshell:ThreadSafePhysicalSignals} still holds.
    We formalize our logic and provide a formal soundness proof in the appendix and in the technical report~\cite{Reinhard2020GhostSignalsTR}.
    The latter contains a more general version of the presented logic that 
    (i) is not restricted to busy-waiting loops of the form \cmdAwait{\mutProgVar}{\cmdVar} and that
    (ii)~is easier to integrate into existing tools like VeriFast~\cite{Jacobs2011Verifast}, as explained in \S~\ref{sec:ToolSupport}.

    \begin{figure}
            \vspace{-1.3cm}
            \begin{subfigure}{\textwidth}
                    $$
                    \begin{array}{l p{0.1cm} l}
                            \progProof{\noObs}
                            \\
                            \keyword{let}\ \x := \cmdAlloc{0}\ \keyword{in}
                            \\
                            \progProof{\noObs \slStar \slPointsTo{\x}{0}}
                            \\
                            \proofHint{new\_ghost\_signal};
                                    &&\proofRuleHint{\vsNewSignalNoBoolName with $\levVar = \meLevS$.}
                            \\
                            \progProof{
                                    \exists \sigProgVar.\
                                    \obsOf{(\sigProgVar, \meLevS)}
                                    \slStar
                                    \slPointsTo{\x}{0}
                                    \slStar
                                    \assSignal{(\sigProgVar, \meLevS)}{\slFalse}
                            }
                                    &&\proofDef{\sigVar := (\sigProgVar, \meLevS)}
                            \\
                            \ghostFont{\forall \sigProgVar.}\
                            \progProof{
                                    \obsOf{\sigVar}
                                    \slStar
                                    \slPointsTo{\x}{0}
                                    \slStar
                                    \assSignal{\sigVar}{\slFalse}
                            }
                                    &&\proofDef{\invM :=  \exists \valVar.\
                                            \slPointsTo{\x}{\valVar}
                                            \slStar
                                            \assSignal{\sigVar}{\valVar = 1}
                                    }
                            \\
                            \phantom{\ghostFont{\forall \meIdS.}\ }
                            \keyword{let}\ \mutProgVar := \cmdNewMut\ \keyword{in}
                                    &&\proofRuleHint{\prRealNonSafeSigsNewMutexName with $\levVar = \levM$}
                            \\
                            \phantom{\ghostFont{\forall \meIdS.}\ }
                            \progProofML{
                                    \obsOf{\sigVar}
                                    \slStar
                                    \assMutex{(\mutProgVar, \levM)}{\invM}
                                    \\
                                    \slStar\,
                                    \assMutex{(\mutProgVar, \levM)}{\invM}
                            }
                                    &&\proofDef{\mutVar := (\mutProgVar, \levM)}
                            \\
                            \phantom{\ghostFont{\forall \meIdS.}\ }
                            \keyword{fork}\ (
                                    \progProof{
                                            \noObs
                                            \slStar
                                            \assMutex{\mutVar}{\invM}
                                    }
                            \\
                            \phantom{\ghostFont{\forall \meIdS.}\ \keyword{fork}\ (}
                                    \keyword{with}\ \mutVar\ \keyword{await}
                                            &&\proofHint{$\mutGetLev{\mutVar}, \sigGetLev{\sigVar} \levObsLt \msEmpty$}
                            \\
                            \phantom{\ghostFont{\forall \meIdS.}\ \keyword{fork}\ (}\quad
                                    \progProof{
                                            \obsOf{\mutVar}
                                            \slStar
                                            \invM
                                    }
                            \\
                            \phantom{\ghostFont{\forall \meIdS.}\ \keyword{fork}\ (}\quad
                                    \ghostFont{\forall \valVar.}\
                                    \progProof{
                                            \obsOf{\mutVar}
                                            \slStar
                                            \slPointsTo{\x}{\valVar}
                                            \slStar
                                            \assSignal{\sigVar}{\valVar = 1}
                                    }
                                    \hspace{-3cm}
                            \\
                            \phantom{\ghostFont{\forall \meIdS.}\ \keyword{fork}\ (\quad \ghostFont{\forall \valVar.\ }}
                                    \cmdHeapLocReadEq{\x}{1}
                            \\
                            \phantom{\ghostFont{\forall \meIdS.}\ \keyword{fork}\ (\quad \ghostFont{\forall \valVar.\ }}
                                    \progProofML[\resVar]{
                                            \obsOf{\mutVar}
                                            \,\slStar
                                            \\
                                            \slIfElseFalseBranchNextLine{\resVar}
                                                    {\invM}
                                                    {
                                                            \slPointsTo{\x}{\valVar}
                                                            \wedge
                                                            \valVar \neq 1
                                                            \slStar
                                                            \assSignal{\sigVar}{\slFalse}
                                                    }
                                    }
                                    \hspace{-3cm}
                            \\
                            \phantom{\ghostFont{\forall \meIdS.}\ \keyword{fork}\ (}
                                    \progProof{
                                            \noObs
                                    });
                            \\
                            \phantom{\ghostFont{\forall \meIdS.}\ }
                            \progProof{
                                    \obsOf{\sigVar}
                                    \slStar
                                    \assMutex{\mutVar}{\invM}
                            }
                            \\
                            \phantom{\ghostFont{\forall \meIdS.}\ }
                            \cmdAcquireMut{\mutProgVar};
                                    &&\proofHint{$\mutGetLev{\mutVar} = \levM < \meLevS = \sigGetLev{\sigVar}$}
                            \\
                            \phantom{\ghostFont{\forall \meIdS.}\ }
                            \ghostFont{\forall \valVar.}\
                            \progProofML{
                                    \obsOf{\sigVar, \mutVar}
                                    \slStar
                                    \assMutLockedNoFrac{\mutVar}{\invM}
                                    \\
                                    \slStar\,
                                    \slPointsTo{\x}{\valVar}
                                    \slStar
                                    \assSignal{\sigVar}{\valVar = 1}
                            }
                            \\
                            \phantom{\ghostFont{\forall \meIdS.}\ \ghostFont{\forall \valVar.}\ }
                            \cmdAssignToHeap{\x}{1};
                            \\
                            \phantom{\ghostFont{\forall \meIdS.}\ \ghostFont{\forall \valVar.}\ }
                            \proofHint{set\_ghost\_signal(\sigVar);}
                            \\
                            \phantom{\ghostFont{\forall \meIdS.}\ \ghostFont{\forall \valVar.}\ }
                            \progProofML{
                                    \obsOf{\mutVar}
                                    \slStar
                                    \assMutLockedNoFrac{\mutVar}{\invM}
                                    \\
                                    \slStar\,
                                    \slPointsTo{\x}{1}
                                    \slStar
                                    \assSignal{\sigVar}{\slTrue}
                            }
                            \\
                            \phantom{\ghostFont{\forall \meIdS.}\ \ghostFont{\forall \valVar.}\ }
                            \cmdReleaseMut{\mutProgVar}
                            \\
                            \phantom{\ghostFont{\forall \meIdS.}\ \ghostFont{\forall \valVar.}\ }
                            \progProof{\noObs}
                    \end{array}
                    $$
                    \caption{
                            Proof outline for the program presented in Fig.~\ref{fig:MinEx:Heap:Code}.
                            Auxiliary commands hinting at view shifts and general hints marked in \proofHint{red}.
                            Applied proof and view shift rules marked in \proofRuleHint{purple}.
                            Abbreviations marked in \proofDef{\text{brown}}.
                    }
                    \label{fig:MinExSignalsErased:ProofSketch}
            \end{subfigure}
            
            \begin{subfigure}{\textwidth}
                    \begin{mathpar}
                            \vsNewSignal
                            \and\hspace{-0.05cm}
                            \vsSetSignal
                    \end{mathpar}
                    \caption{
                            Proof rules.
                    }
                    \label{fig:MinExSignalsErased:ViewShiftRules}
            \end{subfigure}
            \caption{Verifying termination with ghost signals.}
            \vspace{-1.7cm}
    \end{figure}

%% file: CAV_paper_sections/extended_version/realisticExample.tex
	\vspace{-0.3cm}
    To demonstrate the expressiveness of the presented verification approach, we verified the termination of the program presented in Fig.~\ref{fig:RealisticExample}.
    It involves two threads, a consumer and a producer, communicating via a shared bounded FIFO with a maximal capacity of 10.
    The producer enqueues numbers 100, \dots, 1 into the FIFO and the consumer dequeues those.
    Whenever the queue is full, the producer busy-waits for the consumer to dequeue an element.
    Likewise, whenever the queue is empty, the consumer busy-waits for the producer to enqueue the next element.
    Each thread's finishing depends on the other thread's productivity.
    This is, however, no cyclic dependency.
    For instance, in order to prove that the producer eventually pushes number $i$ into the queue, we only need to rely on the consumer to pop $i+10$.
    A similar property holds for the consumer.
    
	\vspace{-0.3cm}

    \begin{figure}
    \vspace{-0.65cm}
    \begin{subfigure}{\textwidth}
    {
            \NewDocumentCommand{\fifo}{}{\progVar{fifo_{10}}}
            \NewDocumentCommand{\fifoLoaded}{}{\progVar{f}}
            \NewDocumentCommand{\m}{}{\progVar{m}}

            \NewDocumentCommand{\counter}{m}{\progVar{c_{#1}}}
            
                    \NewDocumentCommand{\producerID}{}{p}
                    \NewDocumentCommand{\pc}{}{\counter{\producerID}}
                    \NewDocumentCommand{\pcLoaded}{}{\progVar{c}}

                    \NewDocumentCommand{\consumerID}{}{c}
                    \NewDocumentCommand{\cc}{}{\counter{\consumerID}}
                    \NewDocumentCommand{\ccLoaded}{}{\progVar{c}}
                    
                            \NewDocumentCommand{\idGen}{m m}{\metaVar{\idVar_\text{#1}^{#2}}}
                            
                            \NewDocumentCommand{\idPush}{m}{\idGen{push}{#1}}
                            \NewDocumentCommand{\levPush}{m}{\metaVar{\levVar_\text{push}^{#1}}}
                            
                            \NewDocumentCommand{\idPop}{m}{\idGen{pop}{#1}}
                            \NewDocumentCommand{\levPop}{m}{\metaVar{\levVar_\text{pop}^{#1}}}
                            
                            \NewDocumentCommand{\sigPush}{m}{\metaVar{\sigVar_\text{push}^{#1}}}
                            \NewDocumentCommand{\sigPop}{m}{\metaVar{\sigVar_\text{pop}^{#1}}}
            
            $$
            \begin{array}{l l}
                \proofHint{alloc\_ghost\_signal\_IDs(\idPop{i}, \idPush{i})
                        \ \ \text{for}\ \ 
                        $1 \leq i \leq 100$;
                }
                \hspace{-5cm}
                \\
                \proofDef{
                        \levPop{i} := 102 - i,\ \
                        \levPush{i} := 101 - i,\ \
                        \sigVar_x^i := (\idVar_x^i, \levVar_x^i)\ \
                        \text{for}\ \ 1 \leq i \leq 100
                }
                \hspace{-5cm}
                \\
                \proofHint{init\_ghost\_signals(\sigPop{100}, \sigPush{100});}
                \\
                \progProof{\obsOf{\sigPop{100}, \sigPush{100}} \slStar \dots}
                \\
                \keyword{let}\ \fifo := \cmdAlloc{\lstNil}\ \keyword{in}\
                \keyword{let}\ \mutProgVar := \cmdNewMut\ \keyword{in}
                \hspace{-3cm}\\
                \keyword{let}\ \pc := \cmdAlloc{100}\ \keyword{in}\
                \keyword{let}\ \cc := \cmdAlloc{100}\ \keyword{in}\
                \hspace{-3cm}
                \\
                \keyword{fork}\ (
                        \keyword{while}\ (
                                &\hspace{-1.122cm}\proofHint{\pc decreases in each iteration.}
                \\\phantom{\keyword{fork}\ (}\quad
                                \keyword{with}\ \mutProgVar\ \keyword{await}\ (
                                        &\hspace{-1.122cm}\lowlightText{\text{Busy-wait for \fifo not being full.}}
                \\\phantom{\keyword{fork}\ (}\quad\quad
                                        \progProof{\obsOf{\sigPush{\pc}, (\mutProgVar, 0)} \slStar \dots}
                                        &\hspace{-1.122cm}\lowlightText{\rightarrow\text{Wait for consumer to pop.}}
                \\\phantom{\keyword{fork}\ (}\quad\quad
                                        \keyword{let}\ \fifoLoaded := \cmdReadHeapLoc{\fifo}\ \keyword{in}
                \\\phantom{\keyword{fork}\ (}\quad\quad
                                        \keyword{if}\ \lstSize{\fifoLoaded} < 10\ \keyword{then}\ (
                                                &\hspace{-1.122cm}\lowlightText{\text{If \fifo not full, push next element.}}
                \\\phantom{\keyword{fork}\ (}\quad\quad\quad
                                                \keyword{let}\ \pcLoaded := \cmdReadHeapLoc{\pc}\ \keyword{in}\
                                                \cmdAssignToHeap
                                                        {\fifo}
                                                        {\lstAppend
                                                                {\fifoLoaded}
                                                                {\lstOf{\pcLoaded}}
                                                        };\
                                                \cmdAssignToHeap{\pc}{\pcLoaded - 1};
                                                \hspace{-3cm}
                \\\phantom{\keyword{fork}\ (}\quad\quad\quad
                                                \proofHint{set\_ghost\_signal(\sigPush{\pcLoaded});}
                \\\phantom{\keyword{fork}\ (}\quad\quad\quad
                                                \proofHint{\keyword{if} $\pcLoaded-1 \neq 0$ \keyword{then}
                                                        init\_ghost\_signal(\sigPush{\pcLoaded-1})});
                \\\phantom{\keyword{fork}\ (}\quad\quad
                                \lstSize{\fifoLoaded} \neq 10
                                ); 
                                        &\hspace{-1.122cm}\proofHint{\keyword{if} \lstSize{\fifoLoaded} = 10 \keyword{then} wait for \sigPop{\pc+10}}
                \\\phantom{\keyword{fork}\ (}\quad
                                \cmdHeapLocReadNeq{\pc}{0})
                                        &\hspace{-1.122cm}\proofHint{$
                                                \levPop{\pc+10} = 92 - \pc
                                                <
                                                101 - \pc = \levPush{\pc}
                                        $}
                \\\phantom{\keyword{fork}\ (}
                        \keyword{do}\ \cmdSkip); 
                \\
                \keyword{while}\ (
                                &\hspace{-1.122cm}\proofHint{\cc decreases in each iteration.}
                \\\quad
                        \keyword{with}\ \mutProgVar\ \keyword{await}\ (
                                        &\hspace{-1.122cm}\lowlightText{\text{Busy-wait for \fifo not being empty.}}
                \\\quad\quad
                                \progProof{\obsOf{\sigPop{\cc}, (\mutProgVar, 0)} \slStar \dots}
                                        &\hspace{-1.122cm}\lowlightText{\rightarrow\text{Wait for producer to push.}}
                \\\quad\quad
                                \keyword{let}\ \fifoLoaded := \cmdReadHeapLoc{\fifo}\ \keyword{in}
                \\\quad\quad
                                \keyword{if}\ \lstSize{\fifoLoaded} > 0\ \keyword{then}\ (
                                        &\hspace{-1.122cm}\lowlightText{\text{If \fifo not empty, pop next element.}}
                \\\quad\quad\quad
                                        \keyword{let}\ \ccLoaded := \cmdReadHeapLoc{\cc}\ \keyword{in}\
                                        \cmdAssignToHeap
                                                {\fifo}
                                                {\lstTail{\fifoLoaded}};\
                                        \cmdAssignToHeap{\cc}{\ccLoaded - 1};
                                        \hspace{-3cm}
                \\\quad\quad\quad
                                        \proofHint{set\_ghost\_signal(\sigPop{\ccLoaded});}
                \\\quad\quad\quad
                                        \proofHint{\keyword{if} $\ccLoaded-1 \neq 0$ \keyword{then}
                                                init\_ghost\_signal(\sigPop{\ccLoaded-1})});
                \\\quad\quad
                                \lstSize{\fifoLoaded} > 0
                        ); 
                                        &\hspace{-1.122cm}\proofHint{\keyword{if} \lstSize{\fifoLoaded} = 0 \keyword{then} wait for \sigPush{\cc}}
                \\\quad
                        \cmdHeapLocReadNeq{\cc}{0})
                                        &\hspace{-1.122cm}\proofHint{$
                                                \levPush{\cc} = 101 - \cc
                                                <
                                                102 - \cc = \levPush{\cc}
                                        $}
                \\
                \keyword{do}\ \cmdSkip); 
            \end{array}
            $$
            
            \vspace{-0.4cm}
            \caption[Producer-consumer program with bounded FIFO.]
            {
                    Example program with two threads communicating via a shared bounded FIFO with maximal size 10.
                    Auxiliary commands hinting at view shifts 
                    and general hints marked in \proofHint{red}.
                    Abbreviations marked in \proofDef{\text{brown}}.
                    Hints on proof state marked in \ghostFont{blue}.
            }
            \label{fig:RealisticExample}
    }
    \end{subfigure}

	\vspace{-0.2cm}

    \begin{subfigure}{\textwidth}
            \begin{mathpar}
                    \vsAllocSigID
                    \and
                    \vsSigInit
            \end{mathpar}
            \vspace{-0.5cm}
            \caption{Fine-grained view shift rules for signal creation.}
            \label{fig:ViewShiftRules--fineGrainedSignalCreation}
    \end{subfigure}
    
    \vspace{-0.2cm}
    \caption{Realistic example program.}
    \vspace{-1.5cm}
    \end{figure}

{
    \NewDocumentCommand{\m}{}{\progVar{m}}
    \NewDocumentCommand{\idPush}{m}{\metaVar{\idVar_\text{push}^{#1}}}
    \NewDocumentCommand{\levPush}{m}{\metaVar{\levVar_\text{push}^{#1}}}
                            
    \NewDocumentCommand{\idPop}{m}{\metaVar{\idVar_\text{pop}^{#1}}}
    \NewDocumentCommand{\levPop}{m}{\metaVar{\levVar_\text{pop}^{#1}}}
                            
    \NewDocumentCommand{\sigPush}{m}{\metaVar{\sigVar_\text{push}^{#1}}}
    \NewDocumentCommand{\sigPop}{m}{\metaVar{\sigVar_\text{pop}^{#1}}}

    \paragraph{Fine-Tuning Signal Creation}
    To simplify complex proofs involving many signals we refine the process of creating a new ghost signal.
    For simplicity, we combined the allocation of a new signal ID and its association with a level and a boolean in one step.
    For some proofs, such as the one we outline in this section, it can be helpful to fix the IDs of all signals that will be created throughout the proof already at the beginning.
    To realize this, we replace view shift rule \vsNewSignalName by the rules presented in Fig.~\ref{fig:ViewShiftRules--fineGrainedSignalCreation} and adapt our signal chunks accordingly.
    With these more fine-grained view shifts, we start by allocating a signal ID, cf.\@ \vsAllocSigIDName.
    Thereby we obtain an \emph{uninitialized} signal \sigUninit{\idVar} that is not associated with any level or boolean, yet.
    Also, allocating a signal ID does not create any obligation because threads can only wait for \emph{initialized} (and unset) signals.
    When we initialize a signal, we bind its already allocated ID to a level of our choice and associate the signal with \slFalse, cf.~\vsSigInitName.
    This creates an obligation to set the signal.
    

    \paragraph{Loops \& Signals}
    In our program, both threads have a local counter initially set to 100 and run a nested loop.
    The outer loops are controlled by their thread's counter, which is decreased in each iteration until it reaches 0 and the loop stops.
    For such loops, we introduce a conventional proof rule for total correctness of loops, cf.\@ Fig.~\ref{appendix:fig:ProofRulesPartOne} in the appendix.
    Verifying termination of the inner loops is a bit more tricky and requires the use of ghost signals.

    So far, we had to fix a single signal for the verification of every \keyword{await} loop.
    We can relax this restriction to considering a finite set of signals the loop may wait for, cf.~\prAwaitGenName presented in Fig.~\ref{appendix:fig:ProofRulesPartOne} in the appendix.
    Apart from being a generalisation, this rule does not differ from \prRealNonSafeSigsAwaitName introduced in~\S~\ref{sec:Nutshell:NonThreadSafePhysicalSignals}.

    Initially, we allocate 200 signal IDs $\idPush{100}, \dots, \idPush{1}, \idPop{100}, \dots, \idPop{1}$.
    We are going to ensure that always at most one push signal and at most one pop signal are initialized and unset.
    The producer and consumer are going to hold the obligation for the push and pop signal, respectively.
    The producer will hold the obligation for \sigPush{i} while $i$ is the next number to be pushed into the FIFO and it will set \sigPush{i} when it pushes the number $i$ into the FIFO.
    Meanwhile, the consumer will use \sigPush{i} to wait for the number $i$ to arrive in the queue when it is empty.
    Similarly, the consumer will hold the obligation for \sigPop{i} while number $i$ is the next number to be popped from the FIFO and will set \sigPop{i} when it pops the number $i$.
    The producer uses \sigPop{i} to wait for the consumer to pop $i$ from the queue when it is full.
    At any time, we let the mutex \mutProgVar protect the two active signals and thereby make them accessible to both threads.
    
    \paragraph{Choosing the Levels}
    Note that we ignored the levels so far.
    The producer and the consumer both acquire the mutex while holding an obligation for a signal.
    Hence, we choose $\LevelSet = \N$, $\levOf{\m} = 0$ and $\levOf{\sigVar} > 0$ for every signal \sigVar.
    Both threads will justify iterations of their respective \keyword{await} loop by using an unset signal at the end of such an iteration.
    Our proof rules allow us to ignore the mutex obligation during this step.
    Hence, the mutex level does not interfere with the level of the unset signal.
    Whenever the queue is full, the producer waits for the consumer to pop an element and whenever the queue is empty, the consumer waits for the producer to push.
    That is, the producer waits for \sigPop{i+10} while holding an obligation for \sigPush{i} and the consumer waits for \sigPush{i} while holding an obligation for \sigPop{i}.
    So, we have to choose the signal levels such that 
    $\levOf{\sigPop{i+10}} < \levOf{\sigPush{i}}$
    and
    $\levOf{\sigPush{i}} < \levOf{\sigPop{i}}$
    hold.
    We solve this by choosing
    $\levOf{\sigPop{i}} = 102 - i$
    and
    $\levOf{\sigPush{i}} = 101 - i$.

    
    \paragraph{Verifying Termination}
    This setup suffices to verify the example program.
    Via the lock invariant, each thread has access to both active signals.
    Whenever the producer pushes a number $i$ into the queue, it sets \sigPush{i} which discharges the held obligation and decreases its counter.
    Afterwards, if $i > 1$, it uses the uninitialized signal chunk  \sigUninit{\idPush{i-1}} to initialize $\sigPush{i-1} = (\idPush{i-1}, 101 - (i-1))$ and replaces \sigPush{i} in the lock invariant by \sigPush{i-1} before it releases the lock.
    If $i = 1$, the counter reached 0 and the loop ends.
    In this case, the producer holds no obligation.
    The consumer behaves similarly.
    Since we proved that each thread discharged all its obligations, we proved that the program terminates.
    Fig.~\ref{fig:RealisticExample} illustrates the most important proof steps.
    We present the program's verification in full detail in the appendix in \S~\ref{appendix:sec:VerificationOfRealisticExample} and in the technical report~\cite{Reinhard2020GhostSignalsTR}.
    Furthermore, we encoded~\cite{ArtifactJacobs2020VerifastGhostSignalConsumerProducerBoundedFifo} the proof in VeriFast~\cite{Jacobs2011Verifast}.
    
    The number of threads in this program is fixed.
    However, our approach also supports the verification of programs where the number of threads is not even statically bounded.
    In the appendix in \S~\ref{appendix:sec:caseStudy:unbounded} we present and verify such a program.
    It involves $N$ producer and $N$ consumer threads that communicate via a shared buffer of size 1, for a random number $N > 0$ determined during runtime.
	\myComment
	{
		Similarly, we can generalize our proof to support an arbitrary number of elements sent through the buffer as well as an arbitrary buffer size or an unbounded buffer.
	}
}

%% file: CAV_paper_sections/extended_version/fineGrainedConcurrency.tex
    Our approach can be used to verify busy-waiting concurrent objects with respect to abstract specifications. 
    For example, we have verified~\cite{ArtifactJacobs2020VerifastCLHLock} the CLH lock \cite{herlihy-multiprocessor} against a specification that is very similar to our proof rules for built-in mutexes shown in Fig.~\ref{fig:MinEx:RealNonThreadSafeSignals:ProofRules}. 
    The main difference is that it is slightly more abstract: when a lock is initialized, it is associated with a \emph{bounded infinite set} of levels rather than with a single particular level.
    (To make this possible, an appropriate universe of levels should be used, such as the set of lists of natural numbers, ordered lexicographically.) 
    To acquire a lock, the levels of the obligations held by the thread must be above the elements of the set; the new obligation's level is an element of the set.

%% file: CAV_paper_sections/extended_version/toolSupport.tex
        We have extended the VeriFast tool~\cite{verifast2104} for separation logic-based modular verification of C and Java programs so that it supports verifying termination of busy-waiting C or Java programs.
        When verifying termination, VeriFast consumes a \emph{call permission} at each recursive call or loop iteration.
        In the technical report~\cite{Reinhard2020GhostSignalsTR} we define a generalised version of our logic that instead of providing a special proof rule for busy-waiting loops, provides \emph{wait permissions} and a \emph{wait view shift}. 
        A call permission of a \emph{degree} $\delta$ can be turned into a wait permission of a degree $\delta' < \delta$ for a given signal \sigVar. 
        A wait view shift for an unset signal \sigVar for which a wait permission of degree $\delta$ exists produces a call permission of degree $\delta$, which can be used to fuel a busy-waiting loop.
        When busy-waiting for some signal \sigVar, we can generate new permissions to justify each iteration as long as \sigVar remains unset.
        
        VeriFast allows threads to freely exchange permissions.
        This is useful to verify termination of non-blocking algorithms involving compare-and-swap loops~\cite{Jacobs2018ModularTerminationVerification}. 
        However, we must be careful to prevent self-fueling busy-waiting loops.
        Hence, we restrict where a permission can be consumed based on the \emph{thread phase} it was created in.
        The main thread's initial phase is $\epsilon$.
        When a thread in phase $p$ forks a new thread, its phase changes to $p.\mathsf{Forker}$ and the new thread starts in phase $p.\mathsf{Forkee}$.
        We allow a thread in phase $p$ to consume a permission only if it was produced in an \emph{ancestor thread phase} $p' \fpLeq p$.

        The only change we had to make to VeriFast's symbolic execution engine was to enforce the thread phase rule. 
        We encoded the other aspects of the logic simply as axioms in a \emph{trusted header file}. 
        We used this tool support to verify the bounded FIFO (\S~\ref{sec:RealisticExample}) and the CLH lock (\S~\ref{sec:FineGrainedConcurrency}).
		The bounded FIFO proof~\cite{ArtifactJacobs2020VerifastGhostSignalConsumerProducerBoundedFifo} contains 160 lines of proof annotations for 37 lines of code (an annotation overhead of 435\%) and takes 0.08s to verify.
		The CLH lock proof~\cite{ArtifactJacobs2020VerifastCLHLock} contains 343 lines of annotations for 49 lines of code (an overhead of 700\%) and takes 0.1s to verify.
        \vspace{-0.2cm}

%% file: CAV_paper_sections/extended_version/higherOrder.tex
    The logic we presented in this paper does not support higher-order features such as assertions that quantify over assertions, or storing assertions in the (logical) heap as the values of ghost cells.
    While we did not need such features to carry out our example proofs, they are generally useful to verify higher-order program modules against abstract specifications. 
    The typical way to support such features in a program logic is by applying \emph{step indexing} \cite{appel-indexed,BIRKEDAL20104102}, where the domain of logical heaps is indexed by the number of execution steps left in the (partial) program trace under consideration. 
    Assertions stored in a logical heap at index $n+1$ talk about logical heaps at index $n$; i.e., they are meaningful only \emph{later}, after at least one more execution step has been performed.
    
    It follows that such logics apply directly only to \emph{partial} correctness properties. 
    Fortunately, we can reduce a termination property to a safety property by writing our program in a programming language \emph{instrumented} with run-time checks that guarantee termination. 
    Specifically, we can write our program in a programming language that fulfils the following criteria:
    It tracks signals, obligations and permissions at run time and
    has constructs for signal creation, waiting and setting a signal.
    The \keyword{fork} command takes as an extra operand the list of obligations to be transferred to the new thread (and the other constructs similarly take sufficient operands to eliminate any need for angelic choice).
    Programs get stuck when these constructs' preconditions are not satisfied, such as when a thread waits for a signal while holding the obligation for that signal.
    We can then use a step-indexing-based higher-order logic such as Iris \cite{Jung2018IrisGroundUp} to verify that our program never gets stuck. 
    Once we established this, we know none of the instrumentation has any effect and can be safely \emph{erased} from the program.

%% file: CAV_paper_sections/extended_version/relatedAndFutureWork.tex
\NewDocumentCommand{\TadaLive}{}{TaDA Live\xspace}
\NewDocumentCommand{\Lili}{}{LiLi\xspace}

   	In recent work~\cite{Reinhard2020AbruptExitPaper} we propose a separation logic to verify termination of programs where threads busy-wait to be abruptly terminated. 
   	We generalize this work to support busy-waiting for arbitrary conditions.
    
    In~\cite{Jacobs2018ModularTerminationVerification} we propose an approach based on \emph{call permissions} to verify termination of single- and multithreaded programs that involve loops and recursion.
    However, that work does not consider busy waiting loops.
    In the technical report, we present a generalised logic that uses call permissions and allows busy waiting to be implemented using arbitrary looping and/or recursion.
    Furthermore, the use of call permissions allowed us to encode our case studies in our VeriFast tool which also uses call permissions for termination verification.

    Liang and Feng~\cite{Liang2016LiliAPL, Liang2017LiliProgressOC} propose \Lili, a separation logic to verify liveness of blocking constructs implemented via busy-waiting.
    In contrast to our verification approach, theirs is based on the idea of contextual refinement.
    In their approach, client code involving calls of blocking methods of the concurrent object is verified by first applying the contextual refinement result to replace these calls by code involving primitive blocking operations and then verifying the resulting client code using some other approach. 
    In contrast, specifications in our approach are regular Hoare-style triples and proofs are regular Hoare-style proofs.

    In~\cite{Jacobs2020IOLiveness} we propose a Hoare logic to verify liveness properties of the I/O behaviour of programs that do not perform busy waiting. 
    By combining that approach with the one we proposed in this paper, we expect to be able to verify I/O liveness of realistic concurrent programs involving both I/O and busy waiting, such as a server where one thread receives requests and enqueues them into a bounded FIFO, and another one dequeues them and responds.
    To support this claim, we encoded the combined logic in VeriFast and verified a simple server application where the receiver and responder thread communicate via a shared buffer~\cite{ArtifactJacobs2020VerifastIOLivenessServer}.

%% file: CAV_paper_sections/extended_version/conclusion.tex
    We propose what is to the best of our knowledge the first separation logic for verifying termination of programs with busy waiting.
    We offer a soundness proof of the system of the paper in the appendix, and of a more general system in the technical report~\cite{Reinhard2020GhostSignalsTR}.
    Further, we demonstrated its usability by verifying a realistic example.
    We encoded our logic and the realistic example in VeriFast~\cite{ArtifactJacobs2020VerifastGhostSignalConsumerProducerBoundedFifo} and used this encoding also to verify the CLH lock~\cite{ArtifactJacobs2020VerifastCLHLock}.
    Moreover, we expect that our approach can be integrated into other existing concurrent separation logics such as Iris~\cite{Jung2018IrisGroundUp}.

%% file: CAV_paper_sections/extended_version/appendix.tex
In this appendix, we formalize our approach and prove its soundness.
We start in \S~\ref{appendix:sec:General} by defining the notations we use.
In \S~\ref{appendix:sec:Language} we define the simple programming language we consider in this work.
In \S~\ref{appendix:sec:Logic} we define our logic and in particular provide a full overview of all the proof and view shift rules we use and state a soundness theorem.
In \S~\ref{appendix:sec:Soundness} we prove our approach sound.
Our proof utilizes an annotated semantics that keeps track of ghost resources and thereby connects the runtime and the verification level.
We define it in \S~\ref{appendix:sec:Soundness:AnnotatedSemantics}.
Afterwards, in \S~\ref{appendix:sec:Soundness:ModelRelation} we define a model for Hoare triples and prove that every triple that we can derive with our proof rules also holds in our model.
In \S~\ref{appendix:sec:Soundness:Proof} we use this model and the annotated semantics to prove our soundness theorem.
At last, in \S~\ref{appendix:sec:Case Studies} we present two verification case studies.
In \S~\ref{appendix:sec:VerificationOfRealisticExample} we provide a detailed proof outline for the realistic example presented in \S~\ref{sec:RealisticExample}.
In \S~\ref{appendix:sec:caseStudy:unbounded} we present and verify a similar program but with a statically unbounded number of producer and consumer threads.

\section{General}\label{appendix:sec:General}
    \begin{definition}[Projections]
            For any Cartesian product $C = \prod_{i \in I} A_i$ and any index $k \in I$, we denote the $k^\text{th}$ projection by
            $\tupleProj[C]{k}: \prod_{i \in I} A_i \rightarrow A_k$.
            We define
            $$
                    \tupleProj[C]{k}[(a_i)_{i\in I}] 
                    \  := \ 
                    a_k.
            $$
            In case the domain $C$ is clear from the context, we write \tupleProj{k} instead of \tupleProj[C]{k}.
    \end{definition}

    \begin{definition}[Disjoint Union]
            Let $A, B$ be sets.
            We define their disjoint union as
            $$
                    A \disjCup B \ := \ A \cup B
            $$
            if $A \cap B = \emptyset$
            and leave it undefined otherwise.
    \end{definition}

    \begin{definition}[Bags]\label{def:Bags}
            For any set $X$ we define the set of bags \BagsOf{X} and the set of finite bags \FinBagsOf{X} over $X$ as
            $$
            \begin{array}{l l l}
                    \BagsOf{X} \ &:= \
                            &X  \rightarrow  \N,\\
                    \FinBagsOf{X} \ &:= \
                            &\setOf{
                                    B \in \BagsOf{X} 
                                    \ \ | \ \
                                    \setOf{x \in B \ | \ B(x) > 0}\
                                    \text{finite}
                            }.
            \end{array}
            $$
            We define union and subtraction of bags as
            $$
            \begin{array}{l c l}
                    (B_1 \msCup B_2)(x)  \ &:= \
                            &B_1(x) + B_2(x),\\
                    (B_1 \setminus B_2)(x) \ &:= \
                            &\max(0,\ B_1(x) - B_2(x)).
            \end{array}
            $$
            For finite bags where the domain is clear from the context, we define the following set-like notation:
            $$
            \begin{array}{l l l}
                    \msEmpty \ &:= \
                            &x \ \mapsto\ 0,
                    \\
                    \multiset{x} \ &:= \
                            &\left\{
                                    \begin{array}{l c l l}
                                            x &\mapsto & 1\\
                                            y &\mapsto &0 &\text{for}\ y\neq x,
                                    \end{array}
                            \right.
                    \\\vspace{-0.3cm}
                    \\
                    \multiset{x_1, \dots, x_n} \ &:= \
                            &\displaystyle\msBigCup_{i = 1}^n \, \multiset{x_i}.
            \end{array}
            $$
            Further, we define for any $\natVar \in \N$ the notation:
            $$
                    n \cdot \multiset{x}
                    \ := \
                    \multiset{\underbrace{x, \dots, x}_\text{\natVar times}}.
            $$
            We define the following set-like notations for element and subset relationship:
            $$
            \begin{array}{l l l}
                    x \in B \ &\Leftrightarrow \ & B(x) > 0,
                    \\
                    B_1 \subseteq B_2 \ &\Leftrightarrow  \
                            &\forall x \in B_1.\ B_1(x) \leq B_2(x) ,
                    \\
                    B_1 \subset B_2 \ \, &\Leftrightarrow \ \,
                            &\exists C \subseteq B_1.\ 
                                    C \neq \msEmpty
                                    \ \wedge\
                                    B_1 = B_2 \setminus C.
            \end{array}
            $$
            For any bag $B \in \BagsOf{X}$ and predicate $P \subseteq X$ we define the following refinement:
            $$
                    \multiset{x \in B \ | \ P(x)}
                    \ := \ 
                    \left\{
                            \begin{array}{l c l l}
                                    x &\mapsto &B(x)  \ &\text{if}\ P(x),\\
                                    x &\mapsto &0 &\text{otherwise}.
                            \end{array}
                    \right.
            $$
    \end{definition}

    \begin{definition}[Disjoint Union]
            Let $A, B$ be sets.
            We define their disjoint union as
            $$
                    A \disjCup B \ := \ A \cup B
            $$
            if $A \cap B = \emptyset$
            and leave it undefined otherwise.
    \end{definition}

\section{Language}\label{appendix:sec:Language}
        In this section and the next, we present our approach formally. 
        In this section, we define the programming language; 
        in \S~\ref{appendix:sec:Logic} we define the proof system. 

        We consider a simple imperative programming language with support for multi-threading, shared memory and synchronization via mutexes.
        For its definition we assume 
        (i)~an infinite set of program variables $\varVar \in \VarSet$,
        (ii)~an infinite set of heap locations $\hlocVar \in \HeapLocSet$,
        (iii)~a set of values $\valVar \in \ValueSet$ which includes heap locations, booleans $\B = \setOf{\slTrue, \slFalse}$ and the unit value \valUnit,
        (iv)~a set of operations $\opVar \in \OpSet$
        and
        (v)~an infinite, totally ordered and well-founded set of thread IDs $\tidVar \in \ThreadIDSet$.
        
        \begin{definition}[Syntax]\label{def:Syntax}
                We define the sets of commands \CmdSet and expressions \ExpSet according to the syntax presented in Fig.~\ref{appendix:fig:def:Syntax}.
        \end{definition}

    \begin{figure}
            \begin{subfigure}{\textwidth}
                    $$
                    \begin{array}{c c c c c}
                            \varVar \in \VarSet
                            : \text{Program variables}
                            \quad
                            &\hlocVar \in \HeapLocSet
                            : \text{Heap locations}
                            \quad
                            \\
                            \valVar \in \ValueSet
                                    \supseteq \setOf{\valUnit} \cup \B \cup \HeapLocSet
                            : \text{Values}
                            \quad
                            &\opVar \in \OpSet
                            : \text{Operations}
                    \end{array}
                    $$
                    \caption{
                            Assumed sets and variables. 
                            \VarSet and \HeapLocSet infinite.
                    }
                    \label{appendix:fig:AssumedSetsForLanguage}
            \end{subfigure}
            
            \begin{subfigure}{\textwidth}
                    $$
                    \begin{array}{l}
                    \begin{array}{l c l l}        
                            \expVar \in \ExpSet &\ ::= \ \
                                    & \varVar \alt
                                            \valVar \alt
                                            \expEq{\expVar}{\expVar} \alt
                                            \expNot{\expVar} \alt
                                            \opVar(\expListVar)
                            \\
                            \cmdVar \in \CmdSet  &\ ::= \ \
                                    &\expVar \alt
                                            \cmdWhile{\cmdVar}{\cmdSkip} \alt
                                            \cmdFork{\cmdVar}\alt
                                            \cmdLet{\varVar}{\cmdVar}{\cmdVar} \alt
                                            \cmdIf{\cmdVar}{\cmdVar} \alt\\
                                    &&\cmdAlloc{\expVar} \alt
                                            \cmdReadHeapLoc{\expVar} \alt
                                            \cmdAssignToHeap{\expVar}{\expVar} \alt
                                            \cmdNewMut \alt
                                            \cmdAcquireMut{\expVar} \alt
                                            \cmdReleaseMut{\expVar}
                    \end{array}
                    \\
                    \begin{array}{l c l l}
                            \evalCtxt \in \EvalCtxtSet &\ ::=\ \
                                    &\cmdIf{\evalCtxtHole}{\cmdVar} \alt
                                            \cmdLet{\varVar}{\evalCtxtHole}{\cmdVar}
                    \end{array}
                    \end{array}
                    $$
                    \vspace{-0.3cm}
                    \caption{Expressions and commands.}
                    \label{appendix:fig:def:ExpCmdSyntax}
            \end{subfigure}
            
            \begin{subfigure}{\textwidth}
                    $$
                    \begin{array}{l c l }
                            \cmdAwait{\expVar}{\cmdVar}\ \ &:= \ \
                                    &(\cmdWhile
                                            {
                                                    \cmdAcquireMut{\expVar};
                                                    \cmdLet
                                                            {\resVar}
                                                            {\cmdVar}
                                                            {\cmdReleaseMut{\expVar}; \neg\resVar}
                                            }
                                            {\cmdSkip})
                            \\
                            \cmdVar\ ;\, \nextCmdVar &:=\ \
                                    &(\cmdLet{\resVar}{\cmdVar}{\nextCmdVar})
                            \quad\quad\quad\quad
                            \\
                            \expVar \neq \nextExpVar &:=\ \
                                    &\neg(\expVar = \expVar)
                            \\
                            \nextNextExpVar(\cmdReadHeapLoc{\expVar}) &:=\ \
                                    &(\cmdLet{\resVar}{\cmdReadHeapLoc{\expVar}}{\nextNextExpVar(\resVar)})
                    \end{array}
                    $$
                    \vspace{-0.3cm}
                    \caption{Syntactic sugar. $\nextNextExpVar(y)$ expression with free variable $y$. $\resVar \in \VarSet$ not free in \expVar, \nextNextExpVar, \nextCmdVar.}
                    \label{appendix:fig:def:SyntacticSugar}
            \end{subfigure}
            
            \caption{Syntax.}
            \label{appendix:fig:def:Syntax}
    \end{figure}
        
        The language contains standard sets of pure expressions \ExpSet and (potentially) side-effectful commands \CmdSet.
        The latter includes commands for heap allocation and manipulation, forking and loops.
        We define \emph{physical heaps}~\cite{Jacobs2018ModularTerminationVerification} (as opposed to \emph{logical heaps}~\cite{Jacobs2018ModularTerminationVerification} presented in the next section) as a finite set of \emph{physical resource chunks}.
        A points-to chunk \slPointsTo{\hlocVar}{\valVar} expresses that heap location~\hlocVar points to value \valVar~\cite{Reynolds2002SeparationLA, Jacobs2018ModularTerminationVerification}.
        Moreover, we have chunks to represent unlocked and locked mutexes.

        \begin{definition}[Physical Resources]
                We define the set of physical resources \PhysResSet syntactically as follows:
                $$
                \begin{array}{l c l}
                        \presVar \in \PhysResSet &\ ::=\
                                &\slPointsTo{\hlocVar}{\valVar} \alt
                                    \presUnlocked{\hlocVar}\alt
                                    \presLocked{\hlocVar}
                \end{array}
                $$
                $$
                \begin{array}{c p{0.5cm} c}
                        \hlocVar \in \HeapLocSet
                        &&\valVar \in \ValueSet
                \end{array}
                $$
        \end{definition}

        \begin{definition}[Physical Heaps]
                We define the set of physical heaps as 
                $$\PhysHeapSet\ :=\ \finPowerSetOf{\PhysResSet}$$
                and the function 
                $\phGetLocs: \PhysHeapSet \rightarrow \finPowerSetOf{\HeapLocSet}$
                mapping physical heaps to the sets of allocated heap locations as
                $$
                \begin{array}{l c l l}
                        \phGetLocs(\pheapVar) &\ :=\
                                &\setOf{
                                        \hlocVar \in \HeapLocSet \ \ |\
                                        &\presUnlocked{\hlocVar} \in \pheapVar\ \vee\
                                                \presLocked{\hlocVar} \in \pheapVar\ \vee\\
                                        &&&\exists \valVar \in \ValueSet.\
                                                \slPointsTo{\hlocVar}{\valVar} \in \pheapVar
                                }.
                \end{array}
                $$
                We denote physical heaps by \pheapVar.
        \end{definition}

        We represent a program state by a physical heap and a \emph{thread pool}, which we define as a partial function mapping a finite number of thread IDs to threads.
        Thread IDs are unique and never reused.
        Hence, we represent running threads by commands and terminated ones by \thTerminated instead of removing threads from the pool.
       For the following definition, remember that we assume the set of thread IDs \ThreadIDSet to be infinite and well-founded.

        \begin{definition}[Thread Pools]\label{appendix:def:ThreadPools}
                We define the set of thread pools \ThreadPoolSet as the set of finite partial functions mapping thread IDs to threads:
                $$
                        \ThreadPoolSet \ := \
                                \finPartialFunSet
                                        {\ThreadIDSet}
                                        {(\CmdSet \cup \setOf{\thTerminated})}.
                $$
                The symbol \thTerminated represents a terminated thread.
                We denote thread pools by \tpVar, thread IDs by \tidVar and the empty thread pool by \tpEmpty, i.e.,
                $$
                \begin{array}{l}
                        \tpEmpty \ : \ \ThreadIDSet \finPartialFunArrow  (\CmdSet \cup \setOf{\thTerminated}),
                        \\
                        \defDom{\tpEmpty} \, = \, \emptyset.
                \end{array}
                $$
                We define the operation
                $\tpAddSymb : \ThreadPoolSet \times \setOf{C \subset \CmdSet \ |\ \cardinalityOf{C} \leq 1} \rightarrow \ThreadPoolSet$
                as follows:
                $$
                \begin{array}{l c l}
                        \tpAdd{\tpVar}{\emptyset} &\ := \
                                &\tpVar,\\
                        \tpAdd{\tpVar}{\setOf{\cmdVar}} &\ :=\
                                &\tpUpdate{\tpVar}{\newTidVar}{\cmdVar}
                                \quad\text{for}\quad
                                \newTidVar := 
                                        \min(\ThreadIDSet \setminus \defDom{\tpVar}).
                \end{array}
                $$
        \end{definition}

        We define the operational semantics of our language in terms of two small-step reduction relations: \stRedStepSymb for single threads and \tpRedStepSymb for thread pools.
        Since expressions are pure and their evaluation is deterministic we identify closed expressions with their ascribed value.
        (i)~\stRedStep
                        {\cmdVar}
                        [\nextPheapVar]
                        {\nextCmdVar}
                        [\ftsVar]
                expresses that heap \pheapVar and command \cmdVar are reduced in a single step to \nextPheapVar and \nextCmdVar and that this thread forks a set of threads~\ftsVar.
                This set is either empty or a singleton as no step forks more than one thread.
        (ii)~\tpRedStep
                        {\tpVar}
                        {\nextTpVar}
                expresses that heap \pheapVar and thread pool \tpVar are reduced in a single step to \nextPheapVar and \nextTpVar.
                ID \tidVar identifies the thread reduced in this step.

        \begin{definition}[Evaluation of Closed Expressions]\label{def:EvaluationOfClosedExpressions}
                We define a partial evaluation function $\evalExp{\cdot}: \ExpSet \partialFunArrow \ValueSet$ on expressions by recursion on the structure of expressions as follows:
                $$
                \begin{array}{l l l l}
                        \evalExp{\valVar} &:= &\valVar 
                                &\text{if}\quad \valVar \in \ValueSet,\\
                        \evalExp{\expEq{\expVar}{\nextExpVar}} &:= &\valTrue
                                &\text{if}\quad
                                        \evalExp{\expVar} = \evalExp{\nextExpVar} \neq \bot,\\
                        \evalExp{\expEq{\expVar}{\nextExpVar}} \ &:= \ &\valFalse\ \
                                &\text{if}\quad
                                        \evalExp{\expVar} \neq \evalExp{\nextExpVar}\ \wedge\ 
                                        \evalExp{\expVar} \neq \bot\ \wedge\
                                        \evalExp{\nextExpVar} \neq \bot,\\
                        \evalExp{\expNot{\expVar}} &:= &\valFalse
                                &\text{if}\quad
                                        \evalExp{\expVar} = \valTrue,\\
                        \evalExp{\expNot{\expVar}} &:= &\valTrue
                                &\text{if}\quad
                                        \evalExp{\expVar} = \valFalse,\\
                        \evalExp{\expVar} &:= &\bot &\text{otherwise}.
                \end{array}
                $$
                We identify closed expressions \expVar with their ascribed value \evalExp{\expVar}.
        \end{definition}
        
        \begin{definition}[Evaluation Context]
                We define the set of evaluation contexts \EvalCtxtSet syntactically as follows:
                $$
                \begin{array}{l c l}
                        \evalCtxt \in \EvalCtxtSet &\ ::=\
                                &\cmdIf{\evalCtxtHole}{\cmdVar} \alt
                                        \cmdLet{\varVar}{\evalCtxtHole}{\cmdVar}
                \end{array}
                $$
                $$
                \begin{array}{c p{0.5cm} c}
                        \cmdVar \in \CmdSet
                        &&\varVar \in \VarSet
                \end{array}
                $$
                For any $\cmdVar \in \CmdSet$ and $\evalCtxt \in \EvalCtxtSet$, we define 
                $\evalCtxt[\cmdVar] := \subst{\evalCtxt}{\evalCtxtHole}{\cmdVar}$.
        \end{definition}
        
        Note that for every $\cmdVar \in \CmdSet, \evalCtxt \in \EvalCtxtSet$, we have $\evalCtxt[\cmdVar] \in \CmdSet$.

        \begin{definition}[Single Thread Reduction Relation]
                We define a reduction relation \stRedStepSymb for single threads according to the rules presented in Figure~\ref{appendix:fig:SingleThreadReductionRules}.
                A reduction step has the form
                $$\stRedStep
                        {\cmdVar}
                        [\nextPheapVar]
                        {\nextCmdVar}
                        [\ftsVar]
                $$
                for a set of forked threads 
                $\ftsVar \subset \CmdSet$ with 
                $\cardinalityOf{\ftsVar} \leq 1$.
                
                For simplicity of notation, we omit \ftsVar if it is clear from the context that no thread is forked and $\ftsVar = \emptyset$.
        \end{definition}
        
        \begin{definition}[Thread Pool Reduction Relation]
                We define a thread pool reduction relation \tpRedStepSymb according to the rules presented in Figure~\ref{appendix:fig:ThreadPoolReductionRules}.
                A reduction step has the form
                $$\tpRedStep{\tpVar}{\nextTpVar}.$$
        \end{definition}

        \begin{figure}
                \begin{mathpar}
                        \stRedEvalCtxt
                        \and
                        \stRedFork
                        \and
                        \stRedIfTrue
                        \and
                        \stRedIfFalse
                        \and
                        \stRedLet
                        \and
                        \stRedWhile
                        \and
                        \stRedAlloc
                        \and
                        \stRedReadHeapLoc
                        \and
                        \stRedAssign
                        \and
                        \stRedNewMutex
                        \and
                        \stRedAcquire
                        \and
                        \stRedRelease
                \end{mathpar}

        \caption{Single thread reduction rules.}
        \label{appendix:fig:SingleThreadReductionRules}
        \end{figure}
        
        \begin{figure}
                \begin{mathpar}
                        \tpRedLift
                        \and
                        \tpRedTerm
                \end{mathpar}
                \caption{Thread pool reduction rules}
                \label{appendix:fig:ThreadPoolReductionRules}
        \end{figure}

        As thread scheduling is non-deterministic, so is our thread pool reduction relation \tpRedStepSymb.
        Consider the minimal example we presented in Fig.~\ref{fig:MinEx:Heap:Code} in \S~\ref{sec:Nutshell:AbitraryDataStructures} to illustrate busy waiting for a shared memory flag to be set.
        It does not terminate if the main thread is never scheduled after the new thread was forked.
        Hence, our verification approach relies on the assumption of fair scheduling.
        That is, we assume that every thread is always eventually scheduled while it remains running.
        Further, we represent program executions by sequences of reduction steps.
        As we primarily consider infinite sequences in this work, we define reduction sequences to be infinite to simplify our terminology.

        \begin{definition}[Reduction Sequence]
                Let $(\pheapVar_i)_{i\in \N}$ and $(\tpVar_i)_{i \in \N}$ be infinite sequences of physical heaps and thread pools, respectively.
                We call \phtpRSeq a \emph{reduction sequence} if
                there exists a sequence of thread IDs $(\tidVar_i)_{i \in \N}$ such that 
                $
                \tpRedStep
                        [\pheapVar[i]]
                        {\tpVar[i]}
                        [\tidVar_i]
                        [\pheapVar[i+1]]
                        {\tpVar[i+1]}
                $
                holds for every $i \in \N$.
        \end{definition}

        \begin{definition}[Fairness]\label{def:Fairness}
                We call a reduction sequence \phtpRSeq \emph{fair} iff for all $i \in \N$ and $\tidVar \in \defDom{\tpVar[i]}$ with $\tpVar[i](\tidVar) \neq \thTerminated$ there exists some $k \geq i$ with
                $$
                \tpRedStep
                        [\pheapVar[k]]{\tpVar[k]}
                        [\tidVar]
                        [\pheapVar[k+1]]{\tpVar[k+1]}.
                $$
        \end{definition}

\section{Logic}\label{appendix:sec:Logic}
    In this section we formalize the logic we sketched in \S~\ref{sec:ApproachInANutshell}.
    For the definition we assume 
    (i)~an infinite set of ghost signal IDs $\idVar \in \IDSet$
    and
    (ii)~an infinite, partially ordered and well-founded set of levels $\levVar \in \LevelSet$.
    We denote the level order relation by \levLt.

    \begin{definition}[Fractions]
            We define the set of fractions as
            $$\FracSet \ := \
                        \setOf{ \fracVar \in \Q \ \ | \ \ 0 < \fracVar \leq 1}
            $$
            and denote fractions by \fracVar.
    \end{definition}
    
    \begin{definition}[Obligations, Signals \& Mutexes]
            We define the set of obligations \ObSet and signals \SigSetVar as
            $$
            \begin{array}{l c l}
                    \ObSet &\ :=\  &(\HeapLocSet \cup \IDSet) \times \LevelSet,
                    \\
                    \SignalSet &\ :=\ &\IDSet \times \LevelSet.
            \end{array}
            $$
            We denote bags of obligations by $\obBagVar \in \BagsOf{\ObSet}$, signals by $\sigVar \in \SignalSet$ and mutexes by $\mutVar \in \HeapLocSet \times \LevelSet$.
            For convenience of notation, we define the following selector functions:
            $$
            \begin{array}{l c l p{0.5cm} l}
                    \sigGetID{(\idVar, \slWildcard)} &\ :=\ &\idVar 
                            &&\text{for signals},
                    \\
                    \mutGetLoc{(\hlocVar, \slWildcard)} &\ :=\ &\hlocVar
                            &&\text{for mutexes},
                    \\
                    \sigGetLev{(\slWildcard, \levVar)} &\ :=\ &\levVar
                            &&\text{for signals, mutexes and obligations}.
            \end{array}
            $$
    \end{definition}

        \begin{definition}[Assertions]
                We define the set of \emph{assertions} \AssertionSet according to the syntax presented in Figure~\ref{appendix:fig:def:Assertions}.\footnote{
                        That is, we define \AssertionSet as the least fixpoint of $F$ where 
                        $F(\assSetVar) = 
                                \setOf{\slTrue, \slFalse}
                                \cup
                                \setOf{\neg \assVar \ | \
                                                \assVar \in \assSetVar}
                                \cup
                                \setOf{\assVar_1 \wedge \assVar_2 \ | \ 
                                                \assVar_1, \assVar_2 \in \assSetVar} 
                                \cup
                                \dots
                                \cup
                                \setOf{\bigvee \oAssSetVar \ | \
                                                \oAssSetVar \subseteq \assSetVar}
                                \cup
                                \dots
                        $.
                        Since $F$ is a monotonic function over a complete lattice, it has a least fixpoint according to the Knaster-Tarski theorem~\cite{Tarski1955KnasterTarskiTheorem}.
                }
                We omit the index set $I$ in quantifications when its choice becomes clear from the context and write 
                $\exists \assIndexVar.\, \assFam[\assIndexVar]$
                and
                $\forall \assIndexVar.\, \assFam[\assIndexVar]$
                instead of                
                $\exists \assIndexVar \in \assIndexSetVar.\, \assFam[\assIndexVar]$
                and
                $\forall \assIndexVar \in \assIndexSetVar.\, \assFam[\assIndexVar]$,
                respectively.
        \end{definition}

    \begin{figure}
            \begin{subfigure}{\textwidth}
                    $$
                    \begin{array}{c p{0.5cm} c}
                            \assSetVar \subseteq \AssertionSet
                            &
                            &\text{Index set}\ \assIndexSetVar
                    \end{array}
                    $$
                    $$
                    \begin{array}{l}
                    \begin{array}{l c l}
                            \assVar \in \AssertionSet
                                    &\ ::= \
                                    &\slTrue \alt
                                            \slFalse \alt
                                            \neg \assVar \alt
                                            \assVar \wedge \assVar \alt
                                            \assVar \vee \assVar \alt
                                            \assVar \slStar \assVar \alt
                                            \slPointsTo[\fracVar]{\hlocVar}{\valVar} \alt
                                            \bigvee \assSetVar
                                    \\
                                    &&\alt 
                                            \assMutUninit{\hlocVar} \alt
                                            \assMutex
                                                    [\fracVar]
                                                    {(\hlocVar, \levVar)}
                                                    {\assVar} 
                                            \alt
                                            \assMutLocked
                                                    {(\hlocVar, \levVar)}
                                                    {\assVar}
                                                    {\fracVar}
                                    \\
                                    &&\alt
                                            \assSignal{(\idVar, \levVar)}{\slBoolVar}
                                            \alt
                                            \obs{\obBagVar}
                    \end{array}
                    \end{array}
                    $$
                    \caption{
                            Assertion syntax.
                            We omit quantification domain \assIndexSetVar when it is clear from the context.
                    }
                    \label{appendix:fig:def:AssertionSyntax}
            \end{subfigure}
            
            \begin{subfigure}{\textwidth}
                    $$
                    \begin{array}{l c l p{0.65cm} l c l}
                            \assVar_1 \rightarrow \assVar_2
                                    \ \ &:= \ \
                                    &\neg\assVar_1 \vee \assVar_2
                            &
                            &\assVar_1 \leftrightarrow \assVar_2
                                    \ \ &:= \ \
                                    &(\assVar_1 \rightarrow \assVar_2) \wedge
                                        (\assVar_2 \rightarrow \assVar_1)
                            \\
                            \exists \assIndexVar \in \assIndexSetVar.\ \assFam[\assIndexVar]
                                    \ \ &:= \ \
                                    &\bigvee \setOf{\assFam[\assIndexVar] \ | \ \assIndexVar \in \assIndexSetVar}
                            &
                            &\forall \assIndexVar \in \assIndexSetVar.\ \assFam[\assIndexVar] 
                                    \ \ &:=\ \
                                    &\neg\exists \assIndexVar \in \assIndexSetVar.\, \neg \assFam[\assIndexVar]
                    \end{array}
                    $$
                    \caption{Syntactic sugar.}
            \end{subfigure}
            
            \caption{Assertions.}
            \label{appendix:fig:def:Assertions}
    \end{figure}

    In \S~\ref{appendix:sec:Language}, we used physical resources and heaps to model a program's state.
    We use assertions to capture which fraction of a physical resource and which ghost resources a thread owns.
    Therefore, we have to interpret assertions in an extended notion of state.
    We define \emph{logical heaps} and \emph{logical resources}~\cite{Jacobs2018ModularTerminationVerification} which correspond to physical ones but additionally encompass ghost resources and ownership.
    For instance, logical resources include signal chunks and initialized mutex chunks are associated with a lock invariant.
    Rather than being a set of resources, logical heaps map logical resources to fractions.
    This allows us to express which portion of a resource a thread owns.

    \begin{definition}[Logical Resources]
            We define the set of logical resources \LogResSet syntactically as follows:
            $$
            \begin{array}{l c l}
                    \lresVar \in \LogResSet &\ ::=\
                            &\slPointsTo{\hlocVar}{\valVar} \alt
                                    \lresSignal{(\idVar, \levVar)}{\slBoolVar} \alt
                                    \lresUninit{\hlocVar} \alt
                                    \lresMutex{(\hlocVar, \levVar)}{\assVar}
                    \\
                            &&\!\!\alt
                                    \lresLocked{(\hlocVar, \levVar)}{\assVar}{\fracVar} \alt
                                    \lresObs{\obBagVar}
            \end{array}
            $$
            Further, we define the function
            $\lresGetHLocs: \LogResSet \rightarrow \HeapLocSet$
            mapping logical resources to their respective (either empty or singleton) set of involved heap locations as
            $$
            \begin{array}{l c c l}
                    \lresGetHLocs[\slPointsTo{\hlocVar}{\valVar}] 
                            \ \ &:= \ \
                            &\setOf{\hlocVar},
                    \\
                    \lresGetHLocs[\lresUninit{\hlocVar}]
                            \ \ &:= \ \
                            &\setOf{\hlocVar},
                    \\
                    \lresGetHLocs[\lresMutex{(\hlocVar, \levVar)}{\assVar}]
                            \ \ &:= \ \
                            &\setOf{\hlocVar},
                    \\
                    \lresGetHLocs[\lresLocked{(\hlocVar, \levVar)}{\assVar}{\fracVar}]
                            \ \ &:= \ \
                            &\setOf{\hlocVar},
                    \\
                    \lresGetHLocs[\slWildcard]
                            \ \ &:= \ \
                            &\emptyset
                            &\text{otherwise}.
            \end{array}
            $$
    \end{definition}

    \begin{definition}[Logical Heaps]
            We define the set of logical heaps as
            $$\lheapVar \in \LogHeapSet 
                    \ := \
                    \LogResSet \rightarrow \setOf{q \in \Q~|~ q \geq 0}.
            $$
            We define the empty logical heap \lhEmpty as the constant zero function
            $$
                    \lhEmpty : \lresVar \mapsto 0.
            $$
            We denote logical heaps by \lheapVar, point-wise addition by $+$ and multiplication with non-negative rationals by $\cdot$, i.e.,
            $$
            \begin{array}{l c l l}
                    (\lheapVar_1 + \lheapVar_2)(\lresVar) &\ :=\
                            &\lheapVar_1(\lresVar) + \lheapVar_2(\lresVar),\\
                    (q \cdot \lheapVar)(\lresVar) &\ :=\
                            &q \cdot (\lheapVar(\lresVar))
            \end{array}
            $$
            for $q\in \Q$ with $q\geq 0$.
            We give $\cdot$ a higher precedence than $+$.
            For convenience of notation we represent logical heaps containing finitely many resources by sets of resources as follows
            $$
            \begin{array}{l c l}
                    \lheapOf{\lresVar[1],\, \dots,\, \lresVar[n]}
                            &\ :=\
                            &\left\{
                            \begin{array}{l l l}
                                    \lresVar[i] &\mapsto 1\\
                                    x &\mapsto 0 &\text{if}\ x \not\in \setOf{\lresVar[1], \dots, \lresVar[n]}.
                            \end{array}
                            \right.
            \end{array}
            $$
    \end{definition}
    
    \begin{definition}[Logical Heap Predicates]
            Let \lheapVar be a logical heap.
            We call \lheapVar \emph{complete} and write \lhComplete{\lheapVar} if it contains exactly one obligations chunk, i.e.,
            if there exists a bag of obligations \obBagVar with
            $\lheapVar(\lresObs{\obBagVar} ) = 1$
            and if there does not exist any bag of obligations $\obBagVar'$ with 
            $\obBagVar \neq \obBagVar'$ and
            $\lheapVar(\lresObs{\obBagVar'}) > 0$.
            
            We call \lheapVar \emph{finite} and write \lhFinite{\lheapVar} if 
            it contains only finitely many resources, i.e.,
            if the set 
            \setOf{\lresVar \in \LogResSet \ \ | \ \ \lheapVar(\lresVar) > 0}
            is finite.
            
            We call \lheapVar \emph{consistent} and write \lhConsistent{\lheapVar} if 
            (i)~it contains only full obligations chunks, i.e.,
            if
            $$
            \begin{array}{l l l}
                    \lheapVar(\lresObs{\obBagVar}) 
                            &\in &\N
            \end{array}
            $$
            holds for all $\obBagVar \in \BagsOf{\ObSet}$
            and if
            (ii)~heap locations are unique in \lheapVar, i.e.,
            if there are no $\lresVar[1], \lresVar[2] \in \LogResSet$ with 
            $\lresVar[1] \neq \lresVar[2]$,
            $\lheapVar(\lresVar[1]) > 0$,
            $\lheapVar(\lresVar[2]) > 0$ 
            and with
            $\lresGetHLocs[\lresVar[1]] \cap \lresGetHLocs[\lresVar[2]] \neq \emptyset$.
    \end{definition}

    We interpret assertions in terms of a model relation.
    \assModels{\lheapVar}{\assVar} expresses that assertion \assVar holds with respect to logical heap \lheapVar.
    Further, we define the view shift relation~\viewShiftSymb and the proof relation~\htProvesSymb we sketched in \S~\ref{sec:ApproachInANutshell}.

    \begin{definition}[Assertion Model Relation]
            We define a model relation 
            $\assModelsSymb \,\subset \, \LogHeapSet \times \AssertionSet$
            for assertions by recursion on the structure of assertions according to the rules presented in Figure~\ref{appendix:fig:def:AssertionModelRelation}.
            We write \assModels{\lheapVar}{\assVar} to express that logical heap \lheapVar models assertion \assVar and
            \assNotModels{\lheapVar}{\assVar} to express that \assModels{\lheapVar}{\assVar} does not hold.
    \end{definition}

    \begin{figure}
                    $$
                    \begin{array}{l c l}
                            \assModels{\lheapVar}{\slTrue}\\
                            \assNotModels{\lheapVar}{\slFalse}\\
                            \assModels{\lheapVar}{\neg \assVar}
                                    &\text{if}
                                    &\assNotModels{\lheapVar}{\assVar}\\
                            \assModels{\lheapVar}{\assVar_1 \wedge \assVar_2}
                                    &\text{if}
                                    &\assModels{\lheapVar}{\assVar_1}
                                            \wedge
                                            \assModels{\lheapVar}{\assVar_2}\\
                            \assModels{\lheapVar}{\assVar_1 \vee \assVar_2}
                                    &\text{if}
                                    &\assModels{\lheapVar}{\assVar_1}
                                            \vee
                                            \assModels{\lheapVar}{\assVar_2}\\
                            \assModels{\lheapVar}{\assVar_1 \slStar \assVar_2}
                                    &\text{if}
                                    &\exists \lheapVar[1], \lheapVar[2] \in \LogHeapSet.\
                                            \lheapVar = \lheapVar[1] + \lheapVar[2]
                                            \wedge
                                            \assModels{\lheapVar[1]}{\assVar_1}
                                            \wedge
                                            \assModels{\lheapVar[2]}{\assVar_2}\\
                            \assModels
                                {\lheapVar}
                                {\slPointsTo[\fracVar]{\hlocVar}{\valVar}}
                                    &\text{if}
                                    & \lheapVar(\slPointsTo{\hlocVar}{\valVar}) \geq \fracVar\\
                            \assModels
                                {\lheapVar}
                                {\bigvee \assSetVar}
                                    &\text{if}
                                    &\exists \assVar \in \assSetVar.\
                                            \assModels
                                                    {\lheapVar}
                                                    {\assVar}\\
                            \assModels
                                {\lheapVar}
                                {\assMutUninit[\fracVar]{\hlocVar}}
                                    &\text{if}
                                    &\lheapVar(\lresUninit{\hlocVar}) \geq \fracVar\\
                            \assModels
                                {\lheapVar}
                                {\assMutex[\fracVar]{\mutVar}{\assLockInvVar}}
                                    &\text{if}
                                    &\lheapVar(\lresMutex{\mutVar}{\assLockInvVar}) \geq \fracVar\\
                            \assModels
                                {\lheapVar}
                                {\assMutLocked[\fracVar]{\mutVar}{\assLockInvVar}{\fracVar_u}}
                                    &\text{if}
                                    &\lheapVar(\lresLocked{\mutVar}{\assLockInvVar}{\fracVar_u}) 
                                            \geq \fracVar\\
                            \assModels
                                {\lheapVar}
                                {\assSignal[\fracVar]{\sigVar}{\slBoolVar}}
                                    &\text{if}
                                    &\lheapVar(\lresSignal{\sigVar}{\slBoolVar}) \geq \fracVar\\
                            \assModels
                                {\lheapVar}
                                {\obs{\obBagVar}}
                                    &\text{if}
                                    &\lheapVar(\lresObs{\obBagVar}) \geq 1
                    \end{array}
                    $$
                
                    \caption{
                            Assertion model relation.
                            \assNotModels{\lheapVar}{\assVar} expresses that
                            \assModels{\lheapVar}{\assVar} does not hold.
                    }
                    \label{appendix:fig:def:AssertionModelRelation}
    \end{figure}

    \begin{definition}
            Let $\levVar \in \LevelSet$, $\obBagVar \in \BagsOf{\ObSet}$.
            We define $\levObsLt\, \subset \LevelSet \times \ObSet$ as
            $$
                    \levVar \levObsLt \obBagVar
                    \ \Longleftrightarrow \
                    \forall \obVar \in \obBagVar.\
                            \levVar \levLt \sigGetLev{\obVar}.
            $$
    \end{definition}

    \begin{definition}[View Shift]
            We define a view shift relation 
            $\viewShiftSymb \ \subset\ \AssertionSet \times \AssertionSet$
            according to the rules presented in Fig.~\ref{appendix:fig:ViewShiftRules}.
    \end{definition}

    \begin{definition}[Proof Relation]
            We define a proof relation
            $\htProvesSymb \ \subset \ \AssertionSet \times \CmdSet \times (\ValueSet \rightarrow \AssertionSet)$
            according to the rules presented in Fig.~\ref{appendix:fig:ProofRulesPartOne} and \ref{appendix:fig:ProofRulesPartTwo}.
            We state the provability of a Hoare triple in the form of\, \htProves{\assPreVar}{\cmdVar}[\resVar]{\assPostVar(\resVar)} where \resVar captures the value returned by \cmdVar.
            To simplify the notation, we omit the result value if it is clear from the context or irrelevant.
    \end{definition}

    \begin{figure}
            \begin{mathpar}
                    \vsSemImp
                    \and
                    \vsTrans
                    \and
                    \vsOr
                    \and
                    \vsNewSignal
                    \and
                    \vsSetSignal
                    \and
                    \vsMutInit
                    \and
                    \vsGhostLoop
            \end{mathpar}
            
            \caption{View shift rules.}
            \label{appendix:fig:ViewShiftRules}
    \end{figure}

    \begin{figure}
\vspace{-0.9cm}
            \begin{subfigure}{\textwidth}
                    \begin{mathpar}
                            \prFrame
                            \and
                            \prSimpleViewShiftPaper
                            \and
                            \prExp
                            \and
                            \prExists
                            \and
                            \prForkPaperWithResult
                    \end{mathpar}
                    \subcaption{Basic Proof Rules.}
            \end{subfigure}

            \begin{subfigure}{\textwidth}
                    \begin{mathpar}
                            \prIf
                            \and
                            \prAwaitGen
                            \and
                            \prWhileDecStrict
                            \and
                            \prLet
                    \end{mathpar}
                    \subcaption{Control Structures.}
            \end{subfigure}

            \caption{Proof rules (part 1).}
            \label{appendix:fig:ProofRulesPartOne}
    \end{figure}

    \begin{figure}
            \begin{subfigure}{\textwidth}
                    \begin{mathpar}
                            \prAcquire
                            \and
                            \prReleaseSimplifiedPaper
                            \and
                            \prNewMutex
                    \end{mathpar}
                    \subcaption{Mutexes.}
            \end{subfigure}

            \begin{subfigure}{\textwidth}
                    \begin{mathpar}
                            \prAlloc
                            \and
                            \prReadHeapLoc
                            \and
                            \prAssignToHeap
                    \end{mathpar}
                    \subcaption{Heap Access.}
            \end{subfigure}

            \caption{Proof rules (part 2).}
            \label{appendix:fig:ProofRulesPartTwo}
    \end{figure}

    Following the intuition provided in \S~\ref{sec:ApproachInANutshell}, we can prove that a program terminates by proving that it discharges all obligations.
    The following theorem states that this approach is sound.

    \begin{restatable}[Soundness]{theorem}{theoSoundness}\label{theo:Soundness}
            Let \htProves{\noObs}{\cmdVar}{\noObs} hold.
             There exists no fair, infinite reduction sequence \phtpRSeq with
            $\pheapVar[0] = \emptyset$ and 
            $\tpVar[0] = 
                    \setOf{(\tidStartVar, \cmdVar)}
            $
            for any choice of \tidStartVar.
    \end{restatable}

\section{Soundness}\label{appendix:sec:Soundness}

    We already sketched the high-level intuition behind our soundness argument in \S~\ref{sec:Nutshell:ThreadSafePhysicalSignals}.
    In this section, we prove it formally.
    In \S~\ref{appendix:sec:Soundness:AnnotatedSemantics} we define an annotated reduction semantics that tracks which resources threads own (including ghost resources).
    In \S~\ref{appendix:sec:Soundness:ModelRelation} we define a model relation for Hoare triples and prove that every specification we can derive with our proof rules also holds in our model.
    In \S~\ref{appendix:sec:Soundness:Proof} we use the annotated semantics and the model relation to prove the soundness theorem we stated above.

\subsection{Annotated Semantics}\label{appendix:sec:Soundness:AnnotatedSemantics}

    During runtime, all threads share one physical heap where every thread is free to access every resource.
    This does not reflect the notions of ownership and lock invariants which we maintain on the verification level.
    It also does not allow us to restrict actions based on levels, e.g., only allowing the acquisition of a mutex if its level is lower than the level of each held obligation.
    Hence:
    (i)~We annotate every thread by a logical heap to express which resources it owns (including ghost resources) and thereby obtain an \emph{annotated thread pool}.
    (ii)~We represent the program state by an \emph{annotated heap} that keeps track of lock invariants and levels.
    In particular, we associate unlocked mutexes with logical heaps to represent the resources they protect.
    Since annotated heaps keep track of levels, they also keep track of signals.

        \begin{definition}[Intermediate Representation]
                We define an extended set of commands \CmdSetExt according to the syntax presented in Figure~\ref{appendix:fig:ExtendedSyntax}.
        \end{definition}
        
        For the rest of this appendix, commands \cmdVar refer to the extended set of commands, i.e., $\cmdVar \in \CmdSetExt$.
        
        \begin{figure}
                    $$
                    \begin{array}{l c l l}        
                            \cmdVar \in \CmdSetExt  &\ ::= \ \
                                    &\lowlightBack{
                                            \expVar \alt
                                                    \cmdWhile{\cmdVar}{\cmdSkip} \alt
                                                    \cmdFork{\cmdVar}\alt
                                            }\\
                                    &&\lowlightBack{
                                                    \cmdLet{\varVar}{\cmdVar}{\cmdVar} \alt
                                                    \cmdIf{\cmdVar}{\cmdVar} \alt
                                                    \cmdAlloc{\expVar} \alt
                                                    \cmdReadHeapLoc{\expVar} \alt
                                                    \cmdAssignToHeap{\expVar}{\expVar} \alt
                                            }\\
                                    &&\lowlightBack{
                                                    \cmdNewMut \alt
                                                    \cmdAcquireMut{\expVar} \alt
                                                    \cmdReleaseMut{\expVar} \alt
                                            }\\
                                    &&\icmdAwaitStarted{\SigSetVar}{\hlocVar}{\cmdVar} \alt
                                            \icmdWhileDecStarted{\natVar}{\cmdVar}
                        \\
                        \natVar \in \N
                        \\
                        \SigSetVar \finSubset \SignalSet
                                &&\lowlightText{\text{finite set of signals}}
                    \end{array}
                    $$
                    \caption{Extended set of commands for intermediate representation.}
                    \label{appendix:fig:ExtendedSyntax}
        \end{figure}

     \begin{definition}[Annotated Resources]
                We define the set of annotated resources \AnnoResSet as
                $$
                \begin{array}{l l l}
                        \aresVar \in \AnnoResSet \ &::= \
                                &\slPointsTo{\hlocVar}{\valVar} \alt
                                        \aresUninit{\hlocVar} \alt\\
                                &&\aresUnlocked{(\hlocVar, \levVar)}{\assVar}{\lheapVar}
                                        \alt
                                        \aresLocked{(\hlocVar, \levVar)}{\assVar}{\fracVar} \alt\\
                                &&\aresSignal{(\idVar, \levVar)}{\slBoolVar}
                \end{array}
                $$
                where \lheapVar does not contain any obligations chunks.
        \end{definition}

        \begin{definition}[Annotated Heaps]
                We define the set of annotated heaps as
                $$\AnnoHeapSet \ := \ \finPowerSetOf{\AnnoResSet},$$
                the function 
                $\ahGetLocs: \AnnoHeapSet \rightarrow \finPowerSetOf{\HeapLocSet}$
                mapping annotated heaps to the sets of allocated heap locations as
                $$
                \begin{array}{l c l l}
                        \ahGetLocs(\aheapVar) &\ := \
                                &\setOf{ 
                                        \hlocVar \in \HeapLocSet \ \ |\
                                        & \exists \valVar \in \ValueSet.\
                                                \exists \levVar \in \LevelSet.\
                                                \exists \assVar \in \AssertionSet.\\
                                        &&&\exists \lheapVar \in \LogHeapSet.\
                                                \exists \fracVar \in \FracSet.\\
                                        &&&\quad
                                                \slPointsTo{\hlocVar}{\valVar} \in \aheapVar\ \vee\
                                                \aresUninit{\hlocVar} \in \aheapVar\ \vee\\
                                        &&&\quad
                                                \aresUnlocked
                                                        {(\hlocVar, \levVar)}
                                                        {\assVar}
                                                        {\lheapVar}
                                                    \in \aheapVar\ \vee\\
                                        &&&\quad
                                                \aresLocked{(\hlocVar, \levVar)}{\assVar}{\fracVar} \in \aheapVar
                                }
                \end{array}
                $$
                and the function
                $\ahGetIDs : \AnnoHeapSet \rightarrow \finPowerSetOf{\IDSet}$
                mapping annotated heaps to sets of allocated signal IDs as
                $$
                \begin{array}{l c l }
                        \ahGetIDs(\aheapVar) &\ :=\
                                &\setOf{
                                         \idVar \in \IDSet  \ \ | \ \
                                        \exists \levVar \in \LevelSet.\
                                        \exists \slBoolVar \in \B.\
                                                \aresSignal{(\idVar, \levVar)}{\slBoolVar}
                                                \in \aheapVar
                                }.
                \end{array}
                $$
                We denote annotated heaps by \aheapVar.
                
                We call an annotated heap \aheapVar \emph{finite} and write \ahFinite{\aheapVar} if there exists no chunk
                $\aresUnlocked{(\hlocVar, \levVar)}{\assVar}{\lheapVar} \in \aheapVar$
                for which \lhFinite{\lheapVar} does not hold.
        \end{definition}

    Physical, annotated and logical heaps represent program states on different abstraction levels.
    While each level focuses on different aspects of the program state, they also share information.
    For instance, all three kinds of heaps use points-to chunks.
    Also, annotated and logical heaps both have signal chunks.
    Therefore, when referring to multiple kinds of heaps to talk about the same program state, we have to ensure that their contents match and do not contradict each other.
    We call such heaps \emph{compatible}.
    The following definitions make this precise.
        
    \begin{definition}[Compatibility of Annotated and Physical Heaps]
            We inductively define a relation
            $\aphCorrespSymb\ \subset \AnnoHeapSet \times \PhysResSet$
            between annotated and physical heaps such that the following holds:
            $$
            \begin{array}{l c l}
                    \emptyset &\aphCorrespSymb &\emptyset,\\
                    \slPointsTo{\hlocVar}{\valVar}\ \cup\ \aheapVar
                            &\aphCorrespSymb
                            &\slPointsTo{\hlocVar}{\valVar}\ \cup\ \pheapVar,\\
                    \aresUninit{\hlocVar}\ \cup\ \aheapVar
                            &\aphCorrespSymb
                            &\presUnlocked{\hlocVar}\ \cup\ \pheapVar,\\
                    \aresUnlocked
                            {(\hlocVar, \levVar)}
                            {\assLockInvVar}
                            {\lheapLockInvVar}
                        \ \cup\ \aheapVar
                            &\aphCorrespSymb
                            &\presUnlocked{\hlocVar}\ \cup\ \pheapVar,\\
                    \aresLocked{(\hlocVar, \levVar)}{\assLockInvVar}{\fracVar}\ \cup\ \aheapVar
                            &\aphCorrespSymb
                            &\presLocked{\hlocVar}\ \cup\ \pheapVar,\\
                    \aresSignal{\sigVar}{\slBoolVar}\ \cup\ \aheapVar
                            &\aphCorrespSymb
                            &\pheapVar,
            \end{array}
            $$
            where 
            $\aheapVar \in \AnnoHeapSet$ 
            and
            $\pheapVar \in \PhysHeapSet$ 
            are annotated and physical heaps with
            \aphCorresp{\aheapVar}{\pheapVar}.
    \end{definition}

        \begin{definition}[Compatibility of Annotated and Logical Heaps]
                We inductively define a relation
                $\alhCorrespSymb\, \subset\, \AnnoHeapSet \times \LogHeapSet$
                between annotated and logical heaps such that the following holds:
                $$
                \begin{array}{l c l}
                        \emptyset &\alhCorrespSymb &\lhEmpty,
                        \\
                        \ahAddRes
                                {\aheapVar}
                                {\slPointsTo{\hlocVar}{\valVar}}
                        &\alhCorrespSymb
                        &\lhAddRes
                                {\lheapVar}
                                {\slPointsTo{\hlocVar}{\valVar}},
                        \\
                        \ahAddRes
                                {\aheapVar}
                                {\aresUninit{\hlocVar}}
                        &\alhCorrespSymb
                        &\lhAddRes
                                {\lheapVar}
                                {\lresUninit{\hlocVar}},
                        \\
                        \ahAddRes
                                {\aheapVar}
                                {\aresUnlocked
                                        {\mutVar}
                                        {\assLockInvVar}
                                        {\lheapLockInvVar}
                                }
                        &\alhCorrespSymb
                        &\lhAddRes
                                {\lheapVar}
                                {\lresMutex{\mutVar}{\assLockInvVar}}
                            + \lheapLockInvVar,
                        \\
                        \ahAddRes
                                {\aheapVar}
                                {\aresLocked{\mutVar}{\assLockInvVar}{\fracVar}}
                        &\alhCorrespSymb
                        &\lheapVar
                                + \lheapOf{\lresLocked{\mutVar}{\assLockInvVar}{\fracVar}}
                        \\
                                &&\phantom{\lheapVar}
                                        + (1-\fracVar) \cdot \lheapOf{\lresMutex{\mutVar}{\assLockInvVar}},
                        \\
                        \ahAddRes
                                {\aheapVar}
                                {\aresSignal{\sigVar}{\slBoolVar}}
                        &\alhCorrespSymb
                        &\lhAddRes
                                {\lheapVar}
                                {\lresSignal{\sigVar}{\slBoolVar}},
                                
                        \\
                        \aheapVar
                        &\alhCorrespSymb
                        &\lhAddRes
                                {\lheapVar}
                                {\lresObs{\obBagVar}},
                \end{array}
                $$
                where $\aheapVar \in \AnnoHeapSet$ and $\lheapVar \in \LogHeapSet$ are annotated and logical heaps with
                $\hlocVar, \mutGetLoc{\mutVar} \not\in \ahGetLocs[\aheapVar]$,
                $\sigGetID{\sigVar} \not\in \ahGetIDs[\aheapVar]$
                and \alhCorresp.
        \end{definition}

    We define annotated versions \atpRedStepSymb and \astRedStepSymb of the relations \tpRedStepSymb and \stRedStepSymb, respectively.
    The annotated reduction semantics we thereby obtain needs to reflect ghost proof steps implemented by view shifts.
    Hence, we define \atpRedStepSymb in terms of two relations: 
    (i)~\gtpRedStepSymb for ghost steps and 
    (i)~\rtpRedStepSymb for actual program execution steps.
    The annotated semantics ensure that a reduction gets stuck if a thread violates any of the restrictions formulated by our proof rules.

        \begin{definition}[Annotated Single Thread Reduction Relation]
                We define a reduction relation \astRedStepSymb for annotated threads according to the rules presented in Fig.~\ref{appendix:fig:AnnotatedSingleThreadReductionRulesPartOne} and~\ref{appendix:fig:AnnotatedSingleThreadReductionRulesPartTwo}.
                A reduction step has the form
                $$
                        \astRedStep
                                {\cmdVar}
                                [\tidVar]
                                [\nextAheapVar]
                                [\nextLheapVar]
                                {\nextCmdVar}
                                [\aftsVar]
                $$
                for a set of annotated forked threads
                $\aftsVar \subset \LogHeapSet \times \CmdSet$
                with $\cardinalityOf{\aftsVar} \leq 1$.

                It indicates that given annotated heap \aheapVar and a logical heap \lheapVar, 
                command \cmdVar can be reduced to annotated heap \nextAheapVar, logical heap \nextLheapVar and command \nextCmdVar.
                The either empty or singleton set \aftsVar represents whether a new thread is forked in this step.
                
                For simplicity of notation we omit \aftsVar if it is clear from the context that no thread is forked and $\aftsVar = \emptyset$.
        \end{definition}

        \begin{figure}
                \begin{subfigure}{\textwidth}
                        \begin{mathpar}
                                \astRedEvalCtxt
                                \quad
                                \astRedForkPaper
                        \end{mathpar}
                        \caption{Basic constructs.}
                \end{subfigure}

                \begin{subfigure}{\textwidth}
                        \begin{mathpar}
                                \astRedIfTrue
                                \and
                                \astRedIfFalse
                                \and
                                \astRedLet
                        \end{mathpar}
                        \caption{Control structures.}
                \end{subfigure}
                
                \begin{subfigure}{\textwidth}
                        \begin{mathpar}
                                \astRedAlloc
                                \and
                                \astRedReadHeapLoc
                                \and
                                \astRedAssign
                        \end{mathpar}
                        \caption{Heap access.}
                \end{subfigure}
                
                \begin{subfigure}{\textwidth}
                        \begin{mathpar}
                                \astRedNewMutex
                                \and
                                \astRedAcquire
                                \and
                                \astRedReleaseNoPerm
                        \end{mathpar}
                        \caption{Mutexes.}
                \end{subfigure}
                
                \caption{
                        Annotated single thread reduction rules (part 1).
                }
                \label{appendix:fig:AnnotatedSingleThreadReductionRulesPartOne}
        \end{figure}

        \begin{figure}
                \begin{subfigure}{\textwidth}
                        \begin{mathpar}
                                \astRedWhileDecInit
                                \and
                                \astRedWhileDec
                                \and
                                \astRedAwaitInit
                                \and
                                \astRedAwait
                        \end{mathpar}
                        \caption{Loops and auxiliary commands.}
                \end{subfigure}
                
                \caption{
                        Annotated single thread reduction rules (part 2).
                }
                \label{appendix:fig:AnnotatedSingleThreadReductionRulesPartTwo}
        \end{figure}

        \begin{definition}[Annotated Thread Pools]
                We define the set of annotated thread pools \AnnoThreadPoolSet as the set of finite partial functions mapping thread IDs to annotated threads:
                $$
                        \AnnoThreadPoolSet \ := \ 
                                \finPartialFunSet
                                        {\ThreadIDSet}
                                        {
                                                \LogHeapSet \times
                                                (\CmdSetExt \cup \setOf{\thTerminated})
                                        }.
                $$
                We denote annotated thread pools by \atpVar and the empty thread pool by \atpEmpty, i.e.,
                $$
                \begin{array}{l}
                        \atpEmpty \ : \
                                \ThreadIDSet \finPartialFunArrow 
                                \LogHeapSet \times (\CmdSetExt \cup \setOf{\thTerminated}),
                        \\
                        \defDom{\atpEmpty} \, = \, \emptyset.
                \end{array}
                $$
                We define the extension operation \atpAddSymb analogously to \tpAddSymb, cf. Definition~\ref{appendix:def:ThreadPools}.
                
                For convenience of notation we define selector functions for annotated threads as
                $$
                \begin{array}{l c l}
                        \atGetHeap{(\lheapVar, \cmdVar)} &:=
                                &\lheapVar,
                        \\
                        \atGetCmd{(\lheapVar, \cmdVar)} &:=
                                &\cmdVar.
                \end{array}
                $$
        \end{definition}

        \begin{definition}[Ghost Reduction Relation]
                We define a thread pool reduction relation \gtpRedStepSymb according to the rules presented in Fig.~\ref{appendix:fig:GhostThreadPoolReductionRulesPart} to express ghost steps.
                A ghost reduction step has the form
                $$
                        \gtpRedStep
                                {\atpVar}
                                {\nextAtpVar}.
                $$
                We denote its reflexive transitive closure by \gtpRedStepMultSymb.
        \end{definition}

        \begin{figure}
                \begin{mathpar}
                        \gtpRedNewSignal
                        \and
                        \gtpRedSetSignal
                        \and
                        \gtpRedMutInitNoPerm
                \end{mathpar}

                \caption{Ghost thread pool reduction rules.}
                \label{appendix:fig:GhostThreadPoolReductionRulesPart}
        \end{figure}

        \begin{definition}[Non-ghost Thread Pool Reduction Relation]
                We define a thread pool reduction relation \rtpRedStepSymb according to the rules presented in Fig.~\ref{appendix:fig:RealThreadPoolReductionRules} to express real (i.e. non-ghost) reduction steps.
                A reduction step has the form
                $$\rtpRedStep
                        {\atpVar}
                        {\nextAtpVar}.
                $$
        \end{definition}

        \begin{figure}
                \begin{mathpar}
                        \rtpRedLift
                        \and
                        \rtpRedTerm
                \end{mathpar}
                
                \caption{Non-ghost thread pool reduction rules.}
                \label{appendix:fig:RealThreadPoolReductionRules}
        \end{figure}

        \begin{definition}[Annotated Thread Pool Reduction Relation]
                We define the annotated thread pool reduction relation \atpRedStepSymb as
                $$
                \atpRedStepSymb \ \ := \ \ \gtpRedStepSymb \cup \rtpRedStepSymb.
                $$
        \end{definition}

        Note that the  reduction relation \atpRedStepSymb indeed reflects the restrictions our proof rules impose.
        For instance, proof rule \prAwaitGenName only allows a thread to wait for a signal if the signal's level is below the level of all held obligations.
        \astRedAwaitName ensures that any thread that does not comply with this restriction gets stuck.

        \begin{definition}[Annotated Reduction Sequence]\label{appendix:def:AnnotatedRedSeq}
                Let $(\aheapVar_i)_{i\in \N}$ and $(\atpVar_i)_{i \in \N}$ be infinite sequences of annotated heaps and annotated thread pools, respectively.
                Let
                $\sigAnno : \N \partialFunArrow \SignalSet$ be a partial function mapping indices to signals.
                
                We call \ahtpRSeqFull an \emph{annotated reduction sequence} if
                there exists a sequence of thread IDs $(\tidVar_i)_{i \in \N}$ such that the following holds for every $i \in \N$:
                \begin{itemize}
                        \item
                \atpRedStep
                        [\aheapVar[i]]
                        {\atpVar[i]}
                        [\tidVar_i]
                        [\aheapVar[i+1]]
                        {\atpVar[i+1]}
                        \item If this reduction step does not result from an application of \rtpRedLiftName in combination with \astRedAwaitName, then $\sigAnno[i] = \funUndefVal$.
                                If \astRedAwaitName is applied to some signal \sigVar, then $\sigAnno[i] = \sigVar$.
                \end{itemize}
                In case the signal annotation \sigAnno is  clear from the context or not relevant, we omit it and write \ahtpRSeq instead of \ahtpRSeqFull.
                
                We call $(\aheapVar[i], \atpVar[i])$ an \emph{annotated machine configuration}.
        \end{definition}

        \begin{definition}[Fairness of Annotated Reduction Sequences]
                We call an annotated reduction sequence \ahtpRSeq \emph{fair} iff for all $i \in \N$ and $\tidVar \in \defDom{\atpVar[i]}$ with
                $\atGetCmd{\atpVar[i](\tidVar)} \neq \thTerminated$
                there exists some $k \geq i$ with
                $$
                \rtpRedStep
                        [\aheapVar[k]]{\atpVar[k]}
                        [\tidVar]
                        [\aheapVar[k+1]]{\atpVar[k+1]}.
                $$
        \end{definition}

        \begin{lemma}[Preservation of Finiteness]\label{lem:PreservationOfFiniteness}
                Let \ahtpRSeq be an annotated reduction sequence with 
                \ahFinite{\aheapVar[0]}
                and
                \lhFinite{\atGetHeap{\atpVar[0](\tidVar)}}
                for all $\tidVar \in \defDom{\atpVar[0]}$.
                
                Then,
                \lhFinite{\atGetHeap{\atpVar[i](\tidVar)}}
                holds for all $i \in \N$ and all $\tidVar \in \defDom{\atpVar[i]}$.
        \end{lemma}
        \begin{proof}
                Proof by induction on $i$.
        \end{proof}

        \begin{lemma}[Preservation of Completeness]\label{lem:PreservationOfCompleteness}
                Let \ahtpRSeq be an annotated reduction sequence with
                \lhComplete{\atGetHeap{\atpVar[0](\tidVar)}}
                for all $\tidVar \in \defDom{\atpVar[0]}$.
                Then, 
               \lhComplete{\atGetHeap{\atpVar[i](\tidVar)}}
                holds for every $i \in \N$ and every $\tidVar \in \defDom{\atpVar[i]}$.
        \end{lemma}
        \begin{proof}
                Proof by induction on $i$.
        \end{proof}

        Every thread of an annotated thread pool is annotated by a thread-local logical heap that expresses which resources are owned by this thread.
        In the following we define a function to extract the logical heap expressing which resources are owned by threads of a thread pool (i.e. the sum of all thread-local logical heaps).
        
        \begin{definition}
                We define the function 
                $\atpOwnedResHeap : \AnnoThreadPoolSet \rightarrow \LogHeapSet$
                mapping annotated thread pools to logical heaps as follows:
                $$
                \begin{array}{l c l}
                        \atpVar &\ \mapsto
                                &\quadBack
                                        \bigSum[\quad\tidVar\, \in\, \defDom{\atpVar}]
                                        \quadBack\!\!
                                        \atGetHeap{\atpVar(\tidVar)}
                \end{array}
                $$
        \end{definition}

        Annotated resources representing unlocked locks, i.e.,
        \aresUnlocked{\mutVar}{\assVar}{\lheapVar[\assVar]},
        contain a logical heap \lheapVar[\assVar] that expresses which resources are protected by this lock.
        In the following, we define a function that extracts a logical heap from an annotated heap \aheapVar expressing which resources are protected by unlocked locks in \aheapVar.

        \begin{definition}
                We define the function 
                $\ahProtectedResHeap : \AnnoHeapSet \rightarrow$ \phantom{.} $\LogHeapSet$
                mapping  annotated heaps to logical heaps as follows:
                
                For any annotated heap \aheapVar let 
                $$
                \begin{array}{l c l l}
                        \multiLetterSetFont{LockInvs}(\aheapVar) \ &:= \
                                &\leftMsBrace \lheapLockInvVar \in \LogHeapSet \ \ | 
                                        &\exists \mutVar \in \HeapLocSet \times \LevelSet.\
                                            \exists \assLockInvVar \in \AssertionSet. \\
                                        &&&\quad
                                            \aresUnlocked{\mutVar}{\assLockInvVar}{\lheapLockInvVar} \in \aheapVar
                                \rightMsBrace
                \end{array}
                $$
                be the auxiliary bag aggregating all logical heaps corresponding to lock invariants of unlocked locks stored in \aheapVar.
                We define \ahProtectedResHeap as
                $$
                        \aheapVar \ \ \mapsto 
                                \bigSum[\lheapLockInvVar
                                                    \,\in\, \multiLetterSetFont{LockInvs}(\aheapVar)]
                                        \quadBack\quadBack\lheapLockInvVar.
                $$
                
        \end{definition}

        We consider a machine configuration $(\aheapVar, \atpVar)$ to be \emph{consistent} if it fulfils three criteria:
        (i)~Every thread-local logical heap is consistent, i.e.,
        for all used thread IDs \tidVar, \atGetHeap{\atpVar(\tidVar)} only stores full obligations chunks.
        (ii)~Every logical heap protected by an unlocked lock in \aheapVar is consistent.
        (iii)~\aheapVar is compatible with the logical heap that represents
        (a)~the resources owned by threads in \atpVar and
        (b)~the resources protected by unlocked locks stored in \aheapVar.

        \begin{definition}[Consistency of Annotated Machine Configurations]
                We call an annotated machine configuration $(\aheapVar, \atpVar)$ \emph{consistent} and write \\
                \amConfConsistent{\aheapVar}{\atpVar}
                if all of the following hold:
                \begin{itemize}
                        \item \lhConsistent{\atGetHeap{\atpVar(\tidVar)}}
                                    for all $\tidVar \in \defDom{\atpVar}$,
                        \item
                                $
                                        \forall \mutVar.\
                                        \forall \assLockInvVar.\
                                        \forall \lheapLockInvVar.\ 
                                                \aresUnlocked{\mutVar}{\assLockInvVar}{\lheapLockInvVar} \in \aheapVar
                                                \ \rightarrow\
                                                \lhConsistent{\lheapLockInvVar},
                                $
                        \item
                                $
                                        \aheapVar
                                        \alhCorrespSymb\
                                        \atpOwnedResHeap[\atpVar] + 
                                        \ahProtectedResHeap[\aheapVar].
                                $
                \end{itemize}
        \end{definition}

        \begin{lemma}[Preservation of Consistency]
                Let \ahtpRSeq be an annotated reduction sequence with \amConfConsistent{\aheapVar[0]}{\atpVar[0]}.
                Then, \amConfConsistent{\aheapVar[i]}{\atpVar[i]} holds for every $i \in \N$.
        \end{lemma}
        \begin{proof}
                Proof by induction on $i$.
        \end{proof}

\subsection{Hoare Triple Model Relation}\label{appendix:sec:Soundness:ModelRelation}
    
    We interpret program specifications \hoareTriple{\assPreVar}{\cmdVar}{\assPostVar} in terms of a model relation \htModels{\assPreVar}{\cmdVar}{\assPostVar} and an auxiliary safety relation \safe{\lheapVar}{\cmdVar}.
    
    In the annotated semantics we annotate threads by local logical heaps that express which resources they own (including ghost resources) and use an extended intermediate representation for commands.
    We say that an annotated thread pool \atpVar is an \emph{annotation} of a plain thread pool \tpVar if they represent matching information.
    That is, they must store matching threads under the same thread IDs.

    \begin{definition}[Command Annotation]
            We define the predicate
            $\cmdAnnotName \subset \CmdSetExt \times \CmdSet$
            such that \cmdAnnot{\cmdVar^+}{\cmdVar} holds iff \cmdVar results from $\cmdVar^+$ by replacing all subcommands of the form
            (i)~\icmdWhileDecStarted{\natVar}{\nextCmdVar} by
                    \cmdWhileSkip{\nextCmdVar}
            and
            (ii)~\icmdAwaitStarted{\SigSetVar}{\expVar}{\nextCmdVar} by
                    \cmdAwait{\expVar}{\nextCmdVar}.
    \end{definition}

    \begin{definition}[Thread Pool Annotation]
            We define a predicate
            $\tpAnnotName \subset \AnnoThreadPoolSet \times \ThreadPoolSet$
            such that:
            $$
            \begin{array}{c}
                    \tpAnnot{\atpVar}{\tpVar}\\
                    \Longleftrightarrow\\
                    \dom{\atpVar} = \dom{\tpVar}\ \wedge\
                    \forall \tidVar \in \dom{\tpVar}.\
                    \cmdAnnot
                            {\atGetCmd{\atpVar(\tidVar)}}
                            {\tpVar(\tidVar)}
            \end{array}
            $$
    \end{definition}

    Intuitively, a command \cmdVar is \emph{safe} under a logical heap \lheapVar if \lheapVar provides the necessary resources so that for every execution of \cmdVar, there is a corresponding annotated execution of \cmdVar that does not get stuck.
    That is, if we consider a reduction step from \cmdVar to \nextCmdVar in the plain operational semantics, then the resources that thread \cmdVar owns according to \lheapVar, allow us to perform a corresponding sequence of annotated reduction steps leading to \nextCmdVar.
    Specifically, we can perform a finite number of ghost steps to manipulate ghost resources followed by one real step to reduce \cmdVar to \nextCmdVar.
    Furthermore, safety requires \nextCmdVar to be safe under its respective local logical heap.
    The same must hold for any potentially forked thread.

    \begin{definition}[Safety]
            We define the safety predicate
            $\safeName \subseteq \LogHeapSet \times \CmdSet$
            coinductively as the greatest solution (with respect to $\subseteq$) of the following equation:
            $$
            \begin{array}{c}
                    \safe{\lheapVar}{\cmdVar}\\
                    \Longleftrightarrow\\
                    \begin{array}{l c l}
                            \lhComplete{\lheapVar}\ \rightarrow\\
                            \forall \tpVar, \nextTpVar.\,
                            \forall \tidVar \in \dom{\tpVar}.\,
                            \forall \pheapVar, \nextPheapVar.\,
                            \forall \atpVar.\,
                            \forall \aheapVar.\\
                            \quad
                            \amConfConsistent{\aheapVar}{\atpVar}
                            \ \wedge\
                            \aphCorresp{\aheapVar}{\pheapVar}
                            \ \wedge
                            \\
                            \quad
                                    \tpVar(\tidVar) = \cmdVar
                                    \ \wedge\
                                    \atpVar(\tidVar) = (\lheapVar, \cmdVar)
                                    \ \wedge\
                                    \tpAnnot{\atpVar}{\tpVar}
                                    \ \wedge\
                                    \tpRedStep{\tpVar}{\nextTpVar}
                                    \ \rightarrow\\
                            \quad
                            \exists \gtpVar, \nextAtpVar.\
                            \exists \agheapVar, \nextAheapVar.\ \\
                            \quad\quad
                                    \gtpRedStepMult
                                            {\atpVar}
                                            [\tidVar]
                                            [\agheapVar]
                                            {\gtpVar}
                                    \ \wedge\
                                    \rtpRedStep
                                            [\agheapVar]
                                            {\gtpVar}
                                            {\nextAtpVar}
                                    \ \wedge\
                                    \tpAnnot{\nextAtpVar}{\nextTpVar}
                                    \ \wedge\\
                            \quad\quad
                                    \aphCorresp{\nextAheapVar}{\nextPheapVar}
                                    \ \wedge\\
                            \quad\quad
                                    \forall (\lheapForkedVar, \cmdForkedVar) \in
                                        \range{\nextAtpVar} \setminus \range{\atpVar}.\
                                            \safe{\lheapForkedVar}{\cmdForkedVar}
                            .
                    \end{array}
            \end{array}
            $$
    \end{definition}

	We interpret Hoare triples \hoareTriple{\assPreVar}{\cmdVar}{\assPostVar} in terms of safety as follows:
	Let \evalCtxt be an evaluation context that (when instantiated) is safe under any heap \lheapPostVar which fulfills postcondition \assPostVar, i.e., let \evalCtxt be a context for which \phantom{\!\!}
	$
	\forall \valVar.\ \forall \lheapPostVar.\
	\assModelsOR{\lheapPostVar}{\assPostVar(\valVar)}
	\rightarrow
	\safe{\lheapPostVar}{\evalCtxt[\valVar]}
	$
	holds.
	Then, any heap \lheapPreVar that satisfies precondition \assPreVar provides all resources necessary to safely run \evalCtxt[\cmdVar], i.e.,
	$\forall \lheapPreVar.\
		\assModels{\lheapPreVar}{\assPreVar}
		\ \rightarrow\
		\safe
			{\lheapPreVar}
			{\,\evalCtxt[\cmdVar]}
	$.
	Thereby, if the reduction of \cmdVar under \lheapPreVar finishes and returns a value \valVar, then postcondition $\assPostVar(\valVar)$ holds in the final state.
	In the following definition, we also allow \lheapPreVar and \lheapPostVar to be embedded into arbitrary heap frames \lheapFrameVar.

    \begin{definition}[Hoare Triple Model Relation]
            We define the model relation for Hoare triples
            $\htModelsSymb\ \subset \AssertionSet \times \CmdSet \times (\ValueSet \rightarrow \AssertionSet)$
            such that:
            $$
            \begin{array}{c}
                    \htModels{\assPreVar}{\cmdVar}[\resVar]{\assPostVar(\resVar)}\\
                    \Longleftrightarrow\\
                    \begin{array}{l}
                            \forall \lheapFrameVar.\
                            \forall \evalCtxt.\
                            (\forall \valVar.\
                                \forall \lheapPostVar.\
                                \assModelsOR
                                        {\lheapPostVar}
                                        {\assPostVar(\valVar)}
                                \ \rightarrow\
                                \safe
                                        {\lheapPostVar + \lheapFrameVar}
                                        {\,\evalCtxt[\valVar]}
                            )
                            \\
                            \phantom{
                                    \forall \lheapFrameVar.\
                            }
                            \rightarrow\ \
                                \forall \lheapPreVar.\
                                \assModelsOR{\lheapPreVar}{\assPreVar}
                                \ \rightarrow\
                                \safe
                                        {\lheapPreVar + \lheapFrameVar}
                                        {\,\evalCtxt[\cmdVar]}
                                                        
                    \end{array}
            \end{array}
            $$
    \end{definition}
    
    We can instantiate context \evalCtxt in above definition to \cmdLet{\varVar}{\evalCtxtHole}{\valUnit}, 
    which yields the consequent
    \safe
            {\lheapPreVar + \lheapFrameVar}
            {\cmdLet{\varVar}{\cmdVar}{\valUnit}}.
    Note that this implies
    \safe
            {\lheapPreVar + \lheapFrameVar}
            {\cmdVar}.
	Also note that compliance with the frame rule directly follows from above definition, i.e., 
	\htModels{\assPreVar}{\cmdVar}{\assPostVar}
	implies
	\htModels
		{\assPreVar \slStar \assFrameVar}
		{\cmdVar}
		{\assPostVar \slStar \assFrameVar}
	for any frame \assFrameVar.
    Further, every specification we can derive in our proof system also holds in our model.
    
    \begin{lemma}[Hoare Triple Soundness]\label{lem:HoareTripleSoundness}
            Let 
            \htProves{\assPreVar}{\cmdVar}{\assPostVar} hold,
            then also
            \htModels{\assPreVar}{\cmdVar}{\assPostVar}
            holds.
    \end{lemma}
    \begin{proof}
            Proof by induction on the derivation of \htProves{\assPreVar}{\cmdVar}{\assPostVar}.
    \end{proof}

\subsection{Soundness Proof}\label{appendix:sec:Soundness:Proof}
    In this section we prove our soundness theorem.

    \paragraph{Constructing Annotated Executions}
    Given a command \cmdVar which provably discharges all obligations and a fair reduction sequence for \cmdVar,
    we can construct a corresponding annotated reduction sequence \ahtpRSeq. 
    This is a useful tool to analyse program executions as \ahtpRSeq carries much more information than the original sequence, e.g., which obligations a thread holds.
    Note that our definition of fairness forbids \ahtpRSeq to perform ghost steps forever and that we use \alhCorresp[\aheapVar][\lheapVar] to express that \aheapVar is compatible with \lheapVar.

    \begin{restatable}[Construction of Annotated Reduction Sequences]{lemma}{lemConstructionOfAnnotatedRedSeq}\label{lem:ConstructionOfAnnotatedRedSeq}
           Suppose we can prove
           \htModels{\assPreVar}{\cmdVar}{\noObs}.
           Let \lheapPreVar be a logical heap with
           \assModels{\lheapPreVar}{\assPreVar} and \lhComplete{\lheapPreVar}
           and \aheapVar[0] an annotated heap with
           \alhCorresp[\aheapVar[0]][\lheapPreVar].
           Let \phtpRSeq be a fair plain reduction sequence with
           \aphCorresp{\aheapVar[0]}{\pheapVar[0]} and
           $\tpVar[0] = \setOf{(\tidVar_0, \cmdVar)}$
           for some thread ID $\tidVar_0$ and command \cmdVar.

           Then, there exists a fair annotated reduction sequence \ahtpRSeq with
           $\atpVar = \setOf{(\tidVar_0,\, (\lheapPreVar, \cmdVar))}$
           and \amConfConsistent{\aheapVar[i]}{\atpVar[i]} for all $i \in \N$.
    \end{restatable}
    \begin{proof}
            We can construct the annotated reduction sequence inductively from the plain reduction sequence.
    \end{proof}

    \paragraph{Program Order Graph}
    In the remainder of this section, we prove that programs where every thread discharges all obligations terminate.
    For this, we need to analyse each thread's control flow, i.e., the subsequence of execution steps belonging to the thread.
    We do this by taking a sequence \phtpRSeq representing a program execution, constructing an annotation \ahtpRSeq carrying additional information and then analysing its \emph{program order graph} \progOrdGraph[\ahtpRSeq] which represents the execution's control flow.

    \begin{definition}[Program Order Graph]\label{appendix:def:ProgramOrderGraph}
    { 
            \NewDocumentCommand{\RuleNameSet}{}{\metaVar{N^r}}
            \NewDocumentCommand{\AnnoRuleNameSet}{}{\metaVar{N^\contextIdFont{a}}}
            
            \NewDocumentCommand{\ruleNameVar}{o}
                    {\metaVar{
                            n^\contextIdFont{a}
                            \IfValueT{#1}{_{#1}}
                    }}
            \NewDocumentCommand{\stRuleNameVar}{o}
                    {\metaVar{
                            n^\contextIdFont{st}
                            \IfValueT{#1}{_{#1}}
                    }}

            \NewDocumentCommand{\baseNode}{}{\metaVar{a}} 
            \NewDocumentCommand{\succNode}{}{\metaVar{b}} 
            \NewDocumentCommand{\auxNode}{}{\metaVar{k}} 
            \NewDocumentCommand{\genNode}{}{\metaVar{\ell}} 
        
            Let \ahtpRSeqFull be an annotated reduction sequence.
            Let \RuleNameSet be the set of names referring to reduction rules defining the relations \rtpRedStepSymb, \gtpRedStepSymb and \astRedStepSymb.
            We define the set of annotated reduction rule names \AnnoRuleNameSet where \astRedAwaitName is annotated by signals as
            $$
            \begin{array}{l c l}
                    \AnnoRuleNameSet \ &:= \
                            &(\RuleNameSet \setminus \setOf{\astRedAwaitName})\\
                            &&\!\phantom{\RuleNameSet}
                            \ \cup\
                            (\setOf{\astRedAwaitName}  \times \SignalSet)
            \end{array}
            $$
            We define the program order graph
            $\progOrdGraph[\ahtpRSeqFull] = (\N, \EdgeSetVar)$
            with root 0 where 
            $\EdgeSetVar \subset
                    \N \times \ThreadIDSet \times \AnnoRuleNameSet \times \N
            $.

            A node $\baseNode \in \N$ corresponds to the sequence's $\baseNode^\text{th}$ reduction step, i.e., to the step
            \atpRedStep
                    [\aheapVar[\baseNode]]{\atpVar[\baseNode]}
                    [\tidVar]
                    [\aheapVar[\baseNode+1]]{\atpVar[\baseNode+1]}
            for some $\tidVar \in \defDom{\atpVar[\baseNode]}$.
            An edge from node \baseNode to node \succNode expresses that the $\succNode^\text{th}$ reduction step continues the control flow of step $\baseNode$.
            For any $\genNode \in \N$, let $\tidVar_\genNode$ denote the ID of the thread reduced in step \genNode.
            Furthermore, let \ruleNameVar[\genNode] denote the name of the reduction rule applied in the $\genNode^\text{th}$ step, in the following sense:
            \begin{itemize}
                    \item If 
                            \atpRedStep
                                    [\aheapVar[\genNode]]{\atpVar[\genNode]}
                                    [\tidVar]
                                    [\aheapVar[\genNode+1]]{\atpVar[\genNode+1]}
                            results from an application of \rtpRedLiftName in combination with single-thread reduction rule \stRuleNameVar other than \astRedAwaitName,
                            then $\ruleNameVar[\genNode] = \stRuleNameVar$.
                    \item If 
                            \atpRedStep
                                    [\aheapVar[\genNode]]{\atpVar[\genNode]}
                                    [\tidVar]
                                    [\aheapVar[\genNode+1]]{\atpVar[\genNode+1]}
                            results from an application of \rtpRedLiftName in combination with \astRedAwaitName,
                            then $\ruleNameVar[\genNode] = (\astRedAwaitName, \sigAnno[\genNode])$.
                    \item Otherwise, \ruleNameVar denotes the applied (real or ghost) thread pool reduction rule.
            \end{itemize}
            
            An edge 
            $(\baseNode, \tidVar, \ruleNameVar, \succNode) \in 
                    \N \times \ThreadIDSet \times \AnnoRuleNameSet \times \N$
            is contained in \EdgeSetVar if
            $\ruleNameVar = \ruleNameVar[\baseNode]$
            and one of the following conditions applies:
            \begin{itemize}
                    \item $\tidVar = \tidVar_\baseNode = \tidVar_\succNode$ and 
                            $\succNode = \min(\setOf{\auxNode > \baseNode \ | \
                                                    \atpRedStep
                                                            [\aheapVar[\auxNode]]{\atpVar[\auxNode]}
                                                            [\tidVar_\baseNode]
                                                            [\aheapVar[\auxNode+1]]{\atpVar[\auxNode+1]}
                                                })$.
                            \\
                            In this case, the edge expresses that step \succNode marks the first time that thread $\tidVar_\baseNode$ is rescheduled for reduction (after step \baseNode).
                    \item $\defDom{\atpVar[\baseNode+1]} \setminus \defDom{\atpVar[\baseNode]}
                                    = \setOf{\tidVar}$
                            and\\
                            $\succNode = \min{\setOf{ \auxNode \in \N \ | \
                                                    \atpRedStep
                                                            [\aheapVar[\auxNode]]{\atpVar[\auxNode]}
                                                            [\tidVar]
                                                            [\aheapVar[\auxNode+1]]{\atpVar[\auxNode+1]}
                                            }}$.
                            \\
                            In this case, \tidVar identifies the thread forked in step \baseNode.
                            The edge expresses that step \succNode marks the first reduction of the forked thread.
            \end{itemize}
            
            In case the choice of reduction sequence \ahtpRSeqFull is clear from the context, we write \progOrdGraph instead of \progOrdGraph[\ahtpRSeqFull].
    } 
    \end{definition}

    \begin{observation}\label{obs:ProgramOrderGraphBinaryTree}
            Let \ahtpRSeq be an annotated reduction sequence with\\
            $\cardinalityOf{\defDom{\atpVar[0]}} = 1$.
            The sequence's program order graph \progOrdGraph[\ahtpRSeq] is a binary tree.
    \end{observation}

    For any reduction sequence \ahtpRSeq, the paths in its program order graph \progOrdGraph[\ahtpRSeq] represent the sequence's control flow paths.
    Hence, we are going to use program order graphs to analyse reduction sequences' control flows.
    We refer to a program order graph's edges by the kind of reduction step they represent.
    For instance, we call any edge $(a, \tidVar, R, b)$ a \emph{loop edge} where $R$ refers to one of the rules related to the execution loops,
    i.e., \astRedWhileDecInitName, 
    \astRedWhileDecName,
    \astRedAwaitInitName or
    \astRedAwaitName.
    In the following, we prove that any path in a program order graph that does not involve a loop edge is finite.
    This follows from the fact that the size of the command reduced along this path decreases with each non-ghost non-loop step.

    \begin{lemma}\label{lem:InfinitePathsCotainInfinitelyManyLoopEdges}
            Let \ahtpRSeq be a fair annotated reduction sequence.
            Let $\pathVar = (\NodeSetVar, \EdgeSetVar)$ be a path in \progOrdGraph[\ahtpRSeq].
            Let 
            $$
            \begin{array}{l c l}
                    R &\ =\ &\{\,
                            \astRedWhileDecInitName,\
                            \astRedWhileDecName,\
                            \\
                            &&\phantom{\{}\
                            \astRedAwaitInitName\
                    \}
                    \ \cup \
                    (\setOf{\astRedAwaitName} \times \SignalSet)
            \end{array}
            $$
            be the set of names of single-thread reduction rules related to loops and let
            $L = \setOf{\edgeVar \in \EdgeSetVar \ | \ \tupleProj{3}[\edgeVar] \in R }$
            be the set of loop edges contained in \pathVar.
            Then, \pathVar is infinite if and only if $L$ is infinite.
    \end{lemma}
    \begin{proof}
            If $L$ is infinite, \pathVar is obviously infinite as well.
            So, suppose $L$ is finite.
            
            For any command, we consider its size to be the number of nodes contained in its abstract syntax tree.
            By structural induction over the set of commands, it follows that the size of a command $\cmdVar = \atGetCmd{\atpVar(\tidVar)}$ decreases in every non-ghost reduction step
            \atpRedStep
                    [\aheapVar]
                    {\atpVar}
                    [\tidVar]
                    [\nextAheapVar]
                    {\nextAtpVar}            
            that is not an application of \rtpRedLiftName in combination with some $r \in R$.
            
            Since $L$ is finite, there exists a node $x$ such that the suffix $\pathVar_{\geq x}$ starting at node $x$ does not contain any loop edges.
            By fairness of \ahtpRSeq, every non-empty suffix of $\pathVar_{\geq x}$ contains an edge corresponding to a non-ghost reduction step.
            For any edge $\edgeVar = (i, \tidVar, n, j)$ consider the command 
            $\cmdVar_\edgeVar = \atGetCmd{\atpVar[i](\tidVar)}$
            reduced in this edge.
            The size of these commands decreases along $\pathVar_{\geq x}$.
            So, $\pathVar_{\geq x}$ must be finite and thus \pathVar must be finite as well.
    \end{proof}

    \begin{definition}
            Let
            $\progOrdGraph = (\NodeSetVar, \EdgeSetVar)$
            be a subgraph of some program order graph.
            We define the function
            $\awaitEdges[\progOrdGraph] : \SignalSet \rightarrow \powerSetOf{\EdgeSetVar}$
            mapping any signal \sigVar to the set of await edges in \progOrdGraph concerning \sigVar as:
            $$
            \begin{array}{c c l}
                    \awaitEdges[\progOrdGraph][\sigVar] &\ := \
                            &\setOf{
                                    (a, \tidVar, (\astRedAwaitName, \nextSigVar), b) \in \EdgeSetVar 
                                    \ |\
                                    \nextSigVar = \sigVar
                                }.
            \end{array}
            $$
            Furthermore, we define the set
            $\WaitSignalSet[\progOrdGraph] \subset \SignalSet$
            of signals being waited for in \progOrdGraph and its subset
            $\InfWaitSigSet[\progOrdGraph] \subseteq \WaitSignalSet[\progOrdGraph]$
            of signals waited-for infinitely often in \progOrdGraph
            as follows:
            $$
            \begin{array}{c c l}
                    \WaitSignalSet[\progOrdGraph] &\ := \
                            &\setOf{
                                    \sigVar \in \SignalSet
                                    \ | \
                                    \awaitEdges[\progOrdGraph](\sigVar) \neq \emptyset
                                },
                    \\
                    \InfWaitSigSet[\progOrdGraph] &\ := \
                            &\setOf{
                                            \infSig \in \WaitSignalSet[\progOrdGraph] \ | \
                                            \awaitEdges[\progOrdGraph][\infSig]\ \text{infinite}
                                    }.
            \end{array}
            $$
    \end{definition}

    \begin{definition}[Partial Order on Finite Bags]
    {
            \NewDocumentCommand{\baseSet}{}{\metaVar{X}}
            \NewDocumentCommand{\baseLt}{}{\ensuremath{<_\baseSet}}
            
            \NewDocumentCommand{\smallerBag}{}{\metaVar{A}}
            \NewDocumentCommand{\biggerBag}{}{\metaVar{B}}
            
            \NewDocumentCommand{\removedBag}{}{\metaVar{C}}
            \NewDocumentCommand{\addedBag}{}{\metaVar{D}}
            
            \NewDocumentCommand{\elemRemovedBag}{}{\metaVar{c}}
            \NewDocumentCommand{\elemAddedBag}{}{d}
            
            Let \baseSet be a set and $\baseLt \, \subset \baseSet \times \baseSet$ a partial order on \baseSet.
            We define the partial order 
            $\bagLt[\baseSet]\, \subset \FinBagsOf{\baseSet} \times \FinBagsOf{\baseSet}$
            on finite bags over \baseSet
            as the 
            Dershowitz-Manna ordering~\cite{Dershowitz1979MultisetOrder} induced by \baseLt:
            $$
            \begin{array}{c}
                    \smallerBag \bagLt[X] \biggerBag
                    \ \ \Longleftrightarrow
                    \begin{array}[t]{l l}
                    \exists \removedBag, \addedBag \in \FinBagsOf{X}.
                            &\msEmpty \neq \removedBag \subseteq \biggerBag\\
                            &\wedge\ 
                            \smallerBag = 
                                    (\biggerBag \setminus \removedBag) \msCup \addedBag
                            \\
                            &\wedge\
                            \forall \elemAddedBag \in \addedBag.\
                            \exists \elemRemovedBag \in \removedBag.\
                            \elemAddedBag \baseLt \elemRemovedBag.
                    \end{array}
            \end{array}
            $$
            We define $\bagLeq[X]\, \subset \FinBagsOf{X} \times \FinBagsOf{X}$ such that
            $$
                    \smallerBag \bagLeq[\baseSet] \biggerBag
                    \ \ \Longleftrightarrow \ \
                    \smallerBag = \biggerBag \ \vee\  \smallerBag \bagLt[\baseSet] \biggerBag
            $$
            holds.
    }
    \end{definition}

    \begin{corollary}\label{appendix:cor:natBagLtWellFounded}
            The partial order 
            $\bagLt[\N] \ \subset\ \FinBagsOf{\N} \times \FinBagsOf{\N}$
            is well-founded.
    \end{corollary}
    \begin{proof}
            Follows from~\cite{Dershowitz1979MultisetOrder} and and well-foundedness of $<_\N$.
    \end{proof}

    Below we define a metric on commands in a graph that allows us to prove that every control flow path in which no signal is waited-for infinitely often is finite.
    We construct the metric in two steps.
    (i)~We consider the maximal degree for which the command contains a nested loop.
    Here, we consider an uninitialized loop \cmdWhileSkip{\cmdVar} as a doubly nested one since its reduction in the annotated semantics involves its conversion into either a decrease loop or into an await loop, cf.~\astRedWhileDecInitName and~\astRedAwaitInitName.
    (ii)~We use the extracted degree to construct a rank (in form of a finite bag of degrees) that intuitively spoken measures the number of loop iterations the command's execution causes in the graph we consider.

        \begin{definition}[Degree Extraction]\label{appendix:def:DegreeExtraction}
                We define the function
                $\extractDegree{\cdot} : \CmdSet \rightarrow \N$
                mapping augmented commands to degrees by recursion on the structure of commands as follows:
                $$
                \begin{array}{l c l l}
                        \extractDegree{\cmdWhileSkip{\cmdVar}} &\ := \
                                &\extractDegree{\cmdVar} + 2,\\
                        \extractDegree{\icmdWhileDecStarted{\natVar}{\cmdVar}} &:=
                                &\extractDegree{\cmdVar} + 1,\\
                        \extractDegree{\icmdAwaitStarted{\SigSetVar}{\mutVar}{\cmdVar}} &:=
                                &\extractDegree{\cmdVar} + 1,\\
                        \extractDegree{\cmdFork{\cmdVar}} &:=
                                &\extractDegree{\cmdVar},\\
                        \extractDegree{\cmdLet{\varVar}{\cmdVar}{\nextCmdVar}} &:=
                                &\max(\extractDegree{\cmdVar}, \extractDegree{\nextCmdVar}),\\
                        \extractDegree{\cmdIf{\cmdVar}{\nextCmdVar}} &:=
                                &\max(\extractDegree{\cmdVar}, \extractDegree{\nextCmdVar}),\\
                        \extractDegree{\cmdVar} &:= 
                                &0
                                &\ \text{otherwise}.
                \end{array}
                $$
        \end{definition}
        
        \begin{definition}[Rank Extraction]\label{appendix:def:RankExtraction}
                For any subgraph \progOrdGraph of a program order graph with $\InfWaitSigSet[\progOrdGraph] = \msEmpty$, we define the function
                $\extractDegBag{\cdot} : \CmdSet \rightarrow \FinBagsOf{\N}$ mapping commands to finite bags of degrees by recursion on the structure of commands as follows:
                $$
                \begin{array}{l c l}
                        \extractDegBag{\cmdWhileSkip{\cmdVar}} &\ :=\
                                &\multiset{\extractDegree{\cmdWhileSkip{\cmdVar}}},
                        \\
                        \extractDegBag{\icmdWhileDecStarted{\natVar}{\cmdVar}} &\ :=\
                                &\displaystyle
                                        \natVar \cdot \multiset{\extractDegree{\icmdWhileDecStarted{\natVar}{\cmdVar}}},
                        \\
                        \extractDegBag{\icmdAwaitStarted{\SigSetVar}{\mutVar}{\cmdVar}} &\ :=\
                                &\displaystyle
                                        \cardinalityOf{
                                                \bigcup_{\sigVar \in \SigSetVar} \awaitEdges[\progOrdGraph](\sigVar)
                                        }
                                \cdot\ \multiset{\extractDegree{\icmdAwaitStarted{\SigSetVar}{\mutVar}{\cmdVar}}},
                        \\
                        \extractDegBag{\cmdFork{\cmdVar}} &\ :=\
                                &\extractDegBag{\cmdVar},
                        \\
                        \extractDegBag{\cmdLet{\varVar}{\cmdVar}{\nextCmdVar}} &\ :=\
                                &\extractDegBag{\cmdVar} \msCup \extractDegBag{\nextCmdVar},
                        \\
                        \extractDegBag{\cmdIf{\cmdVar}{\nextCmdVar}} &\ :=\
                                &\extractDegBag{\cmdVar} \msCup \extractDegBag{\nextCmdVar},
                        \\
                        \extractDegBag{\cmdVar} &\ :=\
                                &\msEmpty
                                \quad\quad\quad\quad\quad\quad\quad\quad
                                \text{otherwise.}
                \end{array}
                $$
        \end{definition}

    We view paths in a program order graph as single-branched subgraphs.
    This allows us to apply above definition on graphs to paths.
    In the proof of the following lemma, we see that in any control flow path where every signal is waited-for only finitely often, the rank of the reduced command decreases step by step.

    \begin{restatable}{lemma}{lemNoInfSigOnPathImpliesPathFinite}\label{lem:NoInfSigOnPathImpliesPathFinite}
            Let 
            $\progOrdGraph[\ahtpRSeq]$
            be a program order graph and let
            $\pathVar = (\NodeSetVar, \EdgeSetVar)$
            be a path in \progOrdGraph with 
            $\InfWaitSigSet[\pathVar] = \emptyset$.
            For every 
            $\tidVar \in \defDom{\atpVar[0]}$
            let $\atGetHeap{\atpVar[0](\tidVar)}$ be finite and complete.
            Then, \pathVar is finite.
    \end{restatable}
    { 
            \NewDocumentCommand{\rootNode}{}{\metaVar{r}}
            \NewDocumentCommand{\nodeToDegBag}{}{\metaVar{f}}
            \NewDocumentCommand{\pSuff}{m}{\metaVar{\pathVar_{\geq #1}}}
            
            Assume \pathVar is infinite.
            We prove a contradiction by assigning a decreasing metric to every node along the path.
            For every $i \in \NodeSetVar$, let \pSuff{i} be the suffix of \pathVar starting at node $i$.
            Every node $i \in \NodeSetVar$ corresponds to a reduction step
            \atpRedStep[\aheapVar[i]]{\atpVar[i]}[\tidVar_i][\aheapVar[j]]{\atpVar[j]}.
            In the following, we let $\cmdVar_i := \atpVar[i](\tidVar_i)$ denote the command reduced in this step.
            Consider the function
            $\nodeToDegBag : \NodeSetVar \rightarrow \FinBagsOf{\N}$,
            $i \mapsto \extractDegBag{\cmdVar_i}[\pSuff{i}]$
            mapping every node on the path to the rank of the command whose reduction the node represents.
            
            Consider the sequence
            $(\nodeToDegBag(i))_{i \in \NodeSetVar}$.
            Since every element is a finite bag of natural numbers, we can order it by \bagLt[\N].
            We are going to prove a contradiction by proving that the sequence
            is an infinitely descending chain.
            
            Consider any edge
            $\edgeVar = (i, \tidVar, r, j) \in \EdgeSetVar$.
            There are only 4 cases in which 
            $\nodeToDegBag(i) \neq \nodeToDegBag(j)$
            holds.
            \begin{itemize}
                    \item $r = \astRedWhileDecInitName$:\\
                            In this case we have 
                            $\cmdVar_i = \cmdWhileSkip{\cmdVar}$ and 
                            $\cmdVar_j = \icmdWhileDecStarted{\natVar}{\cmdVar}$
                            for some $\cmdVar, \natVar$.
                            We get
                            $$\nodeToDegBag(j) 
                                    \ = \
                                    \natVar \cdot \multiset{\extractDegree{\cmdVar}+1}
                                    \ \bagLt[\N]\
                                    \multiset{\extractDegree{\cmdVar}+2}
                                    \ =\
                                    \nodeToDegBag(i).
                            $$
                    \item $r = \astRedWhileDecName$:\\
                            In this case we have
                            $\cmdVar_i = \icmdWhileDecStarted{\natVar}{\cmdVar}$
                            and 
                            $\cmdVar_j = \cmdIf{\cmdVar}{\icmdWhileDecStarted{\natVar-1}{\cmdVar}}$
                            for some $\natVar, \cmdVar$.
                            We get
                            $$\nodeToDegBag(j)
                                    \ = \
                                    \multiset{\extractDegree{\cmdVar}} 
                                            \msCup
                                            (\natVar-1) \cdot \multiset{\extractDegree{\cmdVar}+1}
                                    \ \bagLt[\N]\
                                    \natVar \cdot \multiset{\extractDegree{\cmdVar}+1}
                                    \ = \
                                    \nodeToDegBag(i).
                            $$
                    \item $r = \astRedAwaitInitName$:\\
                            In this case we have
                            $\cmdVar_i = \cmdAwait{\mutVar}{\cmdVar}$
                            and\\
                            $\cmdVar_j = 
                                    \cmdIf
                                            {
                                                    \cmdAcquireMut{\mutVar};
                                                    \cmdLet{\varVar}{\cmdVar}
                                                            {
                                                                    \cmdReleaseMut{\mutVar};
                                                                    \neg\varVar
                                                            }
                                            }
                                            {\icmdAwaitStarted{\SigSetVar}{\mutVar}{\cmdVar}}
                            $
                            for some $\mutVar, \varVar, \cmdVar, \SigSetVar$.
                            We get
                            $$\nodeToDegBag(j)
                                    \ = \
                                    \multiset{\extractDegree{\cmdVar}}
                                    \msCup
                                    \cardinalityOf{\dots} \cdot \multiset{\extractDegree{\cmdVar} + 1}
                                    \ \bagLt[\N] \
                                    \multiset{\extractDegree{\cmdVar} + 2}
                                    \ = \
                                    \nodeToDegBag(i).
                            $$
                    \item $r = (\astRedAwaitName, \nextSigVar)$ for some \nextSigVar:\\
                            In this case we have 
                            $\cmdVar_i = \icmdAwaitStarted{\SigSetVar}{\mutVar}{\cmdVar}$
                            and\\
                            $\cmdVar_j = 
                                    \cmdIf
                                            {
                                                    \cmdAcquireMut{\mutVar};
                                                    \cmdLet{\varVar}{\cmdVar}
                                                            {
                                                                    \cmdReleaseMut{\mutVar};
                                                                    \varVar
                                                            }
                                            }
                                            {\icmdAwaitStarted{\SigSetVar}{\mutVar}{\cmdVar}}
                            $
                            for some $\mutVar, \varVar, \cmdVar, \SigSetVar$.
                            We get
                            $$\begin{array}{c c c c c}
                                    \displaystyle
                                    \nodeToDegBag(j)
                                    &\ =\
                                    &\multiset{\extractDegree{\cmdVar}}
                                            \msCup
                                            \cardinalityOf{
                                                    \bigcup_{\sigVar \in \SigSetVar} \awaitEdges[\pSuff{j}](\sigVar)
                                            }
                                                    \cdot \multiset{\extractDegree{\cmdVar}+1}
                                    \\
                                    \displaystyle
                                    &\ \bagLt[\N]\
                                            &\left(
                                                    1 + 
                                                    \cardinalityOf{
                                                            \bigcup_{\sigVar \in \SigSetVar} \awaitEdges[\pSuff{j}](\sigVar)
                                                    }
                                            \right)
                                                    \cdot \multiset{\extractDegree{\cmdVar}+1}
                                    \\
                                    \\
                                    \displaystyle
                                    &\ = \
                                            &\cardinalityOf{
                                                    \bigcup_{\sigVar \in \SigSetVar} \awaitEdges[\pSuff{i}](\sigVar)
                                            }
                                                    \cdot \multiset{\extractDegree{\cmdVar}+1}
                                    \ = \
                                    \nodeToDegBag(i)
                            \end{array}
                            $$
            \end{itemize}
            
            By application of Lemma~\ref{lem:InfinitePathsCotainInfinitelyManyLoopEdges} we get that \pathVar contains infinitely many of the loop edges listed above.
            Hence, $(\nodeToDegBag(i))_{i \in \NodeSetVar}$ is an infinitely decreasing chain.
            By Corollary~\ref{appendix:cor:natBagLtWellFounded}, \bagLt[\N] is well-founded.
            A contradiction.
            So, \pathVar is finite.
    } 

    We proceed to prove that no signals are waited for infinitely often.

    \begin{restatable}{lemma}{lemNoSignalWaitedForInfinitelyOften}\label{lem:NoSignalWaitedForInfinitelyOften}
            Let \ahtpRSeq be a fair annotated reduction sequence with
            $\atpVar[0] = \setOf{(\tidStartVar, (\lheapVar[0], \cmdVar))}$,
            \ahFinite{\aheapVar[0]},
            \lhComplete{\lheapVar[0]},
            \lhFinite{\lheapVar[0]}
            and \amConfConsistent{\aheapVar[0]}{\atpVar[0]}.
            Let \lheapVar[0] contain no signal chunks.
            Further, let \aheapVar[0] contain no chunks of the form
            \aresUnlocked{\mutVar}{\assLockInvVar}{\lheapLockInvVar}
            where \lheapLockInvVar contains any signal chunks.
            Let \progOrdGraph be the program order graph of \ahtpRSeq.
            Then, $\InfWaitSigSet[\progOrdGraph] = \emptyset$.
    \end{restatable}
    \begin{proof}
    { 
            \NewDocumentCommand{\obTid}{}{\metaVar{\tidVar_\text{ob}}}
            \NewDocumentCommand{\lowerSig}{}{\metaVar{z}}
            \NewDocumentCommand{\obPath}{}{\metaVar{p_\text{ob}}}
            \NewDocumentCommand{\obExtPath}{}{\metaVar{\bar{\obPath}}}
            
            \NewDocumentCommand{\obPathNodes}{}{\metaVar{V_\text{ob}}}
            
            \NewDocumentCommand{\tmpCapacity}{o}
                    {\ensuremath{
                            \fixedFuncNameFont{cap}
                            \IfValueT{#1}{(#1)}
                    }\xspace}

            Suppose $\InfWaitSigSet[\progOrdGraph] \neq \emptyset$.
            Since \LevelSet is well-founded, the same holds for the set \setOf{\sigGetLev{\sigVar} \ | \ \sigVar \in \InfWaitSigSet}.
            Hence, there is some $\minSig \in \InfWaitSigSet$ for which no $\lowerSig \in \InfWaitSigSet$ with
            $\sigGetLev{\lowerSig} \levLt \sigGetLev{\minSig}$ exists.
            
            Since neither the initial logical heap \lheapVar[0] nor any unlocked lock invariant stored in \aheapVar[0] does contain any signals, \minSig must be created during the reduction sequence.
            The reduction step creating signal \minSig is an application of \gtpRedNewSignalName, which simultaneously creates an obligation to set \minSig.
            By preservation of completeness, Lemma~\ref{lem:PreservationOfCompleteness}, every thread-local logical heap \atGetHeap{\atpVar[i](\tidVar)} annotating some thread \tidVar in some step $i$ is complete.
            According to reduction rule \astRedAwaitName, every await edge
            $(a, \tidVar, (\astRedAwaitName, \minSig), b)$
            implies together with completeness that in step $a$ (i)~thread \tidVar does not hold any obligation for \minSig 
            (i.e.\@ 
                $\atGetHeap{\atpVar[a](\tidVar)}(\lresObs{\obBagVar}) = 1$
                for some bag of obligations \obBagVar with
                $\minSig \not\in \obBagVar$
            )
            and (ii)~\minSig has not been set, yet
            (i.e.\@
                $\aresSignal{\minSig}{\slFalse} \in \aheapVar[a]$).
            Hence, in step $a$ another thread $\obTid \neq \tidVar$ must hold the obligation for \minSig
            (i.e.\@
                $\atGetHeap{\atpVar[a](\obTid)}(\lresObs{\obBagVar}) = 1$
                for some bag of obligations \obBagVar with
                $\minSig \in \obBagVar$).
            Since there are infinitely many await edges concerning \minSig in \progOrdGraph, the signal is never set.
            
            By fairness, for every await edge as above, there must be a non-ghost reduction step 
            \atpRedStep
                    [\aheapVar[k]]{\atpVar[k]}
                    [\obTid]
                    [\aheapVar[k+1]]{\atpVar[k+1]}
            of the thread \obTid holding the obligation for \minSig with $k \geq a$.
            Hence, there exists an infinite path \obPath in \progOrdGraph where each edge $(e, \obTid, n, f) \in \edges{\obPath}$ concerns some thread \obTid holding the obligation for \minSig.
            (Note that this thread ID does not have to be constant along the path, since the obligation can be passed on during fork steps.)
            
            The path \obPath does not contain await edges
            $(e, \obTid, (\astRedAwaitName, \infSig), f)$
            for any $\infSig \in \InfWaitSigSet$,
            since reduction rule \astRedAwaitName would (together with completeness of \atGetHeap{\atpVar[e](\obTid)}) require
            \infSig to be of a lower level than all held obligations.
            This restriction implies
            $\sigGetLev{\infSig} \levLt \sigGetLev{\minSig}$
            and would hence contradict the minimality of \minSig.
            That is, $\InfWaitSigSet[\obPath] = \emptyset$.

            By preservation of finiteness, Lemma~\ref{lem:PreservationOfFiniteness}, we get that every logical heap associated with the root of \obPath is finite.
            This allows us to apply Lemma~\ref{lem:NoInfSigOnPathImpliesPathFinite}, by which we get that \obPath is finite.
            A contradiction.
    } 
    \end{proof}

    Finally, we got everything we need to prove that any program that discharges all its obligations terminates.

    \begin{lemma}\label{lem:NoFairInfiniteAnnotatedRedSeq}
            Let \htModels{\noObs}{\cmdVar}{\noObs} hold.
            There exists no fair, infinite annotated reduction sequence \ahtpRSeq with
            $\atpVar[0] = \setOf{(\tidStartVar, (\lheapVar[0], \cmdVar))}$,
            $\aheapVar[0] = \emptyset$
            and
            $\lheapVar[0]  =  \lheapOf{\lresObs{\msEmpty}}$.
    \end{lemma}
    \begin{proof}
    { 
            \NewDocumentCommand{\infPath}{}{\metaVar{\pathVar_\infty}}
            Suppose a reduction sequence as described above exists.
            We are going to prove a contradiction by considering its infinite program order graph \progOrdGraph.

            Since \atpVar[0] contains only a single thread, \progOrdGraph is a binary tree with an infinite set of vertices.
            By the Weak König's Lemma~\cite{Simpson1999WKL} \progOrdGraph has an infinite branch, i.e. an infinite path \pathVar starting at root 0.

            The initial logical heap \lheapVar[0] is complete and finite and the initial annotated machine configuration $(\aheapVar[0], \atpVar[0])$ is consistent.
            By Lemma~\ref{lem:NoSignalWaitedForInfinitelyOften} we know that $\InfWaitSigSet[\progOrdGraph] = \emptyset$.
            Since
            $\InfWaitSigSet[\pathVar] \subseteq \InfWaitSigSet[\progOrdGraph]$,
            we get
            $\InfWaitSigSet[\pathVar] = \emptyset$.            
            This allows us to apply Lemma~\ref{lem:NoInfSigOnPathImpliesPathFinite}, by which we get that \pathVar is finite, which is a contradiction.
    } 
    \end{proof}

    \theoSoundness*
    \begin{proof}
    { 
            Assume that such a reduction sequence exists.
            By Hoare triple soundness, Lemma~\ref{lem:HoareTripleSoundness},
            we get
            \htModels{\noObs}{\cmdVar}{\noObs}
            from
            \htProves{\noObs}{\cmdVar}{\noObs}.
            Consider the logical heap 
            $\lheapVar[0] = \lheapOf{\lresObs{\msEmpty}}$
            and the annotated heap
            $\aheapVar[0] = \emptyset$.
            It holds
            \assModels
                    {\lheapVar[0]}
                    {\noObs},
            \alhCorresp[\aheapVar[0]][\lheapVar[0]] (since \lheapVar[0] does not contain any logical resources with an annotated counterpart)
            and
            \aphCorresp{\aheapVar[0]}{\pheapVar[0]} (since both heaps are empty).
            This allows us to apply Lemma~\ref{lem:ConstructionOfAnnotatedRedSeq}, by which
            we can construct a corresponding fair annotated reduction sequence \ahtpRSeq that starts with
            $\aheapVar[0] = \emptyset$ and 
            $\atpVar[0] = 
                    \setOf{(\tidStartVar, (\lheapVar[0], \cmdVar))}
            $.
            By Lemma~\ref{lem:NoFairInfiniteAnnotatedRedSeq} \ahtpRSeq does not exist.
            A contradiction.
    } 
    \end{proof}

%% file: CAV_paper_sections/extended_version/realisticExampleTransformation.tex
In this section, we verify the realistic example program from \S~\ref{sec:RealisticExample} presented in Fig.~\ref{fig:RealisticExample}.
We present the full proof outline in Fig.~\ref{fig:VerificationExampleBoundedFifoInitSL} -- \ref{fig:VerificationExampleBoundedFifoConsumptionStepSL}.

    {
                    \NewDocumentCommand{\fifo}{}{\progVar{fifo_{10}}}
                    \NewDocumentCommand{\fifoLoaded}{}{\progVar{f}}
                    \NewDocumentCommand{\m}{}{\mutProgVar}

                    \NewDocumentCommand{\counter}{m}{\progVar{c_{#1}}}
                    
                            \NewDocumentCommand{\producerID}{}{p}
                            \NewDocumentCommand{\pc}{}{\counter{\producerID}}
                            \NewDocumentCommand{\pcLoaded}{}{\progVar{c}}

                            \NewDocumentCommand{\consumerID}{}{c}
                            \NewDocumentCommand{\cc}{}{\counter{\consumerID}}
                            \NewDocumentCommand{\ccLoaded}{}{\progVar{c}}

                    
                            \NewDocumentCommand{\locFifo}{}{\metaVar{\hlocVar_\fifo}}
                            \NewDocumentCommand{\valFifoInv}{}{\metaVar{\valVar_\fifo^\m}}

                            \NewDocumentCommand{\locCC}{}{\metaVar{\hlocVar_{\cc}}}
                            \NewDocumentCommand{\valCC}{}{\metaVar{\valVar_\cc}}
                            \NewDocumentCommand{\nextValCC}{}{\metaVar{\valVar_\cc'}}
                            \NewDocumentCommand{\valCCInv}{}{\metaVar{\valVar_\cc^\m}}
                            \NewDocumentCommand{\cObBag}{}{\metaVar{\obBagVar_\consumerID}}
                            \NewDocumentCommand{\cNextObBag}{}{\metaVar{\obBagVar_\consumerID'}}
                            
                            \NewDocumentCommand{\cLoopInv}{m}{\metaVar{L_\consumerID(#1)}}

                            \NewDocumentCommand{\cPostIf}{}{\metaVar{\fixedPredNameFont{PostIf}_\consumerID}}

                            \NewDocumentCommand{\locPC}{}{\metaVar{\hlocVar_\pc}}
                            \NewDocumentCommand{\valPC}{}{\metaVar{\valVar_\pc}}
                            \NewDocumentCommand{\nextValPC}{}{\metaVar{\valVar_\pc'}}
                            \NewDocumentCommand{\valPCInv}{}{\metaVar{\valVar_\pc^\m}}
                            \NewDocumentCommand{\pObBag}{}{\metaVar{\obBagVar_\producerID}}
                            \NewDocumentCommand{\pNextObBag}{}{\metaVar{\obBagVar_\producerID'}}
                            
                            \NewDocumentCommand{\pLoopInv}{m}{\metaVar{L_\producerID(#1)}}
                            
                            \NewDocumentCommand{\pPostIf}{}{\metaVar{\fixedPredNameFont{PostIf}_\producerID}}
            
                            \NewDocumentCommand{\mut}{}{\mutVar}
                            \NewDocumentCommand{\locM}{}{\metaVar{\hlocVar_\m}}
                            \NewDocumentCommand{\invM}{}{\metaVar{P_\m}}
                            \NewDocumentCommand{\levM}{}{\metaVar{0}}
                            \NewDocumentCommand{\invMNoExFifo}{m}{\metaVar{P_\m'(#1)}}
                            \NewDocumentCommand{\invMNoSigPush}{}
                                    {
                                            \invM^{\substack{\text{no:\sigPush{}}}}
                                    }
                            \NewDocumentCommand{\invMNoSigPop}{}
                                    {
                                            \invM^{\substack{\text{no:\sigPop{}}}}
                                    }
            
                            \NewDocumentCommand{\idGen}{m m}{\metaVar{\idVar_\text{#1}^{#2}}}
                            
                            \NewDocumentCommand{\idPush}{m}{\idGen{push}{#1}}
                            \NewDocumentCommand{\levPush}{m}{\metaVar{\levVar_\text{push}^{#1}}}
                            
                            \NewDocumentCommand{\idPop}{m}{\idGen{pop}{#1}}
                            \NewDocumentCommand{\levPop}{m}{\metaVar{\levVar_\text{pop}^{#1}}}
                            
                            \NewDocumentCommand{\sigPush}{m}{\metaVar{\sigVar_\text{push}^{#1}}}
                            \NewDocumentCommand{\sigPop}{m}{\metaVar{\sigVar_\text{pop}^{#1}}}


    \begin{figure}

            \makebox[\textwidth][c]
            {$
            \hspace{0.06\textwidth}  
            \begin{array}{l | p{0.266\pdfpagewidth}}
                    \progProof{
                            \noObs 
                    }
                    \\
                    \keyword{let}\ \fifo \mathop{:=} \cmdAlloc{\lstNil}\ \keyword{in}
                    \
                    \keyword{let}\ \m \mathop{:=} \cmdNewMut\ \keyword{in}
                            &\ \ \proofRuleHint{\prLetName (2x) \& \prAllocName}
                    \\
                    \ghostFont{\forall \locFifo, \locM.}
                            &\ \ \proofRuleHint{\& \prNewMutexName}
                    \\
                    \progProof{ 
                            \noObs
                            \slStar
                            \progProofNew{
                                    \slPointsTo{\locFifo}{\lstNil}
                                    \slStar
                                    \assMutUninit{\locM}
                            }
                    }
                            &\ \ \proofRuleHintML{\prViewShiftName \& \vsAllocSigIDName\\
                                    \& \prExistsName (200x)
                            }
                    \\
                    \ghostFont{\forall \idPop{1}, \dots \idPop{100}, \idPush{1}, \dots, \idPush{100}.}
                    \\
                    \progProof{
                            \progProofNew{
                                    \slBigStar_{i = 1, \dots, 100}
                                            \quadBack\quadBack\quadBack
                                            \sigUninit{\idPop{i}}
                                    \slStar
                                    \slBigStar_{i = 1, \dots, 100}
                                            \quadBack\quadBack\quadBack
                                            \sigUninit{\idPush{i}}
                            }
                            \slStar \dots
                    }
                    \\
                    \proofDef{
                            \levPop{i} \ := \ 102-i,
                            \quad
                            \levPush{i} :=\ 101-i
                            \quad
                            \text{for}
                            \quad
                            1 \,\leq\, i \,\leq\, 100
                    }
                    \\
                    \proofHint{(Later
                            $\levPop{i+10} < \levPush{i}$
                            and
                            $\levPush{i} < \levPop{i}$
                            must hold, cf. Figures~\ref{fig:VerificationExampleBoundedFifoProducerLoopSL} and~\ref{fig:VerificationExampleBoundedFifoConsumerLoopSL}.)
                    }
                    \\
                    \proofDef{
                            \sigPush{i} \ := \ (\idPush{i}, \levPush{i}),
                            \quad
                            \sigPop{i} \ := \ (\idPop{i}, \levPop{i})
                            \quad
                            \text{for}
                            \quad
                            1 \,\leq\, i \,\leq\, 100
                    }
                            &\ \ \proofRuleHint{\prViewShiftName \& \vsSigInitName}
                    \\
                    \progProofML{
                            \progProofCancel{
                                    \sigUninit{\idPop{100}}
                                    \slStar
                                    \sigUninit{\idPush{100}}
                            }
                            \,
                            \progProofNew{
                                    \assSignal{\sigPop{100}}{\slFalse}
                                    \slStar
                                    \assSignal{\sigPush{100}}{\slFalse}
                            }
                            \\
                            \slStar\,
                            \obsOf{\progProofChanged{
                                    \sigPop{100}, \sigPush{100}
                            }}
                            \slStar \dots
                    }
                    \ 
                    \\
                    \keyword{let}\ \pc \mathop{:=} \cmdAlloc{100}\ \keyword{in}
                    \
                    \keyword{let}\ \cc \mathop{:=} \cmdAlloc{100}\ \keyword{in}
                            &\ \ \proofRuleHint{\prLetName \& \prAllocName (2x)}
                    \\
                    \ghostFont{\forall \locPC, \locCC.}
                    \\
                    \progProof{
                            \progProofNew{
                                    \slPointsTo{\locPC}{100}
                                    \slStar 
                                    \slPointsTo{\locCC}{100}
                            }
                            \slStar \dots
                    }
                            &\ \ \proofRuleHint{\prViewShiftName \& \vsSemImpName}
                    \\
                    \progProofML{
                            \obsOf{\sigPush{100}, \sigPop{100}}
                            \slStar
                            \slPointsTo[\progProofChanged{\frac{1}{2}}]{\locPC}{100}
                            \slStar
                            \slPointsTo[\progProofChanged{\frac{1}{2}}]{\locCC}{100}
                            \slStar
                            \assMutUninit{\locM}
                            \\
                            \slStar
                            \progProofNew{\invM}
                            \\
                            \slStar
                            \slBigStar_{i = 1, \dots, 99}
                                    \quadBack\quadBack\quadBack\!
                                    (\sigUninit{\idPush{i}} \slStar \sigUninit{\idPop{i}})
                    }
                            &\ \ \referenceHintML{For definition of lock invariant \invM\\
                                    cf.\@ Fig.~\ref{fig:VerificationExampleBoundedFifoLockInvariantSL}.
                                    \\
                                    \proofRuleHint{\prViewShiftName \& \vsMutInitName}
                            }
                    \\
                    \proofDef{
                            \mut \ := \ (\locM, \levM)
                    }
                            \quad\quad\quad
                            \proofHintML{(Later
                                    $\mutGetLev{\mut} < \levPush{i}$
                                    and
                                    $\mutGetLev{\mut} < \levPop{i}$ 
                                    must hold
                                    \\
                                    \phantom{(} for all $1 \leq i \leq 100$, 
                                    cf.\@ Figures~\ref{fig:VerificationExampleBoundedFifoProducerLoopSL} and~\ref{fig:VerificationExampleBoundedFifoConsumerLoopSL}.)
                            }
                    \\
                    \progProof{
                            \progProofCancel{
                                    \assMutUninit{\locM}
                                    \slStar
                                    \invM
                            }\
                            \progProofNew{
                                    \assMutex{\mut}{\invM}
                            }
                            \slStar \dots
                    }
                            &\ \ \proofRuleHint{\prViewShiftName \& \vsSemImpName}
                    \\
                    \progProofML{
                            \obsOf{\sigPush{100}, \sigPop{100}}
                            \slStar
                            \slPointsTo[\frac{1}{2}]{\locPC}{100}
                            \slStar
                            \slPointsTo[\frac{1}{2}]{\locCC}{100}
                            \\
                            \slStar
                            \slBigStar_{i = 1, \dots, 99}
                                    \quadBack\quadBack\quadBack\!
                                    (\sigUninit{\idPush{i}} \slStar \sigUninit{\idPop{i}})
                            \\
                            \slStar\,
                            \assMutex[\progProofChanged{\frac{1}{2}}]{\mut}{\invM}
                            \slStar
                            \assMutex[\progProofChanged{\frac{1}{2}}]{\mut}{\invM}
                    }
                    \\
                    \dots &\ \ \referenceHintML{Continued in Fig.~\ref{fig:VerificationExampleBoundedFifoForkingSL}.}
            \end{array}
            $}

            \caption{Verification of realistic example~\ref{fig:RealisticExample}: Initialization.}
            \label{fig:VerificationExampleBoundedFifoInitSL}
    \end{figure}

    \begin{figure}
            \makebox[\textwidth][c]
            {$
            \hspace{0.025\textwidth}  
            \begin{array}{l | p{0.266\pdfpagewidth}}
                    \proofDef{
                            \invMNoExFifo{\valFifoInv} \ := \
                            \exists \valPCInv, \valCCInv.
                    }
                    \\
                    \proofDef{
                            \phantom{\invM \ :=\ }
                            \slPointsTo[\frac{1}{2}]{\locPC}{\valPCInv}
                            \ \slStar\
                            \slPointsTo[\frac{1}{2}]{\locCC}{\valCCInv}
                            \ \slStar \
                            0 \leq \valPCInv \leq 100
                            \ \slStar \
                            0 \leq \valCCInv \leq 100
                    }
                            &\ \ \proofHint{Producer \& consumer counters.}
                    \\
                    \proofDef{
                            \phantom{\invM \ :=\ }
                            \ \slStar\
                            \slPointsTo{\locFifo}{\valFifoInv}
                            \ \slStar\
                            \valCCInv = \valPCInv + \lstSize{\valFifoInv}
                            \ \slStar \
                            0 \leq \lstSize{\valFifoInv} \leq 10
                    }
                    \ \ \
                            &\ \ \proofHint{Bounded FIFO \& its relationship}
                    \\
                    \proofDef{
                            \phantom{\invM \ := \ }
                            \ \slStar\
                            \valFifoInv = 
                                    \lstCons{(\valPC + \lstSize{\valFifoInv})}
                                            {\lstCons{\dots}
                                                    {\lstCons{(\valPC+1)}
                                                            {\lstNil}
                                                    }
                                            }
                    }
                            &\ \ \proofHint{to counters.}
                    \\
                    \proofDef{
                            \phantom{\invM \ := \ }
                            \ \slStar\ 
                            \big(\valPCInv > 0 \ \rightarrow \
                                    \assSignal{(\idPush{\valPCInv}, \levPush{\valPCInv})}{\slFalse}
                            \big)
                    }
                            &\ \ \proofHint{Signal set by producer.}
                    \\
                    \proofDef{
                            \phantom{\invM \ := \ }
                            \ \slStar\ 
                            \big(\valCCInv > 0 \ \rightarrow \
                                    \assSignal{(\idPop{\valCCInv}, \levPop{\valCCInv})}{\slFalse}
                            \big)
                    }
                            &\ \ \proofHint{Signal set by consumer.}
                    \\
                    \\
                    \proofDef{
                            \invM \ := \
                            \exists \valFifoInv.\ \invMNoExFifo{\valFifoInv}
                    }
                    \\
                    \\
                    \proofDef{
                            \invMNoSigPush\ := \
                            \exists \valPCInv, \valCCInv.
                    }
                            &\ \ \proofHint{Shorthand for lock invariant}
                    \\
                    \proofDef{
                            \phantom{\invM \ :=\ }
                            \slPointsTo[\frac{1}{2}]{\locPC}{\valPCInv}
                            \ \slStar\
                            \slPointsTo[\frac{1}{2}]{\locCC}{\valCCInv}
                            \ \slStar \
                            0 \leq \valPCInv \leq 100
                            \ \slStar \
                            0 \leq \valCCInv \leq 100
                    }
                            &\ \ \proofHint{without push-signal chunk.}
                    \\
                    \proofDef{
                            \phantom{\invM \ :=\ }
                            \ \slStar\
                            \slPointsTo{\locFifo}{\valFifoInv}
                            \ \slStar\
                            \valCCInv = \valPCInv + \lstSize{\valFifoInv}
                            \ \slStar \
                            0 \leq \lstSize{\valFifoInv} \leq 10
                    }
                    \\
                    \proofDef{
                            \phantom{\invM \ := \ }
                            \ \slStar\
                            \valFifoInv = 
                                    \lstCons{(\valPC + \lstSize{\valFifoInv})}
                                            {\lstCons{\dots}
                                                    {\lstCons{(\valPC+1)}
                                                            {\lstNil}
                                                    }
                                            }
                    }
                    \\
                    \proofDef{
                            \phantom{\invM \ := \ }
                            \proofDefCancel{
                                    \ \slStar\ 
                                    \big(\valPCInv > 0 \ \rightarrow \
                                            \assSignal{(\idPush{\valPCInv}, \levPush{\valPCInv})}{\slFalse}
                                    \big)
                            }
                    }
                    \\
                    \proofDef{
                            \phantom{\invM \ := \ }
                            \ \slStar\ 
                            \big(\valCCInv > 0 \ \rightarrow \
                                    \assSignal{(\idPop{\valCCInv}, \levPop{\valCCInv})}{\slFalse}
                            \big)
                    }
                    \\
                    \\
                    \proofDef{
                            \invMNoSigPop\ := \
                            \exists \valPCInv, \valCCInv.
                    }
                            &\ \ \proofHint{Shorthand for lock invariant}
                    \\
                    \proofDef{
                            \phantom{\invM \ :=\ }
                            \slPointsTo[\frac{1}{2}]{\locPC}{\valPCInv}
                            \ \slStar\
                            \slPointsTo[\frac{1}{2}]{\locCC}{\valCCInv}
                            \ \slStar \
                            0 \leq \valPCInv \leq 100
                            \ \slStar \
                            0 \leq \valCCInv \leq 100
                    }
                            &\ \ \proofHint{without pop-signal chunk.}
                    \\
                    \proofDef{
                            \phantom{\invM \ :=\ }
                            \ \slStar\
                            \slPointsTo{\locFifo}{\valFifoInv}
                            \ \slStar\
                            \valCCInv = \valPCInv + \lstSize{\valFifoInv}
                            \ \slStar \
                            0 \leq \lstSize{\valFifoInv} \leq 10
                    }
                    \\
                    \proofDef{
                            \phantom{\invM \ := \ }
                            \ \slStar\
                            \valFifoInv = 
                                    \lstCons{(\valPC + \lstSize{\valFifoInv})}
                                            {\lstCons{\dots}
                                                    {\lstCons{(\valPC+1)}
                                                            {\lstNil}
                                                    }
                                            }
                    }
                    \\
                    \proofDef{
                            \phantom{\invM \ := \ }
                            \ \slStar\ 
                            \big(\valPCInv > 0 \ \rightarrow \
                                    \assSignal{(\idPush{\valPCInv}, \levPush{\valPCInv})}{\slFalse}
                            \big)
                    }
                    \\
                    \proofDef{
                            \phantom{\invM \ := \ }
                            \proofDefCancel{
                                    \ \slStar\ 
                                    \big(\valCCInv > 0 \ \rightarrow \
                                            \assSignal{(\idPop{\valCCInv}, \levPop{\valCCInv})}{\slFalse}
                                    \big)
                            }
                    }
            \end{array}
            $}

            \caption
            {Verification of realistic example~\ref{fig:RealisticExample}: Lock invariant.}
            \label{fig:VerificationExampleBoundedFifoLockInvariantSL}
    \end{figure}

    \begin{figure}

            \makebox[\textwidth][c]
            {$
            \hspace{0.025\textwidth}  
            \begin{array}{l | p{0.266\pdfpagewidth}}
                    \proofDef{
                            \pLoopInv{n, \pObBag} \ := \ 
                            0 < n \leq 100
                            \ \slStar \
                            \displaystyle
                            \quadBack
                            \slBigStar_{i = 1, \dots, n-1}
                                    \quadBack
                                    \sigUninit{\idPush{i}}
                    }
                            &\ \ \proofHint{Loop invariant of producer.}
                    \\
                    \proofDef{
                            \phantom{\pLoopInv{n, \pObBag} \ := \ }
                            \ \slStar \
                            (n > 0 \ \leftrightarrow \ \pObBag = \multiset{\sigPush{n}})
                            \ \slStar \
                            (n = 0 \ \leftrightarrow \ \pObBag = \msEmpty)
                    }
                    \ \ \
            \end{array}
            $}

            \caption{Verification of realistic example~\ref{fig:RealisticExample}: Producer's loop invariant.}
            \label{fig:VerificationExampleBoundedFifoProducerLoopInvariantSL}
    \end{figure}

    \begin{figure}

            \makebox[\textwidth][c]
            {$
            \hspace{0.025\textwidth}  
            \begin{array}{l | p{0.266\pdfpagewidth}}
                    \proofDef{
                            \cLoopInv{n, \cObBag} \ := \ 
                            0 < n \leq 100
                            \ \slStar \
                            \displaystyle
                            \quadBack
                            \slBigStar_{i = 1, \dots, n-1}
                                    \quadBack
                                    \sigUninit{\idPop{i}}
                    }
                            &\ \ \proofHint{Loop invariant of consumer.}
                    \\
                    \proofDef{
                            \phantom{\cLoopInv{n, \cObBag} \ := \ }
                            \ \slStar \
                            (n > 0 \ \leftrightarrow \ \cObBag = \multiset{\sigPop{n}})
                            \ \slStar \
                            (n = 0 \ \leftrightarrow \ \cObBag = \msEmpty)
                    }
                    \ \ \
            \end{array}
            $}

            \caption{Verification of realistic example~\ref{fig:RealisticExample}: Consumer's loop invariant,.}
            \label{fig:VerificationExampleBoundedFifoConsumerLoopInvariantSL}
    \end{figure}

    \begin{figure}

            \makebox[\textwidth][c]
            {$
            \hspace{0.025\textwidth}  
            \begin{array}{l | p{0.266\pdfpagewidth}}
                    \dots\ref{fig:VerificationExampleBoundedFifoInitSL}\dots
                            &\ \ \referenceHintML{Continuation of Fig.~\ref{fig:VerificationExampleBoundedFifoInitSL}.}
                    \\
                    \progProofML{
                            \obsOf{\sigPush{100}, \sigPop{100}}
                            \slStar
                            \slPointsTo[\frac{1}{2}]{\locPC}{100}
                            \slStar
                            \slPointsTo[\frac{1}{2}]{\locCC}{100}
                            \\
                            \slStar
                            \slBigStar_{i = 1, \dots, 99}
                                    \quadBack\quadBack\quadBack\!
                                    (\sigUninit{\idPush{i}} \slStar \sigUninit{\idPop{i}})
                            \\
                            \slStar\,
                            \assMutex[\frac{1}{2}]{\mut}{\invM}
                            \slStar
                            \assMutex[\frac{1}{2}]{\mut}{\invM}
                    }
                    \\
                    \keyword{fork}\ (
                            &\ \ \proofRuleHint{\prForkName}
                    \\\quad\quad
                            \progProofML{
                                    \progProofNew{
                                            \obsOf{\sigPush{100}}
                                            \slStar
                                            \slPointsTo[\frac{1}{2}]{\locPC}{100}
                                            \slStar
                                            \slBigStar_{i = 1, \dots, 99}
                                                    \quadBack\quadBack\quadBack\!
                                                    \sigUninit{\idPush{i}}
                                            \slStar
                                            \assMutex[\frac{1}{2}]{\mut}{\invM}
                                    }
                            }
                            \ \ 
                                    &\ \ \proofHintML{Resources transferred to\\producer thread.
                                            \\
                                            \proofRuleHint{\prViewShiftName \& \vsSemImpName}
                                    }
                    \\\quad\quad
                            \progProofML{
                                    \obsOf{\sigPush{100}}
                                    \slStar
                                    \slPointsTo[\frac{1}{2}]{\locPC}{100}
                                    \slStar
                                    \progProofCancel{
                                            \slBigStar_{i=1, \dots, 99}
                                                    \quadBack\quadBack\quadBack
                                                    \sigUninit{\sigPush{i}}
                                    }
                                    \,
                                    \progProofNew{
                                            \pLoopInv{100, \multiset{\sigPush{100}}}
                                    }
                                    \\
                                    \slStar
                                    \assMutex[\frac{1}{2}]{\mut}{\invM}
                            }
                                    &\ \ \referenceHintML{For definition of producer loop
                                            \\
                                            invariant \pLoopInv{n, \obBagVar} cf.\@
                                            Fig.~\ref{fig:VerificationExampleBoundedFifoProducerLoopInvariantSL}.
                                    }
                    \\\quad\quad
                            \dots\ref{fig:VerificationExampleBoundedFifoProducerLoopSL}\dots
                                    &\ \ \referenceHintML{Producer loop in Fig.~\ref{fig:VerificationExampleBoundedFifoProducerLoopSL}.}
                    \\\quad\quad
                            \progProofML{
                                    \progProofCancel{
                                            \obsOf{\sigPush{100}}
                                            \slStar
                                            \slPointsTo[\frac{1}{2}]{\locPC}{100}
                                            \slStar
                                            \pLoopInv{100, \multiset{\sigPush{100}}}
                                            \slStar
                                            \assMutex[\frac{1}{2}]{\mut}{\invM}
                                    }
                                    \\
                                    \progProofNew{
                                            \noObs
                                    }
                            }
                    \\);
                    \\
                    \progProofML{
                            \obsOf{
                                    \progProofCancel{\sigPush{100},}\,
                                    \sigPop{100}
                            }\,
                            \progProofCancel{
                                    \slStar
                                    \slPointsTo[\frac{1}{2}]{\locPC}{100}
                            }
                            \slStar
                            \slPointsTo[\frac{1}{2}]{\locCC}{100}
                            \\
                            \progProofCancel{
                                    \slStar
                                    \slBigStar_{i = 1, \dots, 99}
                                            \quadBack\quadBack\quadBack\!
                                            \sigUninit{\idPush{i}}
                            }
                            \slStar
                            \slBigStar_{i = 1, \dots, 99}
                                    \quadBack\quadBack\quadBack\!
                                    \sigUninit{\idPop{i}}
                            \\
                            \progProofCancel{
                                    \slStar
                                    \assMutex[\frac{1}{2}]{\mut}{\invM}
                            }
                            \slStar
                            \assMutex[\frac{1}{2}]{\mut}{\invM}
                    }
                            &\ \ \proofHintML{Resources remaining with\\consumer thread.
                                    \\
                                    \proofRuleHint{\prViewShiftName \& \vsSemImpName}
                            }
                    \\
                    \progProofML{
                            \obsOf{\sigPop{100}}
                            \slStar
                            \slPointsTo[\frac{1}{2}]{\locCC}{100}
                            \slStar
                            \progProofCancel{
                                    \slStar
                                    \slBigStar_{i = 1, \dots, 99}
                                            \quadBack\quadBack\quadBack\!
                                            \sigUninit{\idPop{i}}
                            }
                            \,
                            \progProofNew{
                                    \cLoopInv{100, \multiset{\sigPop{100}}}
                            }
                            \\
                            \slStar\
                            \assMutex[\frac{1}{2}]{\mut}{\invM}
                    }
                            &\ \ \referenceHintML{For definition of consumer loop
                                    \\
                                    invariant \cLoopInv{n, \obBagVar} cf.\@ Fig.~\ref{fig:VerificationExampleBoundedFifoConsumerLoopInvariantSL}.
                            }
                    \\
                    \dots
                            &\ \ \referenceHintML{Continued in Fig.~\ref{fig:VerificationExampleBoundedFifoConsumerLoopSL}\\(consumer loop).}
                    \\
                    \progProof{
                            \progProofCancel{
                                    \obsOf{\sigPop{100}}
                                    \slStar
                                    \slPointsTo[\frac{1}{2}]{\locCC}{100}
                                    \slStar
                                    \cLoopInv{100, \multiset{\sigPop{100}}}
                                    \slStar
                                    \assMutex[\frac{1}{2}]{\mut}{\invM}
                            }\,
                            \progProofNew{
                                    \noObs
                            }
                    }
            \end{array}
            $}

            \caption{Verification of realistic example~\ref{fig:RealisticExample}: Main thread (consumer) forks producer thread.}
            \label{fig:VerificationExampleBoundedFifoForkingSL}
    \end{figure}

    \begin{figure}

            \makebox[\textwidth][c]
            {$
            \hspace{0.07\textwidth}  
            \begin{array}{l | p{0.266\pdfpagewidth}}
                    \lowlightText{\forall \locFifo, \locM, \locPC, \locCC.}
                    \\
                    \dots &\ \ \referenceHintML{Continuation of Fig.~\ref{fig:VerificationExampleBoundedFifoForkingSL}.}
                    \\
                    \progProof{
                            \obsOf{\sigPush{100}}
                            \slStar
                            \slPointsTo[\frac{1}{2}]{\locPC}{100}
                            \slStar
                            \pLoopInv{100, \multiset{\sigPush{100}}}
                            \slStar
                            \assMutex[\frac{1}{2}]{\mut}{\invM}
                    }
                    \\
                    \keyword{while}\ (
                            &\ \ \proofRuleHint{\prWhileDecStrictName}
                    \\\quad\quad
                            \ghostFont{\forall \valPC.}
                    \\\vspace{-0.6cm}
                    \\\quad\quad
                            \progProof{
                                    \obsOf{\sigPush{\progProofChanged{\valPC}}}
                                    \slStar
                                    \slPointsTo[\frac{1}{2}]{\locPC}{\progProofChanged{\valPC}}
                                    \slStar
                                    \pLoopInv{\progProofChanged{\valPC}, \multiset{\sigPush{\progProofChanged{\valPC}}}}
                                    \slStar
                                    \assMutex[\frac{1}{2}]{\mut}{\invM}
                            }
                    \\\quad\quad
                            \proofHint{$
                                    \mutGetLev{\mut} 
                                    \ = \ 
                                    \levM 
                                    \ <\
                                    101 - \valPC
                                    \ = \
                                    \sigGetLev{\sigPush{\valPC}}
                            $}
                                    &\ \ \proofHint{Justification for application of:}
                    \\\quad\quad
                            \keyword{with}\ \m\ \keyword{await}\ (
                                    &\ \ \proofRuleHint{\prAwaitGenName}
                    \\\vspace{-0.84cm}\\\hspace{3.5cm}\referenceHintML{
                            For definition of producer loop invariant \pLoopInv{n, \obBagVar}, 
                    }
                    \\\quad\quad\quad\quad
                                    \ghostFont{\forall \pObBag.}
                    \\\vspace{-0.84cm}\\\hspace{3.5cm}\referenceHintML{
                            lock invariant \invM
                            and variations cf.\@ Figures~\ref{fig:VerificationExampleBoundedFifoProducerLoopInvariantSL}
                            and~\ref{fig:VerificationExampleBoundedFifoLockInvariantSL}.
                    }
                    \\\quad\quad\quad\quad
                                    \progProofML{
                                            \obs{
                                                    \progProofCancel{\multiset{\sigPush{\valPC}}}
                                                    \,
                                                    \progProofNew{
                                                            \pObBag
                                                            \msCup
                                                            \multiset{\mut}
                                                    }
                                            }
                                            \slStar
                                            \slPointsTo[\frac{1}{2}]{\locPC}{\valPC}
                                            \slStar
                                            \pLoopInv{
                                                    \valPC, 
                                                    \progProofCancel{\multiset{\sigPush{\valPC}}}
                                                    \,
                                                    \progProofNew{\pObBag}
                                            }
                                            \\
                                            \slStar
                                            \progProofCancel{\assMutex[\frac{1}{2}]{\mut}{\invM}}
                                            \,
                                            \progProofNew{\invM}
                                    }
                                            &\ \ \proofRuleHint{\prExistsName }
                    \\\quad\quad\quad\quad
                                    \ghostFont{\forall \valFifoInv.}
                    \\\quad\quad\quad\quad
                                    \progProof{
                                            \progProofCancel{\invM}\,
                                            \progProofNew{\invMNoExFifo{\valFifoInv}}
                                            \slStar \dots
                                    }
                    \\\quad\quad\quad\quad
                                    \keyword{let}\ \fifoLoaded \mathop{:=} \cmdReadHeapLoc{\fifo}\ \keyword{in}
                                            &\ \ \proofRuleHint{\prLetName \& \prReadHeapLocName}
                    \\\quad\quad\quad\quad
                                    \progProof{
                                            \pLoopInv{\valPC, \pObBag}
                                            \slStar
                                            \obs{\pObBag \msCup \multiset{\mut}}
                                            \slStar
                                            \slPointsTo[\frac{1}{2}]{\locPC}{\valPC}
                                            \slStar
                                            \invMNoExFifo{\valFifoInv}
                                    }
                    \\\quad\quad\quad\quad
                                    \keyword{if}\ \lstSize{\fifoLoaded} < 10\ \keyword{then}\ (
                                            &\ \ \proofRuleHint{\prIfName}
                    \\\quad\quad\quad\quad\quad\quad
                                            \progProof{
                                                    \progProofNew{
                                                            \lstSize{\valFifoInv} < 10
                                                    }
                                                    \slStar 
                                                    \pLoopInv{\valPC, \pObBag}
                                                    \slStar
                                                    \obs{\pObBag \msCup \multiset{\mut}}
                                                    \slStar
                                                    \slPointsTo[\frac{1}{2}]{\locPC}{\valPC}
                                                    \slStar
                                                    \invMNoExFifo{\valFifoInv}
                                            }
                    \\\quad\quad\quad\quad\quad\quad
                                            \dots \ref{fig:VerificationExampleBoundedFifoProductionStepSL} \dots
                                                    \quad\quad\quad\quad\quad\quad\quad\quad\quad\quad
                                                    \referenceHintML{Production step presented in Fig.~\ref{fig:VerificationExampleBoundedFifoProductionStepSL}.}
                    \\\quad\quad\quad\quad\quad\quad
                                            \progProofML{
                                                    \exists \pNextObBag.\
                                                    \obs{\pNextObBag \msCup \multiset{\mut}}
                                                    \\
                                                    \slStar
                                                    \slIfElseBranchML{\lstSize{\valFifoInv} \neq 10}
                                                            {
                                                                    \slPointsTo[\frac{1}{2}]{\locPC}{\valPC-1}
                                                                    \slStar
                                                                    \pLoopInv{\valPC-1, \pNextObBag}
                                                                    \slStar
                                                                    \invMNoExFifo{\valFifoInv}
                                                            }
                                                            {
                                                                    \assSignal{\sigPop{\valPC+10}}{\slFalse}
                                                                    \slStar
                                                                    \levPop{\valPC+10} < \levPush{\valPC}
                                                                    \slStar
                                                                    \slPointsTo[\frac{1}{2}]{\locPC}{\valPC}
                                                                    \\
                                                                    \slStar\
                                                                    \invMNoSigPop(\valFifoInv)
                                                                    \slStar
                                                                    \pLoopInv{\valPC, \pNextObBag}
                                                            }
                                            }
                                            &\ \ \proofDef{=: \pPostIf}
                    \\\quad\quad\quad\quad
                                    )\ \ghostFont{\keyword{else}\ (\ }
                    \\\quad\quad\quad\quad\quad\quad
                                            \progProof{
                                                    \progProofNew{\lstSize{\valFifoInv} = 10}
                                                    \slStar 
                                                    \pLoopInv{\valPC, \pObBag}
                                                    \slStar
                                                    \obs{\pObBag \msCup \multiset{\mut}}
                                                    \slStar
                                                    \slPointsTo[\frac{1}{2}]{\locPC}{\valPC}
                                                    \slStar
                                                    \invMNoExFifo{\valFifoInv}
                                            }\
                                                    &\ \ \proofRuleHint{\prViewShiftName \& \vsSemImpName}
                    \\\quad\quad\quad\quad\quad\quad
                                            \progProofML{
                                                    \progProofCancel{
                                                            \lstSize{\valFifoInv} = 10
                                                            \slStar 
                                                            \pLoopInv{\valPC, \pObBag}
                                                            \slStar
                                                            \obs{\pObBag \msCup \multiset{\mut}}
                                                            \slStar
                                                            \slPointsTo[\frac{1}{2}]{\locPC}{\valPC}
                                                            \slStar
                                                            \invMNoExFifo{\valFifoInv}
                                                    }\\
                                                    \progProofNew{
                                                            \pPostIf
                                                    }
                                            }
                                            \ \ 
                    \\\quad\quad\quad\quad
                                    \ghostFont{)};
                    \\\quad\quad\quad\quad
                                    \progProof{
                                            \progProofCancel{
                                                    \pLoopInv{\valPC, \pObBag}
                                                    \slStar
                                                    \obs{\pObBag \msCup \multiset{\mut}}
                                                    \slStar
                                                    \slPointsTo[\frac{1}{2}]{\locPC}{\valPC}
                                                    \slStar
                                                    \invMNoExFifo{\valFifoInv}
                                            }
                                            \,
                                            \progProofNew{\pPostIf}
                                    }
                    \\\vspace{-0.5cm}
                    \\\quad\quad\quad\quad
                                    \lstSize{\fifo} \neq 10
                                            &\ \ \proofRuleHintML{\prExpName\\
                                                    \& \prViewShiftName \& \vsSemImpName
                                            }
                    \\\vspace{-0.65cm}
                    \\\quad\quad\quad\quad
                                    \progProof{
                                            \dots \slStar
                                             \slIfElse{\dots}
                                                    {
                                                            \dots
                                                            \progProofCancel{\invMNoExFifo{\valFifoInv}}
                                                            \,
                                                            \progProofNew{\invM}
                                                    }
                                                    {\dots}
                                    }
                    \\\quad\quad
                            );
                    \\\quad\quad
                            \progProofML{
                                    \progProofNew{
                                            \exists \pNextObBag.
                                    }\
                                    \obs{
                                            \progProofCancel{\multiset{\sigPush{\valPC}}}
                                            \,
                                            \pNextObBag
                                    }
                                    \slStar
                                    \slPointsTo[\frac{1}{2}]
                                            {\locPC}
                                            {\progProofChanged{\valPC-1}}
                                    \slStar
                                    \pLoopInv{
                                            \progProofChanged{\valPC-1}, 
                                            \progProofCancel{\multiset{\sigPush{\valPC}}}
                                            \,
                                            \progProofNew{\pNextObBag}
                                    }
                                    \\
                                    \slStar\
                                    \assMutex[\frac{1}{2}]{\mut}{\invM}
                            }
                    \\\vspace{-0.7cm}
                    \\\quad\quad
                            \cmdHeapLocReadNeq{\pc}{0}
                            \quad\quad\quad\quad\quad\quad\quad\quad\quad\proofHint{Remember that command is syntactic sugar.}
                                    &\ \ \proofRuleHintML{\prLetName \& \prReadHeapLocName
                                            \\
                                            \& \prExpName
                                            \\
                                            \& \prViewShiftName \& \vsSemImpName
                                    }
                    \\\vspace{-0.7cm}
                    \\\quad\quad
                            \progProofML{
                                    \slIfElseML{\valPC-1 \neq 0}
                                            {
                                                    \obsOf{\sigPush{\valPC-1}}
                                                    \slStar
                                                    \slPointsTo[\frac{1}{2}]{\locPC}{\valPC-1}
                                                    \slStar
                                                    \pLoopInv{\valPC-1, \multiset{\sigPush{\valPC-1}}}
                                                    \slStar
                                                    \assMutex[\frac{1}{2}]{\mut}{\invM}
                                            }
                                            {\noObs}
                            }
                    \\
                    )\
                    \\
                    \progProof{
                            \obs{
                                    \progProofCancel{\multiset{\sigPush{100}}}
                                    \,
                                    \progProofNew{\msEmpty}
                            }
                            \slStar
                            \progProofCancel{
                                    \slPointsTo[\frac{1}{2}]{\locPC}{100}
                                    \slStar
                                    \pLoopInv{100, \multiset{\sigPush{100}}}
                                    \slStar
                                    \assMutex[\frac{1}{2}]{\mut}{\invM}
                            }
                    }
                    \\
                    \dots &\ \ \referenceHintML{Continued in Fig.~\ref{fig:VerificationExampleBoundedFifoForkingSL}.}
            \end{array}
            $}

            \caption{Verification of realistic example~\ref{fig:RealisticExample}: Producer loops.}
            \label{fig:VerificationExampleBoundedFifoProducerLoopSL}
    \end{figure}

    \begin{figure}

            \makebox[\textwidth][c]
            {$
            \hspace{0.025\textwidth}  
            \begin{array}{l | p{0.266\pdfpagewidth}}
                    \lowlightText{\forall \locFifo, \locM, \locPC, \locCC, \valPC, \pObBag, \valFifoInv.}
                    \\
                    \dots &\ \ \referenceHintML{Continuation of Fig.~\ref{fig:VerificationExampleBoundedFifoProducerLoopSL}.}
                    \\
                    \referenceHintML{For definition of \invM, \pLoopInv{n, \obBagVar} and variations  cf.\@ 
                            Fig.~\ref{fig:VerificationExampleBoundedFifoLockInvariantSL}
                            and~\ref{fig:VerificationExampleBoundedFifoProducerLoopInvariantSL}.
                    }
                    \\
                    \progProof{
                            \lstSize{\valFifoInv} < 10
                            \slStar 
                            \pLoopInv{\valPC, \pObBag}
                            \slStar
                            \obs{\pObBag \msCup \multiset{\mut}}
                            \slStar
                            \slPointsTo[\frac{1}{2}]{\locPC}{\valPC}
                            \slStar
                            \invMNoExFifo{\valFifoInv}
                    }
                    \ \ \
                            &\ \ \proofRuleHint{\prViewShiftName \& \vsSemImpName}
                    \\
                    \progProofML{
                            \progProofCancel{
                                    \slPointsTo[\frac{1}{2}]{\locPC}{\valPC}
                                    \slStar
                                    \slPointsTo[\frac{1}{2}]{\locPC}{\valPCInv}
                            }\,
                            \progProofNew{
                                    \slPointsTo{\locPC}{\valPC}
                                    \slStar
                                    \valPC = \valPCInv
                            }
                            \\
                            \slStar\,
                            \progProofCancel{
                                    (\valPC > 0 
                                            \ \rightarrow\
                                            \assSignal{\sigPush{\valPCInv}}{\slFalse}
                                    )
                            }
                            \,
                            \progProofNew{
                                    \assSignal{\sigPush{\valPC}}{\slFalse}
                            }
                            \\
                            \slStar\,
                            \progProofCancel{
                                    (\valPC > 0 \ \leftrightarrow\
                                            \pObBag = \multiset{\sigPush{\valPC}}
                                    )
                                    \slStar
                                    (\valPC = 0 \ \leftrightarrow \
                                            \pObBag = \msEmpty
                                    )
                            }\,
                            \progProofNew{
                                    \pObBag = \multiset{\sigPush{\valPC}}
                            }
                            \slStar \dots
                    }
                    \\
                    \keywordFont{let}\ \pcLoaded \mathop{:=} \cmdReadHeapLoc{\pc}\ \keyword{in}
                            &\ \ \proofRuleHint{\prLetName \& \prReadHeapLocName}
                    \\
                    \cmdAssignToHeap
                            {\fifo}
                            {\lstAppend
                                    {\fifoLoaded}
                                    {(\lstCons{\ccLoaded}{\lstNil})}};
                    \
                    \cmdAssignToHeap{\pc}{\pcLoaded-1}
                            &\ \ \proofRuleHint{\prAssignToHeapName (2x)}
                    \\
                    \progProof{
                            \slPointsTo
                                    {\locFifo}
                                    {\progProofChanged{
                                            \lstAppend
                                                    {\valFifoInv}
                                                    {(\lstCons{\valPC}{\lstNil})}
                                    }}
                            \slStar
                            \slPointsTo
                                    {\locPC}
                                    {\progProofChanged{\valPC-1}}
                            \slStar \dots
                    }
                            &\ \ \proofRuleHint{\prViewShiftName \& \vsSetSignalName}
                    \\
                    \progProof{
                            \obsOf{
                                    \progProofCancel{\sigPush{\valPC},}\,
                                    \mut
                            }
                            \slStar
                            \assSignal
                                    {\sigPush{\valPC}}
                                    {\progProofChanged{\slTrue}}
                            \slStar \dots
                    }
                            &\ \ \proofRuleHint{\prViewShiftName \& \vsSemImpName}
                    \\
                    \progProof{
                            \progProofNew{
                                    (\valPC-1 = 0 \vee \valPC > 0)
                            }
                            \slStar \dots
                    }
                            &\ \ \proofRuleHint{\prViewShiftName \& \vsOrName}
                    \\
                    \\
                    \ghostFont{\text{case:}\quad \valPC-1 = 0}
                            &\ \ \proofRuleHint{\prViewShiftName \& \vsSemImpName}
                    \\\vspace{-0.86cm}\\\hspace{5cm}\proofHint{Last iteration, nothing left to do.}
                    \\\quad\quad
                            \progProof{
                                    \obsOf{\mut}
                                    \slStar
                                    \slPointsTo[\frac{1}{2}]{\locPC}{0}
                                    \slStar
                                    \invM
                            }
                                    &\ \ \proofRuleHint{\prViewShiftName \& \vsSemImpName}
                    \\\quad\quad
                            \progProofML{
                                    \exists \pNextObBag.\
                                    \obs{\pNextObBag \msCup \multiset{\mut}}
                                    \\
                                    \slStar
                                    \slIfElseBranchML{\lstSize{\valFifoInv} \neq 10}
                                    {
                                            \slPointsTo[\frac{1}{2}]{\locPC}{\valPC-1}
                                            \slStar
                                            \pLoopInv{\valPC-1, \pNextObBag}
                                            \slStar
                                            \invMNoExFifo{\valFifoInv}
                                    }
                                    {
                                            \assSignal{\sigPop{\valPC+10}}{\slFalse}
                                            \slStar
                                            \levPop{\valPC+10} < \levPush{\valPC}
                                            \slStar
                                            \slPointsTo[\frac{1}{2}]{\locPC}{\valPC}
                                            \\
                                            \slStar\
                                            \invMNoSigPop(\valFifoInv)
                                            \slStar
                                            \pLoopInv{\valPC, \pNextObBag}
                                    }
                            }
                                    &\ \ \proofHintML{= \pPostIf
                                            \\
                                            \referenceHintML{For definition of \pPostIf cf.\@ Fig.~\ref{fig:VerificationExampleBoundedFifoProducerLoopSL}.}
                                    }
                    \\
                    \\
                    \ghostFont{\text{case:}\quad \valPC-1 > 0}
                            &\ \ \proofRuleHintML{\prViewShiftName \& \vsSigInitName}
                    \\\vspace{-0.86cm}\\\hspace{5cm}\proofHint{Must create signal for next iteration.}
                    \\\quad\quad
                            \progProofML{
                                    \obsOf{
                                            \progProofNew{\sigPush{\valPC-1}},
                                            \mut
                                    }
                                    \slStar
                                    \slBigStar_{i = 1, \dots, \valPC-2}
                                            \quadBack\quadBack\quadBack\!\!
                                            \sigUninit{\idPush{i}}
                                    \\
                                    \slStar\,
                                    \progProofCancel{
                                            \sigUninit{\idPush{\valPC-1}}
                                    }
                                    \,
                                    \progProofNew{
                                            \assSignal{\sigPush{\valPC-1}}{\slFalse}
                                    }
                                    \slStar \dots
                            }
                                    &\ \ \proofRuleHint{\prViewShiftName \& \vsSemImpName}
                    \\\quad\quad
                            \progProof{
                                    \pPostIf
                            }
                    \\
                    \\
                    \progProof{
                            \pPostIf
                    }
                    \\
                    \dots &\ \ \referenceHintML{Continued in Fig.~\ref{fig:VerificationExampleBoundedFifoProducerLoopSL}.}
            \end{array}
            $}

            \caption{Verification of realistic example~\ref{fig:RealisticExample}: Producer thread's production step.}
            \label{fig:VerificationExampleBoundedFifoProductionStepSL}
    \end{figure}

    \begin{figure}

            \makebox[\textwidth][c]
            {$
            \hspace{0.07\textwidth}  
            \begin{array}{l | p{0.266\pdfpagewidth}}
                    \lowlightText{\forall \locFifo, \locM, \locPC, \locCC.}
                    \\
                    \dots &\ \ \referenceHintML{Continuation of Fig.~\ref{fig:VerificationExampleBoundedFifoForkingSL}.}
                    \\
                    \progProof{
                            \obsOf{\sigPop{100}}
                            \slStar
                            \slPointsTo[\frac{1}{2}]{\locCC}{100}
                            \slStar
                            \cLoopInv{100, \multiset{\sigPop{100}}}
                            \slStar
                            \assMutex[\frac{1}{2}]{\mut}{\invM}
                    }
                    \\
                    \keyword{while}\ (
                            &\ \ \proofRuleHint{\prWhileDecStrictName}
                    \\\quad\quad
                            \ghostFont{\forall \valCC.}
                    \\\vspace{-0.6cm}
                    \\\quad\quad
                            \progProof{
                                    \obsOf{\sigPop{\progProofChanged{\valCC}}}
                                    \slStar
                                    \slPointsTo[\frac{1}{2}]{\locCC}{\progProofChanged{\valCC}}
                                    \slStar
                                    \cLoopInv{\progProofChanged{\valCC}, \multiset{\sigPop{\progProofChanged{\valCC}}}}
                                    \slStar
                                    \assMutex[\frac{1}{2}]{\mut}{\invM}
                            }
                    \\\quad\quad
                            \proofHint{$
                                    \mutGetLev{\mut} 
                                    \ = \ 
                                    \levM 
                                    \ <\
                                    102 - \valCC
                                    \ = \
                                    \sigGetLev{\sigPop{\valCC}}
                            $}
                                    &\ \ \proofHint{Justification for application of:}
                    \\\quad\quad
                            \keyword{with}\ \m\ \keyword{await}\ (
                                    &\ \ \proofRuleHint{\prAwaitGenName}
                    \\\vspace{-0.84cm}\\\hspace{3.5cm}\referenceHintML{
                            For definition of consumer loop invariant \cLoopInv{n, \obBagVar}, 
                    }
                    \\\quad\quad\quad\quad
                                    \ghostFont{\forall \cObBag.}
                    \\\vspace{-0.84cm}\\\hspace{3.5cm}\referenceHintML{
                            lock invariant \invM
                            and variations cf.\@ Figures~\ref{fig:VerificationExampleBoundedFifoConsumerLoopInvariantSL}
                            and~\ref{fig:VerificationExampleBoundedFifoLockInvariantSL}.
                    }
                    \\\quad\quad\quad\quad
                                    \progProofML{
                                            \obs{
                                                    \progProofCancel{\multiset{\sigPop{\valCC}}}
                                                    \,
                                                    \progProofNew{
                                                            \cObBag
                                                            \msCup
                                                            \multiset{\mut}
                                                    }
                                            }
                                            \slStar
                                            \slPointsTo[\frac{1}{2}]{\locCC}{\valCC}
                                            \slStar
                                            \cLoopInv{
                                                    \valCC, 
                                                    \progProofCancel{\multiset{\sigPop{\valCC}}}
                                                    \,
                                                    \progProofNew{\cObBag}
                                            }
                                            \\
                                            \slStar
                                            \progProofCancel{\assMutex[\frac{1}{2}]{\mut}{\invM}}
                                            \,
                                            \progProofNew{\invM}
                                    }
                                            &\ \ \proofRuleHint{\prExistsName }
                    \\\quad\quad\quad\quad
                                    \ghostFont{\forall \valFifoInv.}
                    \\\quad\quad\quad\quad
                                    \progProof{
                                            \progProofCancel{\invM}\,
                                            \progProofNew{\invMNoExFifo{\valFifoInv}}
                                            \slStar \dots
                                    }
                    \\\quad\quad\quad\quad
                                    \keyword{let}\ \fifoLoaded \mathop{:=} \cmdReadHeapLoc{\fifo}\ \keyword{in}
                                            &\ \ \proofRuleHint{\prLetName \& \prReadHeapLocName}
                    \\\quad\quad\quad\quad
                                    \progProof{
                                            \cLoopInv{\valCC, \cObBag}
                                            \slStar
                                            \obs{\cObBag \msCup \multiset{\mut}}
                                            \slStar
                                            \slPointsTo[\frac{1}{2}]{\locCC}{\valCC}
                                            \slStar
                                            \invMNoExFifo{\valFifoInv}
                                    }
                    \\\quad\quad\quad\quad
                                    \keyword{if}\ \lstSize{\fifoLoaded} > 0\ \keyword{then}\ (
                                            &\ \ \proofRuleHint{\prIfName}
                    \\\quad\quad\quad\quad\quad\quad
                                            \progProof{
                                                    \progProofNew{
                                                            \lstSize{\valFifoInv} > 0
                                                    }
                                                    \slStar 
                                                    \cLoopInv{\valCC, \cObBag}
                                                    \slStar
                                                    \obs{\cObBag \msCup \multiset{\mut}}
                                                    \slStar
                                                    \slPointsTo[\frac{1}{2}]{\locCC}{\valCC}
                                                    \slStar
                                                    \invMNoExFifo{\valFifoInv}
                                            }
                    \\\quad\quad\quad\quad\quad\quad
                                            \dots \ref{fig:VerificationExampleBoundedFifoConsumptionStepSL} \dots
                                                    \quad\quad\quad\quad\quad\quad\quad\quad\quad\quad
                                                    \referenceHintML{Consumption step presented in Fig.~\ref{fig:VerificationExampleBoundedFifoConsumptionStepSL}.}
                    \\\quad\quad\quad\quad\quad\quad
                                            \progProofML{
                                                    \exists \cNextObBag.\
                                                    \obs{\cNextObBag \msCup \multiset{\mut}}
                                                    \\
                                                    \slStar
                                                    \slIfElseBranchML{\lstSize{\valFifoInv} \neq 0}
                                                            {
                                                                    \slPointsTo[\frac{1}{2}]{\locCC}{\valCC-1}
                                                                    \slStar
                                                                    \cLoopInv{\valCC-1, \cNextObBag}
                                                                    \slStar
                                                                    \invMNoExFifo{\valFifoInv}
                                                            }
                                                            {
                                                                    \assSignal{\sigPush{\valCC}}{\slFalse}
                                                                    \slStar
                                                                    \levPush{\valCC} < \levPop{\valCC}
                                                                    \slStar
                                                                    \slPointsTo[\frac{1}{2}]{\locCC}{\valCC}
                                                                    \\
                                                                    \slStar\
                                                                    \invMNoSigPush(\valFifoInv)
                                                                    \slStar
                                                                    \cLoopInv{\valCC, \cNextObBag}
                                                            }
                                            }
                                            &\ \ \proofDef{=: \cPostIf}
                    \\\quad\quad\quad\quad
                                    )\ \ghostFont{\keyword{else}\ (\ }
                    \\\quad\quad\quad\quad\quad\quad
                                            \progProof{
                                                    \progProofNew{\lstSize{\valFifoInv} = 0}
                                                    \slStar 
                                                    \cLoopInv{\valCC, \cObBag}
                                                    \slStar
                                                    \obs{\cObBag \msCup \multiset{\mut}}
                                                    \slStar
                                                    \slPointsTo[\frac{1}{2}]{\locCC}{\valCC}
                                                    \slStar
                                                    \invMNoExFifo{\valFifoInv}
                                            }\
                                                    &\ \ \proofRuleHint{\prViewShiftName \& \vsSemImpName}
                    \\\quad\quad\quad\quad\quad\quad
                                            \progProofML{
                                                    \progProofCancel{
                                                            \lstSize{\valFifoInv} = 10
                                                            \slStar 
                                                            \cLoopInv{\valCC, \cObBag}
                                                            \slStar
                                                            \obs{\cObBag \msCup \multiset{\mut}}
                                                            \slStar
                                                            \slPointsTo[\frac{1}{2}]{\locCC}{\valCC}
                                                            \slStar
                                                            \invMNoExFifo{\valFifoInv}
                                                    }\\
                                                    \progProofNew{
                                                            \cPostIf
                                                    }
                                            }
                                            \ \
                    \\\quad\quad\quad\quad
                                    \ghostFont{)};
                    \\\quad\quad\quad\quad
                                    \progProof{
                                            \progProofCancel{
                                                    \cLoopInv{\valCC, \cObBag}
                                                    \slStar
                                                    \obs{\cObBag \msCup \multiset{\mut}}
                                                    \slStar
                                                    \slPointsTo[\frac{1}{2}]{\locCC}{\valCC}
                                                    \slStar
                                                    \invMNoExFifo{\valFifoInv}
                                            }
                                            \,
                                            \progProofNew{\cPostIf}
                                    }
                    \\\vspace{-0.5cm}
                    \\\quad\quad\quad\quad
                                    \lstSize{\fifo} > 0
                                            &\ \ \proofRuleHintML{\prExpName\\
                                                    \& \prViewShiftName \& \vsSemImpName
                                            }
                    \\\vspace{-0.65cm}
                    \\\quad\quad\quad\quad
                                    \progProof{
                                            \dots \slStar
                                             \slIfElse{\dots}
                                                    {
                                                            \dots
                                                            \progProofCancel{\invMNoExFifo{\valFifoInv}}
                                                            \,
                                                            \progProofNew{\invM}
                                                    }
                                                    {\dots}
                                    }
                    \\\quad\quad
                            );
                    \\\quad\quad
                            \progProofML{
                                    \progProofNew{
                                            \exists \cNextObBag.
                                    }\
                                    \obs{
                                            \progProofCancel{\multiset{\sigPop{\valCC}}}
                                            \,
                                            \cNextObBag
                                    }
                                    \slStar
                                    \slPointsTo[\frac{1}{2}]
                                            {\locCC}
                                            {\progProofChanged{\valCC-1}}
                                    \slStar
                                    \cLoopInv{
                                            \progProofChanged{\valCC-1}, 
                                            \progProofCancel{\multiset{\sigPop{\valCC}}}
                                            \,
                                            \progProofNew{\cNextObBag}
                                    }
                                    \\
                                    \slStar\
                                    \assMutex[\frac{1}{2}]{\mut}{\invM}
                            }
                    \\\vspace{-0.7cm}
                    \\\quad\quad
                            \cmdHeapLocReadNeq{\cc}{0}
                            \quad\quad\quad\quad\quad\quad\quad\quad\quad\proofHint{Remember that command is syntactic sugar.}
                                    &\ \ \proofRuleHintML{\prLetName \& \prReadHeapLocName
                                            \\
                                            \& \prExpName
                                            \\
                                            \& \prViewShiftName \& \vsSemImpName
                                    }
                    \\\vspace{-0.7cm}
                    \\\quad\quad
                            \progProofML{
                                    \slIfElseML{\valCC-1 \neq 0}
                                            {
                                                    \obsOf{\sigPop{\valCC-1}}
                                                    \slStar
                                                    \slPointsTo[\frac{1}{2}]{\locCC}{\valCC-1}
                                                    \slStar
                                                    \cLoopInv{\valCC-1, \multiset{\sigPop{\valCC-1}}}
                                                    \slStar
                                                    \assMutex[\frac{1}{2}]{\mut}{\invM}
                                            }
                                            {\noObs}
                            }
                    \\
                    )\
                    \\
                    \progProof{
                            \obs{
                                    \progProofCancel{\multiset{\sigPop{100}}}
                                    \,
                                    \progProofNew{\msEmpty}
                            }
                            \slStar
                            \progProofCancel{
                                    \slPointsTo[\frac{1}{2}]{\locCC}{100}
                                    \slStar
                                    \cLoopInv{100, \multiset{\sigPop{100}}}
                                    \slStar
                                    \assMutex[\frac{1}{2}]{\mut}{\invM}
                            }
                    }
                    \\
                    \dots &\ \ \referenceHintML{Continued in Fig.~\ref{fig:VerificationExampleBoundedFifoForkingSL}.}
            \end{array}
            $}

            \caption{Verification of realistic example~\ref{fig:RealisticExample}: Consumer loops.}
            \label{fig:VerificationExampleBoundedFifoConsumerLoopSL}
    \end{figure}

    \begin{figure}

            \makebox[\textwidth][c]
            {$
            \hspace{0.04\textwidth}  
            \begin{array}{l | p{0.266\pdfpagewidth}}
                    \lowlightText{\forall \locFifo, \locM, \locPC, \locCC, \valCC, \cObBag, \valFifoInv.}
                    \\
                    \dots &\ \ \referenceHintML{Continuation of Fig.~\ref{fig:VerificationExampleBoundedFifoConsumerLoopSL}.}
                    \\
                    \referenceHintML{For definition of \invM, \cLoopInv{n, \obBagVar}  and variations cf.\@ 
                            Figures~\ref{fig:VerificationExampleBoundedFifoLockInvariantSL}
                            and~\ref{fig:VerificationExampleBoundedFifoConsumerLoopInvariantSL}.
                    }
                    \\
                    \progProof{
                            \lstSize{\valFifoInv} > 0
                            \slStar 
                            \cLoopInv{\valCC, \cObBag}
                            \slStar
                            \obs{\cObBag \msCup \multiset{\mut}}
                            \slStar
                            \slPointsTo[\frac{1}{2}]{\locCC}{\valCC}
                            \slStar
                            \invMNoExFifo{\valFifoInv}
                    }
                    \ \ \ \
                            &\ \ \proofRuleHint{\prViewShiftName \& \vsSemImpName}
                    \\
                    \progProofML{
                            \progProofCancel{
                                    \slPointsTo[\frac{1}{2}]{\locCC}{\valCC}
                                    \slStar
                                    \slPointsTo[\frac{1}{2}]{\locCC}{\valCCInv}
                            }\,
                            \progProofNew{
                                    \slPointsTo{\locCC}{\valCC}
                                    \slStar
                                    \valCC = \valCCInv
                            }
                            \\
                            \slStar\,
                            \progProofCancel{
                                    (\valCC > 0 
                                            \ \rightarrow\
                                            \assSignal{\sigPop{\valCCInv}}{\slFalse}
                                    )
                            }
                            \,
                            \progProofNew{
                                    \assSignal{\sigPop{\valCC}}{\slFalse}
                            }
                            \\
                            \slStar\,
                            \progProofCancel{
                                    (\valCC > 0 \ \leftrightarrow\
                                            \cObBag = \multiset{\sigPop{\valCC}}
                                    )
                                    \slStar
                                    (\valCC = 0 \ \leftrightarrow \
                                            \cObBag = \msEmpty
                                    )
                            }\,
                            \progProofNew{
                                    \cObBag = \multiset{\sigPop{\valCC}}
                            }
                            \slStar \dots
                    }
                    \\
                    \keywordFont{let}\ \ccLoaded \mathop{:=} \cmdReadHeapLoc{\cc}\ \keyword{in}
                            &\ \ \proofRuleHint{\prLetName \& \prReadHeapLocName}
                    \\
                    \cmdAssignToHeap
                            {\fifo}
                            {\lstTail{\fifoLoaded}};
                    \
                    \cmdAssignToHeap{\cc}{\ccLoaded-1}
                            &\ \ \proofRuleHint{\prAssignToHeapName (2x)}
                    \\
                    \progProof{
                            \slPointsTo
                                    {\locFifo}
                                    {\progProofChanged{
                                            \lstTail{\valFifoInv}
                                    }}
                            \slStar
                            \slPointsTo
                                    {\locCC}
                                    {\progProofChanged{\valCC-1}}
                            \slStar \dots
                    }
                            &\ \ \proofRuleHint{\prViewShiftName \& \vsSetSignalName}
                    \\
                    \progProof{
                            \obsOf{
                                    \progProofCancel{\sigPop{\valCC},}\,
                                    \mut
                            }
                            \slStar
                            \assSignal
                                    {\sigPop{\valCC}}
                                    {\progProofChanged{\slTrue}}
                            \slStar \dots
                    }
                            &\ \ \proofRuleHint{\prViewShiftName \& \vsSemImpName}
                    \\
                    \progProof{
                            \progProofNew{
                                    (\valCC-1 = 0 \vee \valCC > 0)
                            }
                            \slStar \dots
                    }
                            &\ \ \proofRuleHint{\prViewShiftName \& \vsOrName}
                    \\
                    \\
                    \ghostFont{\text{case:}\quad \valCC-1 = 0}
                            &\ \ \proofRuleHint{\prViewShiftName \& \vsSemImpName}
                    \\\vspace{-0.86cm}\\\hspace{5cm}\proofHint{Last iteration, nothing left to do.}
                    \\\quad\quad
                            \progProof{
                                    \obsOf{\mut}
                                    \slStar
                                    \slPointsTo[\frac{1}{2}]{\locCC}{0}
                                    \slStar
                                    \invM
                            }
                                    &\ \ \proofRuleHint{\prViewShiftName \& \vsSemImpName}
                    \\\quad\quad
                            \progProofML{
                                    \exists \cNextObBag.\
                                    \obs{\cNextObBag \msCup \multiset{\mut}}
                                    \\
                                    \slStar
                                    \slIfElseBranchML{\lstSize{\valFifoInv} \neq 0}
                                    {
                                            \slPointsTo[\frac{1}{2}]{\locCC}{\valCC-1}
                                            \slStar
                                            \cLoopInv{\valCC-1, \cNextObBag}
                                            \slStar
                                            \invMNoExFifo{\valFifoInv}
                                    }
                                    {
                                            \assSignal{\sigPush{\valCC}}{\slFalse}
                                            \slStar
                                            \levPush{\valCC} < \levPop{\valCC}
                                            \slStar
                                            \slPointsTo[\frac{1}{2}]{\locCC}{\valCC}
                                            \\
                                            \slStar\
                                            \invMNoSigPush(\valFifoInv)
                                            \slStar
                                            \cLoopInv{\valCC, \cNextObBag}
                                    }
                            }
                                    &\ \ \proofHintML{= \cPostIf
                                            \\
                                            \referenceHintML{For definition of \cPostIf cf.\@ Fig.~\ref{fig:VerificationExampleBoundedFifoConsumerLoopSL}.}
                                    }
                    \\
                    \\
                    \ghostFont{\text{case:}\quad \valCC-1 > 0}
                            &\ \ \proofRuleHintML{\prViewShiftName \& \vsSigInitName}
                    \\\vspace{-0.86cm}\\\hspace{5cm}\proofHint{Must create signal for next iteration.}
                    \\\quad\quad
                            \progProofML{
                                    \obsOf{
                                            \progProofNew{\sigPop{\valCC-1}},
                                            \mut
                                    }
                                    \slStar
                                    \slBigStar_{i = 1, \dots, \valCC-2}
                                            \quadBack\quadBack\quadBack\!\!
                                            \sigUninit{\idPop{i}}
                                    \\
                                    \slStar\,
                                    \progProofCancel{
                                            \sigUninit{\idPop{\valCC-1}}
                                    }
                                    \,
                                    \progProofNew{
                                            \assSignal{\sigPop{\valCC-1}}{\slFalse}
                                    }
                                    \slStar \dots
                            }
                                    &\ \ \proofRuleHint{\prViewShiftName \& \vsSemImpName}
                    \\\quad\quad
                            \progProof{
                                    \cPostIf
                            }
                    \\
                    \\
                    \progProof{
                            \cPostIf
                    }
                    \\
                    \dots &\ \ \referenceHintML{Continued in Fig.~\ref{fig:VerificationExampleBoundedFifoConsumerLoopSL}.}
            \end{array}
            $}

            \caption{Verification of realistic example~\ref{fig:RealisticExample}: Consumer thread's consumption step.}
            \label{fig:VerificationExampleBoundedFifoConsumptionStepSL}
    \end{figure}

    }

%% file: CAV_paper_sections/extended_version/caseStudyUnboundedMutliParty.tex
{   
    \NewDocumentCommand{\numThs}{}{\progVar{N}}
    
    
    \NewDocumentCommand{\buffer}{}{\progVar{buf}}
    \NewDocumentCommand{\bufferLoaded}{}{\progVar{b}}
    \NewDocumentCommand{\locBuffer}{}{\metaVar{\hlocVar_\buffer}}
    \NewDocumentCommand{\valBuffer}{}{\metaVar{\valVar_\buffer}}
    \NewDocumentCommand{\nextValBuffer}{}{\metaVar{\nextValVar_\buffer}}
    
    \NewDocumentCommand{\locMut}{}{\metaVar{\hlocVar_\mutProgVar}}
    \NewDocumentCommand{\levMut}{}{0}
    
    
    \NewDocumentCommand{\flCnt}{}{\progVar{c}}
    \NewDocumentCommand{\locFlCnt}{}{\metaVar{\hlocVar_\flCnt}}
    \NewDocumentCommand{\valFlCnt}{}{\metaVar{\valVar_\flCnt}}
    
    \NewDocumentCommand{\sigPush}{m m}{\metaVar{\sigVar_{\text{push}, #1}^{#2}}}
    \NewDocumentCommand{\sigPop}{m m}{\metaVar{\sigVar_{\text{pop}, #1}^{#2}}}
    \NewDocumentCommand{\idPush}{m m}{\metaVar{\idVar_{\text{push}, #1}^{#2}}}
    \NewDocumentCommand{\idPop}{m m}{\metaVar{\idVar_{\text{pop}, #1}^{#2}}}
    \NewDocumentCommand{\levPush}{m}{\metaVar{\levVar_\text{push}^{#1}}}
    \NewDocumentCommand{\levPop}{m}{\metaVar{\levVar_\text{pop}^{#1}}}
    
    \NewDocumentCommand{\numWritten}{}{\metaVar{W}}
    \NewDocumentCommand{\numRead}{}{\metaVar{R}}
    
    \NewDocumentCommand{\prodFinished}{m}{\metaVar{w_{#1}}}
    \NewDocumentCommand{\consFinished}{m}{\metaVar{r_{#1}}}
    \NewDocumentCommand{\glocProdFin}{m}{\metaVar{\glocVar_{#1}^w}}
    \NewDocumentCommand{\glocConsFin}{m}{\metaVar{\glocVar_{#1}^r}}

    \NewDocumentCommand{\flInv}{m}{\metaVar{L_f(#1)}}
    \NewDocumentCommand{\pInv}{m m}{\metaVar{L_p^{#1}(#2)}}
    \NewDocumentCommand{\cInv}{m m}{\metaVar{L_c^{#1}(#2)}}

    \NewDocumentCommand{\nextPush}{}{\metaVar{n_\text{push}}}
    \NewDocumentCommand{\nextPop}{}{\metaVar{n_\text{pop}}}

    \NewDocumentCommand{\lockInvNoQuant}{}{\metaVar{\widehat{\assLockInvVar}}}
    \NewDocumentCommand{\pLockInvRest}{}{\metaVar{\assLockInvVar_p^\text{rest}}}
    \NewDocumentCommand{\cLockInvRest}{}{\metaVar{\assLockInvVar_c^\text{rest}}}

    \NewDocumentCommand{\pPostIf}{}{\metaVar{\fixedPredNameFont{PostIf}_p}}
    \NewDocumentCommand{\cPostIf}{}{\metaVar{\fixedPredNameFont{PostIf}_c}}

    In this section we present and verify a program similar to the one from the previous section, but where the number of producer and consumer threads is not statically bounded.
    Fig.~\ref{appendix:caseStudy:undbounded:code} presents this program.
    It involves three parties: the main thread, producer threads and consumer threads.
    The main thread creates a shared buffer of size 1, generates a random number $\numThs > 0$ and spawns \numThs producer and \numThs consumer threads., which communicate via the shared buffer.
    Each producer tries to push a single random number into the buffer.
    In case the buffer is full, the producer busy-waits for it to become empty, i.e., it busy-waits for a consumer to pop the number currently stored in the buffer.
    After it pushed, the producer terminates.
    Each consumer tries to pop a single number from the buffer.
    In case the buffer is empty, it waits for some producer to push a number into the buffer.
    After it popped, the consumer terminates.

    \begin{figure}
            $$
            \begin{array}{l p{0.5cm} l}
                    \keyword{let}\ \numThs := \cmdRandomNat + 1\ \keyword{in}
                            &&\lowlightText{\text{Random number of threads.}}
                    \\
                    \keyword{let}\ \buffer := \cmdAlloc{-1}\ \keyword{in}
                            &&\lowlightText{\text{Shared buffer. Negative value $\leftrightarrow$ empty.}}
                    \\
                    \keyword{let}\ \mutProgVar := \cmdNewMut\ \keyword{in}
                    \\
                    \keyword{let}\ \flCnt := \cmdAlloc{\numThs}\ \keyword{in}
                    \\
                    \keyword{while}\ (
                            &&\lowlightText{\text{Main thread forks \numThs producer and}}
                    \\&&\lowlightText{\text{\numThs consumer threads.}}
                    \\\quad\quad
                            \keyword{fork}\ (
                                    &&\lowlightText{\text{Producer}}
                    \\\quad\quad\quad\quad
                                    \keyword{with}\ \mutProgVar\ \keyword{await}\ (
                                            &&\lowlightText{\text{Producer busy-waits for empty buffer.}}
                    \\\quad\quad\quad\quad\quad\quad
                                                    \keyword{let}\ \bufferLoaded := \cmdReadHeapLoc{\buffer}\ \keyword{in}
                    \\\quad\quad\quad\quad\quad\quad
                                                    \keyword{if}\ \bufferLoaded < 0\ \keyword{then}\
                    \\\quad\quad\quad\quad\quad\quad\quad\quad
                                                            \cmdAssignToHeap{\buffer}{\cmdRandomNat};
                                                                    &&\lowlightText{\text{Push random number to buffer.}}
                    \\\quad\quad\quad\quad\quad\quad
                                                    \bufferLoaded < 0
                    \\\quad\quad\quad\quad
                                    ) 
                    \\\quad\quad
                            ); 
                    \\\quad\quad
                            \keyword{fork}\ (
                                    &&\lowlightText{\text{Consumer}}
                    \\\quad\quad\quad\quad
                                    \keyword{with}\ \mutProgVar\ \keyword{await}\ (
                                            &&\lowlightText{\text{Consumer busy-waits for non-empty buffer.}}
                    \\\quad\quad\quad\quad\quad\quad
                                                    \keyword{let}\ \bufferLoaded := \cmdReadHeapLoc{\buffer}\ \keyword{in}
                    \\\quad\quad\quad\quad\quad\quad
                                                    \keyword{if}\ \bufferLoaded \geq 0\ \keyword{then}\
                    \\\quad\quad\quad\quad\quad\quad\quad\quad
                                                            \cmdAssignToHeap{\buffer}{-1};
                                                                    &&\lowlightText{\text{Pop buffer value.}}
                    \\\quad\quad\quad\quad\quad\quad
                                                    \bufferLoaded \geq 0
                    \\\quad\quad\quad\quad
                                    ) 
                    \\\quad\quad
                            ); 
                    \\\quad\quad
                            \cmdAssignToHeap{\flCnt}{\cmdReadHeapLoc{\flCnt}-1};
                    \\\quad\quad
                            \flCnt > 0 
                    \\
                    )\ \keyword{do}\ \cmdSkip
            \end{array}
            $$
            \caption{Program with statically unbounded number of threads communicating via a shared buffer of size 1.}
            \label{appendix:caseStudy:undbounded:code}
    \end{figure}

    \paragraph{Threads Racing for Buffer}
    Since the program is conceptionally similar to the one from the previous section, the same holds for its termination proof.
    There is, however, one fundamental difference.
    The previous program involved one producer and one consumer.
    So, it was clear which thread would push the $k^\text{th}$ element, set the $k^\text{th}$ push signal and discharge the corresponding obligation.
    The same was true for the consumer and the pop signals.
    Now, \numThs producers and \numThs consumers race for the buffer.
    We cannot statically determine which ones will win.
    Therefore, we cannot statically decide which threads we should delegate the push and pop obligations to.
    
    \paragraph{Signals}
    To solve this, we create \numThs variations of every push and pop signal.
    That is, we create signals  \sigPush{i}{k} and \sigPop{i}{k} for $1 \leq i,k \leq \numThs$ where $i$ refers to the $i$'s producer or consumer thread and $k$ to the $k^\text{th}$ element pushed to or popped from the buffer.
    We delegate the obligations for $\sigPush{i}{1}, \dots, \sigPush{i}{\numThs}$ to the $i^\text{th}$ producer and the obligations for $\sigPop{i}{1}, \dots, \sigPop{i}{\numThs}$ to the $i^\text{th}$ consumer.

    When the $i^\text{th}$ producer waits because the buffer is full and contains the $k^\text{th}$ number, it sets all its remaining unset signals up to the $k^\text{th}$ one.
    That is, signals $\sigPush{i}{k}, \dots, \sigPush{i}{\numThs}$ remain unset.
    (Reducing the number of obligations that a producer holds while waiting for a consumer to pop, reduces the potential for level conflicts and simplifies ordering the signals.)
    When the $i^\text{th}$ producer pushes any number to the buffer, it sets all its remaining signals, before it terminates.
    A consumer can use any unset \sigPush{j}{k} for any $j$ to wait for the $k^\text{th}$ element to arrive in the buffer.
    We proceed analogously with the consumers and their pop signals.
    Similarly to the proof from the previous section, we include our signals in the lock invariant associated with the mutex that protects the shared buffer.
    Thereby, we allow all threads to share the signals in a synchronised fashion.
    
    \paragraph{Levels}
    The $i^\text{th}$ producer uses some unset \sigPop{j}{k} to wait for the $k^\text{th}$ element to be popped from the buffer.
    Meanwhile it holds obligations $\sigPush{i}{k}, \dots, \sigPush{i}{\numThs}$.
    Analogously, the $i^\text{th}$ consumer holds obligations $\sigPop{i}{k}, \dots, \sigPop{i}{\numThs}$ while using some unset \sigPush{j}{k} to wait for the $k^\text{th}$ element to be pushed.
    We assign one common level \levPush{k} to all $\sigPush{1}{k}, \dots, \sigPush{\numThs}{k}$ and one common level \levPop{k} to all $\sigPop{1}{k}, \dots, \sigPop{\numThs}{k}$.
    It must hold
    $\levPop{k} < \levPush{j}$ for all $j > k$ and
    $\levPush{k} < \levPop{h}$ for all $h\geq k$.
    Hence, we chose
    $\levPush{k} = 2\cdot k - 1$ and
    $\levPop{k} = 2\cdot k$.

    \paragraph{Ghost Variables}
    In order to simplify the verification of this program, we introduce \emph{ghost variables}, which are standard~\cite{Owicki1976VerifyingPropiertiesGhostVariables, Jung2016HigherorderGS, Jacobs2018ModularTerminationVerification}.
    Ghost variables are ghost resources that behave like regular program variables, but since they only exist on the verification level, they cannot affect the runtime behaviour.
    We assume an infinite set of ghost locations \GhostLocSet with $\GhostLocSet \cap \HeapLocSet = \emptyset$ and a set of ghost values \GhostValueSet.
    We implement ghost variables in the form of \emph{ghost heap cells} and represent them by \emph{ghost points-to chunks} \slPointsTo{\glocVar}{\gvalVar}.
    Fig.~\ref{appendix:caseStudy:undbounded:ghostVariables} presents two new view shift rules for ghost variables.
    Rule \vsNewGhostCellName allows to create a new ghost variables and \vsMutateGhostCellName allows to change the value of an existing one.
    Adding ghost variables does not affect the soundness of our verification approach.
    In fact, the generalised logic we present and prove sound in the technical report~\cite{Reinhard2020GhostSignalsTR} includes ghost variables.
    In our termination proof, we introduce ghost variables \prodFinished{i} and \consFinished{i}.
    Variable \prodFinished{i} tracks whether the $i^\text{th}$ producer has already pushed ("written") an element to the buffer and 
    \consFinished{i} tracks whether the $i^\text{th}$ consumer has already popped ("read") an element from the buffer.
    These variables conveniently allow us to to refer to the number of elements written to and read from the buffer in the lock invariant.
    In Fig.~\ref{appendix:caseStudy:unbounded:init} -- \ref{appendix:caseStudy:unbounded:consumerLoop}, we present the full proof outline for our program from Fig.~\ref{appendix:caseStudy:undbounded:code}.

    \begin{figure}
            \begin{mathpar}
                    \vsNewGhostCell
                    \and
                    \vsMutateGhostCell
            \end{mathpar}
            \caption{View shift rules for ghost variables.}
            \label{appendix:caseStudy:undbounded:ghostVariables}
    \end{figure}

    \begin{figure}

            \makebox[\textwidth][c]
            {$
            \hspace{0.04\textwidth}  
            \begin{array}{l p{0.1cm} | p{0.3cm} p{0.3\pdfpagewidth}}
                    \progProof{\noObs}&
                    \\
                    \keyword{let}\ \numThs := \cmdRandomNat+1\ \keyword{in}\
                    \keyword{let}\ \buffer := \cmdAlloc{-1}\ \keyword{in}\
                            &&&\proofRuleHint{\prLetName (2x) \& \prAllocName}
                    \\
                    \keyword{let}\ \mutProgVar := \cmdNewMut\ \keyword{in}\
                    \keyword{let}\ \flCnt := \cmdAlloc{\numThs}\ \keyword{in}\
                            &&&\proofRuleHint{\prLetName (2x) \& \prNewMutexName \& \prAllocName}
                            \hspace{-1cm}
                    \\
                    \ghostFont{\forall \locBuffer, \locMut, \locFlCnt.}&
                    \\
                    \progProof{
                            \noObs 
                            \slStar
                            \progProofNew{
                                    \numThs > 0 
                                    \slStar
                                    \slPointsTo{\locBuffer}{-1}
                                    \slStar
                                    \assMutUninit{\locMut}
                                    \slStar
                                    \slPointsTo{\locFlCnt}{\numThs}
                            }
                    }
                            &&&\proofRuleHintML{
                                    \prViewShiftName\\
                                    \& \vsGhostLoopName \& \vsNewSignalName
                            }
                    \\
                    \proofDef{
                            \levPush{k} := 2 \cdot k - 1,\ \
                            \levPop{k} := 2 \cdot k
                            \quad\text{for}\quad
                            1 \leq k \leq \numThs
                    }
                            &&&\proofHint{Creating $\numThs\cdot\numThs$ push and $\numThs\cdot\numThs$ pop signals.}
                    \\
                    \proofHint{Later 
                            $\levPop{k} < \levPush{j}$ and $\levPush{k} < \levPop{h}$
                            must hold for $j > k$ and $h \geq k$.
                    }
                            &&&\referenceHint{cf. Fig.~\ref{appendix:caseStudy:unbounded:producerLoop} and~\ref{appendix:caseStudy:unbounded:consumerLoop}.}
                    \\
                    \progProofML{
                            \progProofNew{
                                    \exists \idPush{1}{1}, \idPop{1}{1}, \dots, \idPush{\numThs}{\numThs}, \idPop{\numThs}{\numThs}.
                            }
                            \\
                            \obsOf{\progProofChanged{
                                    (\idPush{1}{1}, \levPush{1}), (\idPop{1}{1}, \levPop{1}),
                                    \dots,
                                    (\idPush{\numThs}{\numThs}, \levPush{\numThs}), (\idPop{\numThs}{\numThs}, \levPop{\numThs}),
                            }}
                            \\
                            \slStar\,
                            \numThs > 0
                            \slStar
                            \slPointsTo{\locBuffer}{-1}
                            \slStar
                            \assMutUninit{\locMut}
                            \slStar
                            \slPointsTo{\locFlCnt}{\numThs}
                            \\
                            \slStar\,
                            \progProofNew{
                                    \displaystyle
                                    \slBigStar_{i, k = 1, \dots, \numThs}
                                            \!\!\!\!
                                            \assSignal{(\idPush{i}{k}, \levPush{k})}{\slFalse}
                                            \slStar
                                            \assSignal{(\idPop{i}{k}, \levPop{k}}{\slFalse}
                            }
                    }
                            &&&\proofRuleHint{\prExistsName}
                    \\
                    \ghostFont{\forall \idPush{1}{1}, \idPop{1}{1}, \dots, \idPush{\numThs}{\numThs}, \idPop{\numThs}{\numThs}.}&
                    \\
                    \proofDef{
                            \sigPush{i}{k} := (\idPush{i}{k}, \levPush{k}),\ \
                            \sigPop{i}{k} := (\idPop{i}{k}, \levPop{k})
                            \quad\text{for}\quad
                            1 \leq i, k \leq \numThs
                    }&
                    \\
                    \progProofML{
                            \progProofCancel{
                                    \exists \idPush{1}{1}, \idPop{1}{1}, \dots, \idPush{\numThs}{\numThs}, \idPop{\numThs}{\numThs}.
                            }\
                            \obsOf{\sigPush{1}{1}, \sigPop{1}{1}, \dots, \sigPush{\numThs}{\numThs}, \sigPop{\numThs}{\numThs}}
                            \\
                            \slStar\,
                            \numThs > 0
                            \slStar
                            \slPointsTo{\locBuffer}{-1}
                            \slStar
                            \assMutUninit{\locMut}
                            \slStar
                            \slPointsTo{\locFlCnt}{\numThs}
                            \\
                            \slStar\,
                            \displaystyle
                            \!\!\!\!
                            \slBigStar_{i, k = 1, \dots, \numThs}
                                    \!\!\!\!
                                    \assSignal{\sigPush{i}{k}}{\slFalse}
                                    \slStar
                                    \assSignal{\sigPop{i}{k}}{\slFalse}
                    }
                            &&&\proofRuleHintML{\prViewShiftName\\
                                    \& \vsGhostLoopName \& \vsNewGhostCellName
                             }
                    \\
                    \ \proofHintML{
                            A consumer can use any unset \sigPush{i}{k} to wait for the $k^\text{th}$ number to be pushed.\\
                            When producer $i$ pushes, it sets all \sigPush{i}{j}. When it waits for a consumer to pop\\
                            the $k^\text{th}$ number, it sets all \sigPush{i}{j} for $j\leq k$ and keeps obligations for $j> k$.\\
                            So, if $k < \numThs$ numbers have been pushed, there exists some $i$ s.t.\@ \sigPush{i}{k+1} is unset.
                    }&
                    \\
                    \progProofML{
                            \progProofNew{\exists \glocProdFin{1},\glocConsFin{1}, \dots, \glocProdFin{\numThs},\glocConsFin{\numThs}.}\
                            \obsOf{
                                    \sigPush{1}{1}, \sigPop{1}{1}, \dots, 
                                    \sigPush{\numThs}{\numThs}, \sigPop{\numThs}{\numThs}
                            }
                            \slStar
                            \numThs > 0
                            \\
                            \slStar\,
                            \slPointsTo{\locBuffer}{-1}
                            \slStar
                            \assMutUninit{\locMut}
                            \slStar
                            \slPointsTo{\locFlCnt}{\numThs}
                            \slStar
                            \progProofNew{
                                    \displaystyle
                                    \slBigStar_{i=1, \dots, \numThs}
                                            \!\!
                                            \slPointsTo{\glocProdFin{i}}{\slFalse}
                                            \slStar
                                            \slPointsTo{\glocConsFin{i}}{\slFalse}
                            }
                            \\
                            \slStar
                            \displaystyle
                            \slBigStar_{i,k=1,\dots,\numThs}
                                    \!\!\!\!
                                    \assSignal{\sigPush{i}{k}}{\slFalse}
                                    \slStar
                                    \assSignal{\sigPop{i}{k}}{\slFalse}
                    }
                            &&&\proofRuleHint{\prExistsName}
                    \\
                    \proofHint{Ghost heap cell \glocProdFin{i}/ \glocConsFin{i} records whether prod./cons.\@ $i$ has already pushed/popped.}&
                    \\
                    \ghostFont{\forall \glocProdFin{1},\glocConsFin{1}, \dots, \glocProdFin{\numThs},\glocConsFin{\numThs}.}&
                    \\
                    \progProofML{
                            \progProofCancel{\exists \glocProdFin{1},\glocConsFin{1}, \dots, \glocProdFin{\numThs},\glocConsFin{\numThs}.}\
                            \obsOf{
                                    \sigPush{1}{1}, \sigPop{1}{1}, \dots, 
                                    \sigPush{\numThs}{\numThs}, \sigPop{\numThs}{\numThs}
                            }
                            \slStar
                            \numThs > 0
                            \\
                            \slStar\,
                            \slPointsTo{\locBuffer}{-1}
                            \slStar
                            \assMutUninit{\locMut}
                            \slStar
                            \slPointsTo{\locFlCnt}{\numThs}
                            \slStar
                            \displaystyle
                            \slBigStar_{i = 1,\dots, \numThs}
                                    \!\!
                                    \slPointsTo{\glocProdFin{i}}{\slFalse}
                                    \slStar
                                    \slPointsTo{\glocConsFin{i}}{\slFalse}
                            \\
                            \slStar\,
                            \displaystyle
                            \slBigStar_{i,k=1,\dots,\numThs}
                                    \!\!\!\!
                                    \assSignal{\sigPush{i}{k}}{\slFalse}
                                    \slStar
                                    \assSignal{\sigPop{i}{k}}{\slFalse}
                    }
                            &&&\proofRuleHintML{\prViewShiftName\\ \& \vsSemImpName \& \vsMutInitName}
                    \\
                    \proofDef{
                            \mutVar := (\mutProgVar, \levMut)
                    }&
                    \\
                    \proofHint{Later 
                            $\mutGetLev{\mutVar} < \levPush{k}$
                            and
                            $\mutGetLev{\mutVar} < \levPop{k}$
                            must hold.
                    }
                            &&&\referenceHint{cf.\@ Fig.~\ref{appendix:caseStudy:unbounded:producerLoop} and~\ref{appendix:caseStudy:unbounded:consumerLoop}.}
                    \\
                    \progProofML{
                            \obsOf{
                                    \sigPush{1}{1}, \sigPop{1}{1}, \dots, 
                                    \sigPush{\numThs}{\numThs}, \sigPop{\numThs}{\numThs}
                            }\,
                            \slStar
                            \progProofNew{
                                    \assMutex{\mutVar}{\assLockInvVar}
                            }
                            \\
                            \progProofCancel{
                                    \numThs > 0
                                    \slStar
                                    \slPointsTo{\locBuffer}{-1}
                                    \slStar
                                    \assMutUninit{\locMut}
                            }
                            \slStar
                            \slPointsTo{\locFlCnt}{\numThs}
                            \slStar
                            \quadBack
                            \displaystyle
                            \slBigStar_{i=1, \dots, \numThs}
                                    \!\!\!\!
                                    \slPointsTo[\textstyle\progProofChanged{\frac{1}{2}}]{\glocProdFin{i}}{\slFalse}
                                    \slStar
                                    \slPointsTo[\textstyle\progProofChanged{\frac{1}{2}}]{\glocConsFin{i}}{\slFalse}
                            \\
                            \progProofCancel{
                                    \slStar
                                    \!\!\!
                                    \displaystyle
                                    \slBigStar_{i,k=1,\dots,\numThs}
                                            \!\!\!\!
                                            \assSignal{\sigPush{i}{k}}{\slFalse}
                                            \slStar
                                            \assSignal{\sigPop{i}{k}}{\slFalse}
                            }
                    }
                            &&&\referenceHintML{
                                    For definition of lock invariant \assLockInvVar\\
                                    cf. Fig.~\ref{appendix:caseStudy:unbounded:lockInv}.
                            }
                    \\
                    \dots&&&\referenceHint{Continued in Fig.~\ref{appendix:caseStudy:unbounded:forkLoop}}.
            \end{array}
            $}
            
            \caption{Verification of program~\ref{appendix:caseStudy:undbounded:code}: Initialisation.}
            \label{appendix:caseStudy:unbounded:init}
    \end{figure}

    \begin{figure}

            \makebox[\textwidth][c]
            {$
            \hspace{0.04\textwidth}  
            \begin{array}{l c l p{0.5cm} | p{0.5cm} p{0.3\pdfpagewidth}}
                    \proofDef{
                            \nextAssLockInvVar(
                                    \valBuffer, \numWritten, \numRead, 
                                    \prodFinished{1}, \dots, \prodFinished{\numThs},
                                    \consFinished{1}, \dots, \consFinished{\numThs}
                            )
                    } \hspace{-6.5cm}
                    &&&
                    \\
                    &\proofDef{\ := \ }
                    &\proofDef{
                            \slPointsTo{\locBuffer}{\valBuffer}
                            \ \wedge\
                            \valBuffer \in \Z
                            \ \wedge\ 
                            \numThs > 0
                    }
                            &&&\proofHint{Shared buffer. Negative $\leftrightarrow$ empty.}
                    \\
                    &&\proofDef{
                            \displaystyle
                            \slStar\
                            \slBigStar_{i= 1, \dots, \numThs}
                                    \!\!
                                    \slPointsTo[\textstyle\frac{1}{2}]{\glocProdFin{i}}{\prodFinished{i}}
                            \ \slStar\
                            \numWritten = \displaystyle\cardinalityOf{\setOf{i \in \setOf{1, \dots, \numThs}\ | \ \prodFinished{i} = \slTrue}}
                    }
                            &&&\proofHintML{Ghost value \prodFinished{i} tracks whether \\
                                    producer $i$ already pushed.\\
                                    \numWritten: Number of elements written to buffer.
                            }
                    \\
                    &&\proofDef{
                            \displaystyle
                            \slStar\
                            \slBigStar_{i= 1, \dots, \numThs}
                                    \!\!
                                    \slPointsTo[\textstyle\frac{1}{2}]{\glocConsFin{i}}{\consFinished{i}}
                            \ \slStar\
                            \numRead = \displaystyle\cardinalityOf{\setOf{i \in \setOf{1, \dots, \numThs}\ | \ \consFinished{i} = \slTrue}}
                    }
                            &&&\proofHintML{Ghost value \consFinished{i} tracks whether \\
                                    consumer $i$ already popped.\\
                                    \numRead: Number of elements read from buffer.
                            }
                    \\
                    &&\proofDef{
                            \slStar\
                            (\valBuffer \geq 0 
                                    \ \rightarrow\ 
                                    \numWritten > 0
                                    \ \wedge\
                                    \numRead < \numThs
                            )
                    }
                            &&&\proofHint{}
                    \\
                    &&\proofDef{
                            \slStar\
                            (\valBuffer < 0 \ \leftrightarrow\ \numRead = \numWritten)
                            \ \slStar\
                            (\valBuffer \geq 0 \ \leftrightarrow\ \numRead = \numWritten -1)
                    }
                            &&&\proofHint{}
                    \\
                    &&\proofDef{
                            \displaystyle
                            \slStar\
                            \slBigStar_{\substack{i= 1, \dots, \numThs\\ k=\numWritten+1, \dots, \numThs}}
                                    \quadBack
                                    \assSignal{\sigPush{i}{k}}{\prodFinished{i}}
                            \ \slStar\
                            \slBigStar_{\substack{i=1, \dots, \numThs\\ k=1, \dots, \numWritten}}
                                    \!\!
                                    \assSignal{\sigPush{i}{k}}{\slWildcard}
                    }
                            &&&\proofHintML{
                            Signals set by producers \& used by
                            \\
                            consumers to wait when buffer is empty.}
                    \\
                    &&\proofDef{
                            \displaystyle
                            \slStar\
                            \slBigStar_{\substack{i= 1, \dots, \numThs\\ k=\numRead+1, \dots, \numThs}}
                                    \quadBack
                                    \assSignal{\sigPop{i}{k}}{\consFinished{i}}
                            \ \slStar\
                            \slBigStar_{\substack{i=1, \dots, \numThs\\ k=1, \dots, \numRead}}
                                    \!\!
                                    \assSignal{\sigPop{i}{k}}{\slWildcard}
                    }
                            &&&\proofHintML{
                            Signals set by consumers \& used by
                            \\
                            producers to wait when buffer is empty.}
                    \\&&&
                    \\
                    \proofDef{\assLockInvVar} &\ \proofDef{:=} \
                            &\proofDef{
                                    \exists \valBuffer \in \Z.\
                                    \exists \numWritten, \numRead  \in \N.\
                                    \exists \prodFinished{1}, \dots, \prodFinished{\numThs},
                                            \consFinished{1}, \dots, \consFinished{\numThs} 
                                            \in \B.\
                            }&
                    \\
                            &&\proofDef{
                                    \quad\quad
                                    \nextAssLockInvVar(\valBuffer, \numWritten, \numRead, 
                                            \prodFinished{1}, \dots, \prodFinished{\numThs},
                                            \consFinished{1}, \dots, \consFinished{\numThs})
                            }
            \end{array}
            $}
            
            \caption{Verification of program~\ref{appendix:caseStudy:undbounded:code}: Lock invariant.}
            \label{appendix:caseStudy:unbounded:lockInv}
    \end{figure}

    \begin{figure}

            \makebox[\textwidth][c]
            {$
            \hspace{0.04\textwidth}  
            \begin{array}{l c l p{0.3cm} | p{0.3cm} p{0.3\pdfpagewidth}}
                    \proofDef{\flInv{\flCnt}} &\ \proofDef{:=} \
                            &\proofDef{
                                            \slPointsTo{\locFlCnt}{\flCnt}
                                            \ \wedge\
                                            0 \leq \flCnt \leq \numThs
                                            \ \wedge\
                                            \numThs > 0
                            }
                            &&&\proofHintML{
                                    Decreasing counter.\\
                                    Prods.\@ \& cons.\@ $\flCnt+1, \dots, \numThs$ already forked.
                            }
                \\
                            &&\proofDef{
                                    \slStar\
                                    \obs{
                                            \multiset{\sigPush{i}{k}, \sigPop{i}{k} \ | \
                                                    1 \leq i \leq \flCnt
                                                    \ \wedge \
                                                    1 \leq k \leq \numThs
                                            }
                                    }
                            }
                            &&&\proofHintML{
                                    Remaining obligations for threads\\
                                    that have not been forked, yet.
                            }
                    \\
                    &&\proofDef{
                                    \slStar
                                    \displaystyle
                                    \slBigStar_{i = 1, \dots, \flCnt}
                                            \slPointsTo[\textstyle\frac{1}{2}]{\glocProdFin{i}}{\slFalse}
                                            \ \slStar\
                                            \slPointsTo[\textstyle\frac{1}{2}]{\glocConsFin{i}}{\slFalse}
                    }
                            &&&\proofHintML{
                                    Ghost heap cells for unforked consumers.
                            }
                    \\
                    &&\proofDef{
                            \slStar\
                            \assMutex[\frac{2\cdot\flCnt}{2\cdot\numThs}]{\mutVar}{\assLockInvVar}
                    }
                            &&&\proofHintML{
                                    Partial mutex chunk for unforked producers\\
                                    \& consumers.
                            }
            \end{array}
            $}
            
            \caption{Verification of program~\ref{appendix:caseStudy:undbounded:code}: Fork loop invariant.}
            \label{appendix:caseStudy:unbounded:forkLoopInv}
    \end{figure}

    \begin{figure}

            \makebox[\textwidth][c]
            {$
            \hspace{0.04\textwidth}  
            \begin{array}{l p{0.5cm} | p{0.5cm} p{0.3\pdfpagewidth}}
                    \dots&&&\referenceHint{Continuation of Fig.~\ref{appendix:caseStudy:unbounded:init}}.
                    \\
                    \lowlightText{\forall \locBuffer, \locMut, \locFlCnt,\
                            \idPush{1}{1}, \idPop{1}{1}, \dots, \idPush{\numThs}{\numThs}, \idPop{\numThs}{\numThs},\
                            \glocProdFin{1},\glocConsFin{1}, \dots, \glocProdFin{\numThs},\glocConsFin{\numThs}.
                    }&
                    \\
                    \progProofML{
                            \obsOf{
                                    \sigPush{1}{1}, \sigPop{1}{1}, \dots, 
                                    \sigPush{\numThs}{\numThs}, \sigPop{\numThs}{\numThs}
                            }\,
                            \slStar
                            \assMutex{\mutVar}{\assLockInvVar}
                            \\
                            \slStar\,
                            \slPointsTo{\locFlCnt}{\numThs}
                            \slStar
                            \displaystyle
                            \slBigStar_{i=1, \dots, \numThs}
                                    \!\!
                                    \slPointsTo[\textstyle\frac{1}{2}]{\glocProdFin{i}}{\slFalse}
                                    \slStar
                                    \slPointsTo[\textstyle\frac{1}{2}]{\glocConsFin{i}}{\slFalse}
                    }
                            &&&\proofRuleHint{\prViewShiftName \& \vsSemImpName}
                    \\
                    \progProofML{
                            \progProofCancel{
                                    \obsOf{
                                            \sigPush{1}{1}, \sigPop{1}{1}, \dots, 
                                            \sigPush{\numThs}{\numThs}, \sigPop{\numThs}{\numThs}
                                    }
                                    \slStar
                                    \assMutex{\mutVar}{\assLockInvVar}
                            }
                            \\
                            \progProofCancel{
                                    \slStar\,
                                    \slPointsTo{\locFlCnt}{\numThs}
                                    \slStar
                                    \displaystyle
                                    \slBigStar_{i=1, \dots, \numThs}
                                            \!\!
                                            \slPointsTo[\textstyle\frac{1}{2}]{\glocProdFin{i}}{\slFalse}
                                            \slStar
                                            \slPointsTo[\textstyle\frac{1}{2}]{\glocConsFin{i}}{\slFalse}
                        }\
                        \progProofNew{\flInv{\numThs}}
                    }
                            &&&\referenceHintML{
                                    For definition of fork loop invariant\\
                                    cf.\@ Fig.~\ref{appendix:caseStudy:unbounded:forkLoopInv}.
                            }
                    \\
                    \keyword{while}\ (
                            &&&\proofRuleHint{\prWhileDecStrictName}
                    \\\quad\quad
                            \ghostFont{\forall \valFlCnt.}&
                    \\\quad\quad
                            \progProof{
                                    \flInv{\progProofChanged{\valFlCnt}}
                            }&
                    \\\quad\quad
                            \keyword{fork}\ (       
                                    &&&\proofRuleHint{\prForkName}
                    \\\quad\quad\quad\quad
                                    \progProofML{
                                            \progProofNew{
                                                    \obsOf{\sigPush{\valFlCnt}{1}, \dots, \sigPush{\valFlCnt}{\numThs}}
                                                    \slStar
                                                    \assMutex[\frac{1}{\numThs}]{\mutVar}{\assLockInvVar}
                                            }
                                            \\
                                            \progProofNew{
                                                    \slStar\,
                                                    \slPointsTo[\textstyle\frac{1}{2}]{\glocProdFin{\valFlCnt}}{\slFalse}
                                            }
                                    }&
                    \\\quad\quad\quad\quad
                                    \dots&&&\referenceHint{Producer loop on Fig.~\ref{appendix:caseStudy:unbounded:producerLoop}.}
                    \\\quad\quad\quad\quad
                                    \progProof{
                                            \obs{\progProofChanged{\msEmpty}}
                                            \,
                                            \progProofCancel{
                                                    \slStar
                                                    \assMutex[\frac{1}{\numThs}]{\mutVar}{\assLockInvVar}
                                            }\
                                    }&
                    \\\quad\quad
                            ); & 
                    \\\quad\quad
                            \progProofML{
                                    \slPointsTo{\locFlCnt}{\valFlCnt}
                                    \slStar
                                    \quadBack\!\!\!\!
                                    \displaystyle
                                    \slBigStar_{i = 1, \dots, \progProofChanged[\scriptsize]{\valFlCnt-1}}
                                            \quadBack\!\!\!\!\!
                                            \slPointsTo[\textstyle\frac{1}{2}]{\glocProdFin{i}}{\slFalse}
                                    \slStar
                                    \displaystyle
                                    \slBigStar_{i = 1, \dots, \valFlCnt}
                                            \!\!\!
                                            \slPointsTo[\textstyle\frac{1}{2}]{\glocConsFin{i}}{\slFalse}
                                    \slStar
                                    \assMutex[\progProofChanged{\frac{2\cdot\valFlCnt-1}{2\cdot\numThs}}]{\mutVar}{\assLockInvVar}
                                    \\
                                    \slStar\,
                                    \fixedPredNameFont{obs}
                                    \!\left(\begin{array}{l}
                                            \progProofCancel{\multiset{\sigPush{\valFlCnt}{1}, \dots, \sigPush{\valFlCnt}{\numThs}}\, \msCup}
                                            \,
                                            \multiset{\sigPop{\valFlCnt}{1}, \dots, \sigPop{\valFlCnt}{\numThs}}       
                                            \\
                                            \msCup\,
                                            \multiset{
                                                    \sigPush{i}{k}, \sigPop{i}{k} \ | \
                                                    1 \leq i \leq \valFlCnt-1
                                                    \wedge
                                                    1 \leq k \leq \numThs
                                            }
                                    \end{array}\right)
                            }&
                    \\\quad\quad
                            \keyword{fork}\ (       
                                    &&&\proofRuleHint{\prForkName}
                    \\\quad\quad\quad\quad
                                    \progProofML{
                                            \progProofNew{
                                                    \obsOf{\sigPop{\valFlCnt}{1}, \dots, \sigPop{\valFlCnt}{\numThs}}
                                                    \slStar
                                                    \assMutex[\frac{1}{\numThs}]{\mutVar}{\assLockInvVar}
                                            }
                                            \\
                                            \progProofNew{
                                                    \slStar\,
                                                    \slPointsTo[\frac{1}{2}]{\glocConsFin{\valFlCnt}}{\slFalse}
                                            }
                                    }&
                    \\\quad\quad\quad\quad
                                    \dots&&&\referenceHint{Consumer loop on Fig.~\ref{appendix:caseStudy:unbounded:consumerLoop}.}
                    \\\quad\quad\quad\quad
                                    \progProof{
                                            \obs{\progProofChanged{\msEmpty}}
                                            \,
                                            \progProofCancel{
                                                    \slStar
                                                    \assMutex[\frac{1}{\numThs}]{\mutVar}{\assLockInvVar}
                                            }\
                                    }&
                    \\\quad\quad
                            ); & 
                    \\\quad\quad
                            \progProofML{
                                    \slPointsTo{\locFlCnt}{\valFlCnt}
                                    \slStar
                                    \quadBack\!\!\!\!\!
                                    \displaystyle
                                    \slBigStar_{ i = 1, \dots, \progProofChanged[\scriptsize]{\valFlCnt-1}}
                                            \!\!\!\quadBack
                                            \slPointsTo[\textstyle\frac{1}{2}]{\glocProdFin{i}}{\slFalse}
                                            \slStar
                                            \slPointsTo[\textstyle\frac{1}{2}]{\glocConsFin{i}}{\slFalse}
                                    \slStar
                                    \assMutex[\progProofChanged{\frac{2\cdot\valFlCnt-2}{2\cdot\numThs}}]{\mutVar}{\assLockInvVar}
                                    \\
                                    \slStar\,
                                    \fixedPredNameFont{obs}
                                    \!\left(\begin{array}{l}
                                            \progProofCancel{\multiset{\sigPop{\valFlCnt}{1}, \dots, \sigPop{\valFlCnt}{\numThs}}\, \msCup}
                                            \\
                                            \multiset{
                                                    \sigPush{i}{k}, \sigPop{i}{k} \ | \
                                                    1 \leq i \leq \valFlCnt-1
                                                    \wedge
                                                    1 \leq k \leq \numThs
                                            }
                                    \end{array}\right)
                            }&
                    \\\quad\quad
                    \cmdAssignToHeap{\flCnt}{\cmdReadHeapLoc{\flCnt}-1};
                            &&&\proofHintML{
                                    \proofRuleHintML{
                                    \prLetName \& \prReadHeapLocName\\
                                    \& \prAssignToHeapName
                                    }
                                    \\
                                    Remember that command is syntactic sugar.
                            }
                    \\\quad\quad
                            \progProofML{
                                    \slPointsTo{\locFlCnt}{\progProofChanged{\valFlCnt-1}}
                                    \slStar
                                    \quadBack\!
                                    \displaystyle
                                    \slBigStar_{i = 1, \dots, \valFlCnt-1}
                                            \!\!\!\!\!\!
                                            \slPointsTo[\textstyle\frac{1}{2}]{\glocProdFin{i}}{\slFalse}
                                            \slStar
                                            \slPointsTo[\textstyle\frac{1}{2}]{\glocConsFin{i}}{\slFalse}
                                    \, \slStar
                                    \assMutex[\textstyle\frac{2\cdot\valFlCnt-2}{2\cdot\numThs}]{\mutVar}{\assLockInvVar}
                                    \\
                                    \slStar\,
                                    \obs{\multiset{
                                            \sigPush{i}{k}, \sigPop{i}{k} \ | \
                                            1 \leq i \leq \valFlCnt-1
                                            \wedge
                                            1 \leq k \leq \numThs
                                    }}
                            }&
                    \\\quad\quad
                            \progProof{\flInv{\valFlCnt-1}}&
                    \\\quad\quad
                            \flCnt > 0
                                    &&&\proofRuleHint{\prExpName}
                    \\\quad\quad
                            \progProof{\flInv{\valFlCnt-1}}
                                    &&&\proofRuleHint{\prViewShiftName \& \vsSemImpName}
                    \\\quad\quad
                            \progProof{
                                    \slIfElse{\valFlCnt-1 > 0}
                                            {\flInv{\valFlCnt-1}}
                                            {\noObs}
                            }&
                    \\
                    )\ \keyword{do}\ \cmdSkip & 
                    \\
                    \progProof{
                            \progProofCancel{\flInv{\numThs}}
                            \,
                            \progProofNew{\noObs}
                    }&
            \end{array}
            $}
            
            \caption{Verification of program~\ref{appendix:caseStudy:undbounded:code}: Fork loop.}
            \label{appendix:caseStudy:unbounded:forkLoop}
    \end{figure}

    \begin{figure}

            \makebox[\textwidth][c]
            {$
            \hspace{0.04\textwidth}  
            \begin{array}{l c l p{0.3cm} | p{0.3cm} p{0.3\pdfpagewidth}}
                    \proofDef{\pInv{i}{\obBagVar}} &\ \proofDef{:=} \
                            &\proofDef{
                                            \exists \nextPush.\
                                            1 \leq \nextPush \leq \numThs
                            }
                            &&&\proofHintML{
                                    Lower bound for index of next element\\
                                    that will be pushed by any thread.
                            }
                \\
                            &&\proofDef{
                                    \slStar\
                                    \obBagVar = \multiset{\sigPush{i}{\nextPush}, \dots, \sigPush{i}{\numThs}}
                            }
                            &&&\proofHintML{}
                \\
                            &&\proofDef{
                                    \slStar\
                                    \slPointsTo[\frac{1}{2}]{\glocProdFin{i}}{\slFalse}
                            }
                            &&&\proofHintML{}
            \end{array}
            $}
            
            \caption{Verification of program~\ref{appendix:caseStudy:undbounded:code}: Loop invariant for producer $i$.}
            \label{appendix:caseStudy:unbounded:producerLoopInv}
    \end{figure}

    \begin{figure}

            \makebox[\textwidth][c]
            {$
            \hspace{0.04\textwidth}  
            \begin{array}{l p{0.5cm} | p{0.5cm} p{0.3\pdfpagewidth}}
                    \dots&&&\referenceHint{Continuation of Fig.~\ref{appendix:caseStudy:unbounded:forkLoop}}.
                    \\
                    \lowlightText{\forall \locBuffer, \locMut, \locFlCnt,
                            \idPush{1}{1}, \idPop{1}{1}, \dots, \idPush{\numThs}{\numThs}, \idPop{\numThs}{\numThs},\
                            \glocProdFin{1},\glocConsFin{1}, \dots, \glocProdFin{\numThs},\glocConsFin{\numThs}.
                            \valFlCnt.
                    }&
                    \\
                    \progProof{
                            \obsOf{\sigPush{\valFlCnt}{1}, \dots, \sigPush{\valFlCnt}{\numThs}}
                            \slStar
                            \assMutex[\frac{1}{\numThs}]{\mutVar}{\assLockInvVar}
                            \slStar
                            \slPointsTo[\frac{1}{2}]{\glocProdFin{\valFlCnt}}{\slFalse}
                    }
                            &&&\proofRuleHint{\prViewShiftName \& \vsSemImpName}
                    \\
                    \progProofML{
                            \obsOf{\sigPush{\valFlCnt}{1}, \dots, \sigPush{\valFlCnt}{\numThs}}
                            \slStar
                            \pInv{\valFlCnt}{\multiset{\sigPush{\valFlCnt}{1}, \dots, \sigPush{\valFlCnt}{\numThs}}}
                            \\
                            \slStar\,
                            \assMutex[\frac{1}{\numThs}]{\mutVar}{\assLockInvVar}
                    }
                            &&&\referenceHintML{
                                    For definition of this producer's\\
                                    loop invariant \pInv{\valFlCnt}{\obBagVar} cf.\@ Fig.~\ref{appendix:caseStudy:unbounded:producerLoopInv}.
                            }
                    \\
                    \proofHint{$
                            \mutGetLev{\mutVar} \ = \ \levMut 
                                \ <\ 2\cdot k-1 \ =\ \sigGetLev{\sigPush{\valFlCnt}{k}}$
                            \quad
                            for
                            \quad
                            $1 \leq k \leq \numThs$
                    }
                            &&&\proofHint{Justification for application of:}
                    \\
                    \keyword{with}\ \mutProgVar\ \keyword{await}\ (
                            &&&\proofRuleHint{\prAwaitGenName}
                    \\\quad\quad
                            \ghostFont{\forall \obBagVar.}
                                    &&&\referenceHintML{
                                            For definition of lock invariant \assLockInvVar \& \lockInvNoQuant\\
                                            cf.\@ Fig.~\ref{appendix:caseStudy:unbounded:lockInv}.
                                    }
                    \\\quad\quad
                            \progProof{
                                    \obs{
                                            \progProofChanged{\obBagVar} 
                                            \msCup 
                                            \progProofNew{\multiset{\mutVar}}
                                    }
                                    \slStar
                                    \pInv{\valFlCnt}{\progProofChanged{\obBagVar}}
                                    \slStar
                                    \progProofNew{\assLockInvVar}
                            }
                                    &&&\proofRuleHint{\prExistsName}
                    \\\quad\quad
                            \ghostFont{
                                    \forall
                                    \nextPush,
                                    \valBuffer,
                                    \numWritten, \numRead,
                                    \prodFinished{1}, \dots, \prodFinished{\numThs},
                                    \consFinished{1}, \dots, \consFinished{\numThs}.
                            }&
                    \\\quad\quad
                            \proofDef{
                                    \lockInvNoQuant := 
                                    \nextAssLockInvVar(\valBuffer, \numWritten, \numRead, 
                                            \prodFinished{1}, \dots, \prodFinished{\numThs},
                                            \consFinished{1}, \dots, \consFinished{\numThs})
                            }&
                    \\\quad\quad
                            \progProof{
                                    \obsOf{
                                            \progProofChanged{
                                                    \sigPush{\valFlCnt}{\nextPush}, \dots, \sigPush{\valFlCnt}{\numThs},
                                            }
                                            \,
                                            \mutVar
                                    }
                                    \slStar
                                    \slPointsTo[\frac{1}{2}]{\glocProdFin{\valFlCnt}}{\slFalse}
                                    \slStar
                                    \progProofChanged{\lockInvNoQuant}
                            }
                                    &&&\proofRuleHintML{
                                            \prViewShiftName\\
                                            \& \vsGhostLoopName \& \vsSetSignalName
                                    }
                    \\\quad\quad
                            \progProof{
                                    \obsOf{
                                            \sigPush{\valFlCnt}{\progProofChanged[\scriptsize]{\max(\nextPush, \numWritten+1)}}, 
                                            \dots, 
                                            \sigPush{\valFlCnt}{\numThs},
                                            \mutVar
                                    }
                                    \slStar
                                    \slPointsTo[\frac{1}{2}]{\glocProdFin{\valFlCnt}}{\slFalse}
                                    \slStar
                                    \lockInvNoQuant
                            }&
                    \\\quad\quad
                            \keyword{if}\ \bufferLoaded < 0\ \keyword{then}
                                    &&&\proofRuleHint{\prIfName}
                    \\\quad\quad\quad\quad
                                    \progProof{
                                            \progProofNew{\valBuffer < 0}
                                            \slStar\dots
                                    }&
                    \\\quad\quad\quad\quad
                                    \cmdAssignToHeap{\buffer}{\cmdRandomNat}
                                            &&&\proofRuleHintML{
                                                    \prAssignToHeapName\\
                                                    \& \prViewShiftName \& \vsSemImpName
                                            }
                    \\\quad\quad\quad\quad
                                    \progProofML{
                                            \progProofNew{
                                                    \exists \nextValBuffer.\
                                                    \slPointsTo{\locBuffer}{\nextValBuffer}
                                                    \wedge
                                                    \nextValBuffer \geq 0
                                            }
                                            \\
                                            \slStar\,
                                            \progProofCancel{
                                                    \slPointsTo[\frac{1}{2}]{\glocProdFin{\valFlCnt}}{\slFalse}
                                                    \slStar
                                                    \slPointsTo[\frac{1}{2}]{\glocProdFin{\valFlCnt}}{\prodFinished{\valFlCnt}}
                                            }\,
                                            \progProofNew{
                                                    \slPointsTo{\glocProdFin{\valFlCnt}}{\slFalse}
                                    }
                                            \slStar\dots
                                    }
                                            &&&\proofRuleHint{\prViewShiftName \& \vsMutateGhostCellName}
                    \\\quad\quad\quad\quad
                                    \progProof{
                                            \slPointsTo{\glocProdFin{\valFlCnt}}{\progProofChanged{\slTrue}}
                                            \slStar\dots
                                    }
                                            &&&\proofRuleHintML{
                                                    \prViewShiftName\\
                                                    \& \vsGhostLoopName \& \vsSetSignalName
                                            }
                    \\\quad\quad\quad\quad
                                    \progProofML{
                                            \obsOf{
                                                    \progProofCancel{
                                                            \sigPush{\valFlCnt}{\max(\nextPush, \numWritten+1)},
                                                            \dots,
                                                            \sigPush{\valFlCnt}{\numThs},
                                                    }
                                                    \
                                                    \mutVar
                                            }
                                            \\
                                            \slStar
                                            \displaystyle
                                            \slBigStar_{k=\max(\nextPush, \numWritten+1), \dots, \numThs}
                                                    \quadBack\quadBack\!\!\!\!
                                                    \assSignal{\sigPush{\valFlCnt}{k}}{\progProofChanged{\slTrue}}
                                            \slStar\dots
                                    }
                                            &&&\ \proofDefML{
                                                    \proofRuleHint{\prViewShiftName \& \vsSemImpName}\\
                                                    \text{Define}\ \pLockInvRest\ \text{such that}\\
                                                    \viewShift
                                                            {
                                                                    \exists i.\ \assSignal{\sigPop{i}{\numRead+1}}{\slFalse}
                                                                    \slStar \pLockInvRest
                                                            }
                                                            {\assLockInvVar}.
                                            }
                    \\\quad\quad\quad\quad
                                    \progProofML{
                                            \exists \nextObBagVar.\
                                            \obs{\nextObBagVar \msCup \multiset{\mutVar}}
                                            \\
                                            \slStar\,
                                            \slIfElseBranchML{\valBuffer < 0}
                                                    {
                                                            \assLockInvVar 
                                                            \slStar
                                                            \obBagVar = \msEmpty
                                                    }
                                                    {
                                                            \pInv{\valFlCnt}{\nextObBagVar}
                                                            \slStar
                                                            \exists i.\ \assSignal{\sigPop{i}{\numRead+1}}{\slFalse}
                                                            \slStar
                                                            \pLockInvRest
                                                            \\
                                                            \slStar\,
                                                            \levPop{\numRead+1} \levObsLt \nextObBagVar
                                                            \slStar
                                                            \mutGetLev{\mutVar} \levObsLt \nextObBagVar
                                                    }
                                    }
                                            &&&\proofDef{=: \pPostIf}
                    \\\quad\quad
                            \ghostFont{\keyword{else}}
                                            &&&\proofRuleHint{\prViewShiftName \& \vsSemImpName}
                    \\\quad\quad\quad\quad
                                    \progProof{
                                            \progProofNew{
                                                    \valBuffer \geq 0
                                                    \wedge
                                                    \numRead = \numWritten-1
                                            }
                                            \slStar
                                    \obsOf{
                                            \sigPush{\valFlCnt}{\max(\nextPush, \numWritten+1)}, 
                                            \dots, 
                                            \sigPush{\valFlCnt}{\numThs},
                                            \mutVar
                                    }
                                            \slStar\dots
                                    }
                                            &&&\proofRuleHint{\prViewShiftName \& \vsSemImpName}
                    \\\quad\quad\quad\quad
                                    \proofHintML{$
                                            \levPop{\numRead+1} \ =\ 2\cdot (\numRead+1)
                                                    \ =\  2\cdot \numWritten
                                            $
                                            \\
                                            $\phantom{\levPop{\numRead+1}} 
                                                    \ <\  2\cdot \max(\nextPush, \numWritten+1) - 1
                                                    \ \leq\ \levPush{k}
                                            $
                                            \quad 
                                            for $\max(\nextPush, \numWritten+1) \leq k \leq \numThs$.
                                            \\
                                            $\mutGetLev{\mutVar} \ = \ 0 \ < \ \levPush{k}$
                                            \quad
                                            for any $k$.
                                            \\
                                            $\numRead < \numThs \wedge \lockInvNoQuant 
                                                    \ \Rightarrow\
                                                    \exists i.\ 
                                                            \consFinished{i} = \slFalse
                                                            \,\wedge
                                                            \!\!\!
                                                            \displaystyle
                                                            \slBigStar_{j=1, \dots, \numThs}
                                                                    \!\!\!
                                                                    \assSignal{\sigPop{j}{\numRead+1}}{\consFinished{j}}
                                            $
                                            \\
                                            $\phantom{\numRead < \numThs \wedge \lockInvNoQuant}
                                                    \ \Rightarrow\
                                                    \exists i.\ \assSignal{\sigPop{i}{\numRead+1}}{\slFalse}
                                            $
                                    }
                                            &&&\proofHintML{Justification of \keyword{else} branch in \pPostIf.\\\\\\\\\\\\}
                    \\\quad\quad\quad\quad
                                    \progProof{\pPostIf}&
                    \\\quad\quad
                            \progProof{
                                    \progProofCancel{
                                            \obsOf{
                                                    \sigPush{\valFlCnt}{\max(\nextPush, \numWritten+1)}, 
                                                    \dots, 
                                                    \sigPush{\valFlCnt}{\numThs},
                                                    \mutVar
                                            }
                                            \slStar
                                            \slPointsTo[\frac{1}{2}]{\glocProdFin{\valFlCnt}}{\slFalse}
                                            \slStar
                                            \lockInvNoQuant
                                    }\,
                                    \progProofNew{\pPostIf}
                            }&
                    \\\quad\quad
                            \bufferLoaded < 0
                                    &&&\proofRuleHint{\prExpName}
                    \\\quad\quad
                            \progProof{\pPostIf}&
                    \\
                    ) &
                    \\
                    \progProofML{
                            \progProofNew{\exists \nextObBagVar.}\
                            \obs{
                                    \progProofCancel{\multiset{
                                            \sigPush{\valFlCnt}{1}, \dots, \sigPush{\valFlCnt}{\numThs}
                                    }}\,
                                    \progProofNew{\nextObBagVar}
                            }\,
                            \progProofCancel{
                                    \slStar\,
                                    \pInv{\valFlCnt}{\multiset{\sigPush{\valFlCnt}{1}, \dots, \sigPush{\valFlCnt}{\numThs}}}
                            }
                            \\
                            \slStar\,
                            \assMutex[\frac{1}{\numThs}]{\mutVar}{\assLockInvVar}
                            \,
                            \progProofNew{
                                    \slStar\,
                                    \nextObBagVar = \msEmpty
                            }
                    }
                            &&&\proofRuleHint{\prViewShiftName \& \vsSemImpName}
                    \\
                    \progProof{\noObs}
            \end{array}
            $}
            
            \caption{Verification of program~\ref{appendix:caseStudy:undbounded:code}: Producer loop.}
            \label{appendix:caseStudy:unbounded:producerLoop}
    \end{figure}

    \begin{figure}

            \makebox[\textwidth][c]
            {$
            \hspace{0.04\textwidth}  
            \begin{array}{l c l p{0.3cm} | p{0.3cm} p{0.3\pdfpagewidth}}
                    \proofDef{\cInv{i}{\obBagVar}} &\ \proofDef{:=} \
                            &\proofDef{
                                            \exists \nextPop.\
                                            1 \leq \nextPop \leq \numThs
                            }
                            &&&\proofHintML{
                                    Lower bound for index of next element\\
                                    that will be popped by any thread.
                            }
                \\
                            &&\proofDef{
                                    \slStar\
                                    \obBagVar = \multiset{\sigPop{i}{\nextPop}, \dots, \sigPush{i}{\numThs}}
                            }
                            &&&\proofHintML{}
                \\
                            &&\proofDef{
                                    \slStar\
                                    \slPointsTo[\frac{1}{2}]{\glocConsFin{i}}{\slFalse}
                            }
                            &&&\proofHintML{}
            \end{array}
            $}
            
            \caption{Verification of program~\ref{appendix:caseStudy:undbounded:code}: Loop invariant for consumer $i$.}
            \label{appendix:caseStudy:unbounded:consumerLoopInv}
    \end{figure}

    \begin{figure}

            \makebox[\textwidth][c]
            {$
            \hspace{0.04\textwidth}  
            \begin{array}{l p{0.5cm} | p{0.5cm} p{0.3\pdfpagewidth}}
                    \dots&&&\referenceHint{Continuation of Fig.~\ref{appendix:caseStudy:unbounded:forkLoop}}.
                    \\
                    \lowlightText{\forall \locBuffer, \locMut, \locFlCnt,
                            \idPush{1}{1}, \idPop{1}{1}, \dots, \idPush{\numThs}{\numThs}, \idPop{\numThs}{\numThs},\
                            \glocProdFin{1},\glocConsFin{1}, \dots, \glocProdFin{\numThs},\glocConsFin{\numThs}.
                            \valFlCnt.
                    }&
                    \\
                    \progProof{
                            \obsOf{\sigPop{\valFlCnt}{1}, \dots, \sigPop{\valFlCnt}{\numThs}}
                            \slStar
                            \assMutex[\frac{1}{\numThs}]{\mutVar}{\assLockInvVar}
                            \slStar
                            \slPointsTo[\frac{1}{2}]{\glocConsFin{\valFlCnt}}{\slFalse}
                    }
                            &&&\proofRuleHint{\prViewShiftName \& \vsSemImpName}
                    \\
                    \progProofML{
                            \obsOf{\sigPop{\valFlCnt}{1}, \dots, \sigPop{\valFlCnt}{\numThs}}
                            \slStar
                            \cInv{\valFlCnt}{\multiset{\sigPop{\valFlCnt}{1}, \dots, \sigPop{\valFlCnt}{\numThs}}}
                            \\
                            \slStar\,
                            \assMutex[\frac{1}{\numThs}]{\mutVar}{\assLockInvVar}
                    }
                            &&&\referenceHintML{
                                    For definition of this consumer's\\
                                    loop invariant \cInv{\valFlCnt}{\obBagVar} cf.\@ Fig.~\ref{appendix:caseStudy:unbounded:consumerLoopInv}.
                            }
                    \\
                    \proofHint{$
                            \mutGetLev{\mutVar} \ = \ \levMut 
                                \ <\ 2\cdot k-1 \ =\ \sigGetLev{\sigPush{\valFlCnt}{k}}$
                            \quad
                            for
                            \quad
                            $1 \leq k \leq \numThs$
                    }
                            &&&\proofHint{Justification for application of:}
                    \\
                    \keyword{with}\ \mutProgVar\ \keyword{await}\ (
                            &&&\proofRuleHint{\prAwaitGenName}
                    \\\quad\quad
                            \ghostFont{\forall \obBagVar.}
                                    &&&\referenceHintML{
                                            For definition of lock invariant \assLockInvVar \& \lockInvNoQuant\\
                                            cf.\@ Fig.~\ref{appendix:caseStudy:unbounded:lockInv}.
                                    }
                    \\\quad\quad
                            \progProof{
                                    \obs{
                                            \progProofChanged{\obBagVar} 
                                            \msCup 
                                            \progProofNew{\multiset{\mutVar}}
                                    }
                                    \slStar
                                    \cInv{\valFlCnt}{\progProofChanged{\obBagVar}}
                                    \slStar
                                    \progProofNew{\assLockInvVar}
                            }
                                    &&&\proofRuleHint{\prExistsName}
                    \\\quad\quad
                            \ghostFont{
                                    \forall
                                    \nextPop,
                                    \valBuffer,
                                    \numWritten, \numRead,
                                    \prodFinished{1}, \dots, \prodFinished{\numThs},
                                    \consFinished{1}, \dots, \consFinished{\numThs}.
                            }&
                    \\\quad\quad
                            \proofDef{
                                    \lockInvNoQuant := 
                                    \nextAssLockInvVar(\valBuffer, \numWritten, \numRead, 
                                            \prodFinished{1}, \dots, \prodFinished{\numThs},
                                            \consFinished{1}, \dots, \consFinished{\numThs})
                            }&
                    \\\quad\quad
                            \progProof{
                                    \obsOf{
                                            \progProofChanged{
                                                    \sigPop{\valFlCnt}{\nextPop}, \dots, \sigPop{\valFlCnt}{\numThs},
                                            }
                                            \,
                                            \mutVar
                                    }
                                    \slStar
                                    \slPointsTo[\frac{1}{2}]{\glocConsFin{\valFlCnt}}{\slFalse}
                                    \slStar
                                    \progProofChanged{\lockInvNoQuant}
                            }
                                    &&&\proofRuleHintML{
                                            \prViewShiftName\\
                                            \& \vsGhostLoopName \& \vsSetSignalName
                                    }
                    \\\quad\quad
                            \progProof{
                                    \obsOf{
                                            \sigPop{\valFlCnt}{\progProofChanged[\scriptsize]{\max(\nextPop, \numRead+1)}}, 
                                            \dots, 
                                            \sigPop{\valFlCnt}{\numThs},
                                            \mutVar
                                    }
                                    \slStar
                                    \slPointsTo[\frac{1}{2}]{\glocConsFin{\valFlCnt}}{\slFalse}
                                    \slStar
                                    \lockInvNoQuant
                            }
                                    &&&\proofRuleHint{\prViewShiftName \& \vsSemImpName}
                    \\\quad\quad
                            \progProof{
                                    \progProofCancel{
                                            \slPointsTo[\frac{1}{2}]{\glocConsFin{\valFlCnt}}{\slFalse}
                                            \slStar
                                            \slPointsTo[\frac{1}{2}]{\glocConsFin{\valFlCnt}}{\consFinished{\valFlCnt}}
                                    }\,
                                    \progProofNew{
                                            \slPointsTo{\glocConsFin{\valFlCnt}}{\slFalse}
                                            \wedge
                                            \consFinished{\valFlCnt} = \slFalse
                                            \wedge
                                            \numRead < \numThs
                                    }
                                    \slStar\dots
                            }&
                    \\\quad\quad
                            \keyword{if}\ \bufferLoaded \geq 0\ \keyword{then}
                                    &&&\proofRuleHint{\prIfName}
                    \\\quad\quad\quad\quad
                                    \progProof{
                                            \progProofNew{\valBuffer \geq 0}
                                            \slStar\dots
                                    }&
                    \\\quad\quad\quad\quad
                                    \cmdAssignToHeap{\buffer}{-1}
                                            &&&\proofRuleHintML{
                                                    \prAssignToHeapName
                                            }
                    \\\quad\quad\quad\quad
                                    \progProofML{
                                            \slPointsTo{\locBuffer}{\progProofChanged{-1}}
                                            \slStar\dots
                                    }
                                            &&&\proofRuleHint{\prViewShiftName \& \vsMutateGhostCellName}
                    \\\quad\quad\quad\quad
                                    \progProof{
                                            \slPointsTo{\glocConsFin{\valFlCnt}}{\progProofChanged{\slTrue}}
                                            \slStar\dots
                                    }
                                            &&&\proofRuleHintML{
                                                    \prViewShiftName\\
                                                    \& \vsGhostLoopName \& \vsSetSignalName
                                            }
                    \\\quad\quad\quad\quad
                                    \progProofML{
                                            \obsOf{
                                                    \progProofCancel{
                                                            \sigPop{\valFlCnt}{\max(\nextPop, \numRead+1)},
                                                            \dots,
                                                            \sigPop{\valFlCnt}{\numThs},
                                                    }
                                                    \
                                                    \mutVar
                                            }
                                            \\
                                            \slStar
                                            \displaystyle
                                            \slBigStar_{k=\max(\nextPop, \numRead+1), \dots, \numThs}
                                                    \quadBack\quadBack\!\!\!\!
                                                    \assSignal{\sigPop{\valFlCnt}{k}}{\progProofChanged{\slTrue}}
                                            \slStar\dots
                                    }
                                            &&&\ \proofDefML{
                                                    \proofRuleHint{\prViewShiftName \& \vsSemImpName}\\
                                                    \text{Define}\ \cLockInvRest\ \text{such that}\\
                                                    \viewShift
                                                            {
                                                                    \exists i.\ \assSignal{\sigPush{i}{\numWritten+1}}{\slFalse}
                                                                    \slStar \cLockInvRest
                                                            }
                                                            {\assLockInvVar}.
                                            }
                    \\\quad\quad\quad\quad
                                    \progProofML{
                                            \exists \nextObBagVar.\
                                            \obs{\nextObBagVar \msCup \multiset{\mutVar}}
                                            \\
                                            \slStar\,
                                            \slIfElseBranchML{\valBuffer \geq 0}
                                                    {
                                                            \assLockInvVar 
                                                            \slStar
                                                            \obBagVar = \msEmpty
                                                    }
                                                    {
                                                            \cInv{\valFlCnt}{\nextObBagVar}
                                                            \slStar
                                                            \exists i.\ \assSignal{\sigPush{i}{\numWritten+1}}{\slFalse}
                                                            \slStar
                                                            \cLockInvRest
                                                            \\
                                                            \slStar\,
                                                            \levPush{\numWritten+1} \levObsLt \nextObBagVar
                                                            \slStar
                                                            \mutGetLev{\mutVar} \levObsLt \nextObBagVar
                                                    }
                                    }
                                            &&&\proofDef{=: \cPostIf}
                    \\\quad\quad
                            \ghostFont{\keyword{else}}
                                            &&&\proofRuleHint{\prViewShiftName \& \vsSemImpName}
                    \\\quad\quad\quad\quad
                                    \progProof{
                                            \progProofNew{
                                                    \valBuffer < 0
                                                    \wedge
                                                    \numRead = \numWritten < \numThs
                                            }
                                            \slStar
                                    \obsOf{
                                            \sigPop{\valFlCnt}{\max(\nextPop, \numRead+1)}, 
                                            \dots, 
                                            \sigPop{\valFlCnt}{\numThs},
                                            \mutVar
                                    }
                                            \slStar\dots
                                    }
                                            &&&\proofRuleHint{\prViewShiftName \& \vsSemImpName}
                    \\\quad\quad\quad\quad
                                    \proofHintML{$
                                            \levPush{\numWritten+1} \ =\ 2\cdot (\numWritten+1) - 1
                                                    \ =\  2\cdot (\numRead+1) - 1
                                            $
                                            \\
                                            $\phantom{\levPop{\numRead+1}} 
                                                    \ <\  2\cdot \max(\nextPop, \numRead+1) 
                                                    \ \leq\ \levPop{k}
                                            $
                                            \quad 
                                            for $\max(\nextPop, \numRead+1) \leq k \leq \numThs$.
                                            \\
                                            $\mutGetLev{\mutVar} \ = \ 0 \ < \ \levPop{k}$
                                            \quad
                                            for any $k$.
                                            \\
                                            $\numWritten < \numThs \wedge \lockInvNoQuant 
                                                    \ \Rightarrow\
                                                    \exists i.\ 
                                                            \prodFinished{i} = \slFalse
                                                            \,\wedge
                                                            \!\!\!
                                                            \displaystyle
                                                            \slBigStar_{j=1, \dots, \numThs}
                                                                    \!\!\!
                                                                    \assSignal{\sigPush{j}{\numWritten+1}}{\prodFinished{j}}
                                            $
                                            \\
                                            $\phantom{\numRead < \numThs \wedge \lockInvNoQuant}
                                                    \ \Rightarrow\
                                                    \exists i.\ \assSignal{\sigPush{i}{\numWritten+1}}{\slFalse}
                                            $
                                    }
                                            &&&\proofHintML{Justification of \keyword{else} branch in \cPostIf.\\\\\\\\\\\\}
                    \\\quad\quad\quad\quad
                                    \progProof{\cPostIf}&
                    \\\quad\quad
                            \progProof{
                                    \progProofCancel{
                                            \obsOf{
                                                    \sigPop{\valFlCnt}{\max(\nextPop, \numRead+1)}, 
                                                    \dots, 
                                                    \sigPop{\valFlCnt}{\numThs},
                                                    \mutVar
                                            }
                                            \slStar\dots
                                    }\,
                                    \progProofNew{\cPostIf}
                            }&
                    \\\quad\quad
                            \bufferLoaded \geq 0
                                    &&&\proofRuleHint{\prExpName}
                    \\\quad\quad
                            \progProof{\cPostIf}&
                    \\
                    ) &
                    \\
                    \progProofML{
                            \progProofNew{\exists \nextObBagVar.}\
                            \obs{
                                    \progProofCancel{\multiset{
                                            \sigPop{\valFlCnt}{1}, \dots, \sigPop{\valFlCnt}{\numThs}
                                    }}\,
                                    \progProofNew{\nextObBagVar}
                            }\,
                            \progProofCancel{
                                    \slStar\,
                                    \cInv{\valFlCnt}{\multiset{\sigPop{\valFlCnt}{1}, \dots, \sigPop{\valFlCnt}{\numThs}}}
                            }
                            \\
                            \slStar\,
                            \assMutex[\frac{1}{\numThs}]{\mutVar}{\assLockInvVar}
                            \,
                            \progProofNew{
                                    \slStar\,
                                    \nextObBagVar = \msEmpty
                            }
                    }
                            &&&\proofRuleHint{\prViewShiftName \& \vsSemImpName}
                    \\
                    \progProof{\noObs}
            \end{array}
            $}
            
            \caption{Verification of program~\ref{appendix:caseStudy:undbounded:code}: Consumer loop.}
            \label{appendix:caseStudy:unbounded:consumerLoop}
    \end{figure}

}   

%% file: ghostSignals--CAV-paper--extended.bbl
\begin{thebibliography}{10}
\providecommand{\url}[1]{\texttt{#1}}
\providecommand{\urlprefix}{URL }
\providecommand{\doi}[1]{https://doi.org/#1}

\bibitem{BIRKEDAL20104102}
The category-theoretic solution of recursive metric-space equations.
  Theoretical Computer Science  \textbf{411}(47),  4102 -- 4122 (2010).
  \doi{10.1016/j.tcs.2010.07.010}

\bibitem{appel-indexed}
Appel, A.W., McAllester, D.: An indexed model of recursive types for
  foundational proof-carrying code. ACM Trans. Program. Lang. Syst.
  \textbf{23}(5),  657–683 (Sep 2001). \doi{10.1145/504709.504712}

\bibitem{finite-blocking}
Bostr{\"{o}}m, P., M{\"{u}}ller, P.: Modular verification of finite blocking in
  non-terminating programs. In: Boyland, J.T. (ed.) 29th European Conference on
  Object-Oriented Programming, {ECOOP} 2015, July 5-10, 2015, Prague, Czech
  Republic. LIPIcs, vol.~37, pp. 639--663. Schloss Dagstuhl - Leibniz-Zentrum
  f{\"{u}}r Informatik (2015). \doi{10.4230/LIPIcs.ECOOP.2015.639},
  \url{https://doi.org/10.4230/LIPIcs.ECOOP.2015.639}

\bibitem{Boyapati2002OwnershipTF}
Boyapati, C., Lee, R., Rinard, M.: Ownership types for safe programming:
  preventing data races and deadlocks. In: OOPSLA '02 (2002)

\bibitem{Dershowitz1979MultisetOrder}
Dershowitz, N., Manna, Z.: Proving termination with multiset orderings. In:
  ICALP (1979). \doi{10.1007/3-540-09510-1\_15}

\bibitem{Flanagan2002ExtendedStaticCheckingJava}
Flanagan, C., Leino, K.R.M., Lillibridge, M., Nelson, G., Saxe, J.B., Stata,
  R.: Extended static checking for java. In: Proceedings of the ACM SIGPLAN
  2002 Conference on Programming Language Design and Implementation. p.
  234–245. PLDI '02, Association for Computing Machinery, New York, NY, USA
  (2002). \doi{10.1145/512529.512558},
  \url{https://doi.org/10.1145/512529.512558}

\bibitem{Hamin2018DeadlockFreeM}
Hamin, J., Jacobs, B.: Deadlock-free monitors. In: ESOP (2018).
  \doi{10.1007/978-3-319-89884-1\_15}

\bibitem{Hamin2019TransferringOT}
Hamin, J., Jacobs, B.: {Transferring Obligations Through Synchronizations}. In:
  Donaldson, A.F. (ed.) 33rd European Conference on Object-Oriented Programming
  (ECOOP 2019). Leibniz International Proceedings in Informatics (LIPIcs),
  vol.~134, pp. 19:1--19:58. Schloss Dagstuhl--Leibniz-Zentrum fuer Informatik,
  Dagstuhl, Germany (2019). \doi{10.4230/LIPIcs.ECOOP.2019.19},
  \url{http://drops.dagstuhl.de/opus/volltexte/2019/10811}

\bibitem{herlihy-multiprocessor}
Herlihy, M., Shavit, N.: The Art of Multiprocessor Programming, Revised
  Reprint. Morgan Kaufmann Publishers Inc., San Francisco, CA, USA, 1st edn.
  (2012)

\bibitem{Hoare1968HoareLogic}
Hoare, C.A.R.: An axiomatic basis for computer programming. Commun. ACM
  \textbf{12},  576--580 (1968). \doi{10.1145/363235.363259}

\bibitem{Jacobs2020IOLiveness}
Jacobs, B.: Modular verification of liveness properties of the {I/O} behavior
  of imperative programs. In: ISoLA (2020). \doi{10.1007/978-3-030-61362-4\_29}

\bibitem{verifast2104}
Jacobs, B. (ed.): {VeriFast 21.04}. Zenodo (2021). \doi{10.5281/zenodo.4705416}

\bibitem{Jacobs2018ModularTerminationVerification}
Jacobs, B., Bosnacki, D., Kuiper, R.: Modular termination verification of
  single-threaded and multithreaded programs. ACM Trans. Program. Lang. Syst.
  \textbf{40},  12:1--12:59 (2018). \doi{10.1145/3210258}

\bibitem{Jacobs2011Verifast}
Jacobs, B., Smans, J., Philippaerts, P., Vogels, F., Penninckx, W., Piessens,
  F.: Verifast: A powerful, sound, predictable, fast verifier for {C} and
  {Java}. In: Bobaru, M., Havelund, K., Holzmann, G., Joshi, R. (eds.) NASA
  Formal Methods (NFM 2011). vol.~6617, pp. 41--55. Springer (2011).
  \doi{10.1007/978-3-642-20398-5\_4}, \url{https://lirias.kuleuven.be/95720}

\bibitem{Jung2016HigherorderGS}
Jung, R., Krebbers, R., Birkedal, L., Dreyer, D.: Higher-order ghost state.
  Proceedings of the 21st ACM SIGPLAN International Conference on Functional
  Programming  (2016). \doi{10.1145/2951913.2951943}

\bibitem{Jung2018IrisGroundUp}
Jung, R., Krebbers, R., Jourdan, J.H., Bizjak, A., Birkedal, L., Dreyer, D.:
  Iris from the ground up: A modular foundation for higher-order concurrent
  separation logic. J. Funct. Program.  \textbf{28}, ~e20 (2018).
  \doi{10.1017/S0956796818000151}

\bibitem{Kim2017LayerByLayer}
Kim, J., Sj{\"o}berg, V., Gu, R., Shao, Z.: Safety and liveness of mcs
  lock—layer by layer. In: Asian Symposium on Programming Languages and
  Systems. pp. 273--297. Springer (2017)

\bibitem{Kobayashi2006ANT}
Kobayashi, N.: A new type system for deadlock-free processes. In: CONCUR
  (2006). \doi{10.1007/11817949\_16}

\bibitem{Leino2009MultithreadedPrograms}
Leino, K., M{\"u}ller, P.: A basis for verifying multi-threaded programs. In:
  ESOP (2009)

\bibitem{Leino2010DeadlockFreeCA}
Leino, K.R.M., M{\"u}ller, P., Smans, J.: Deadlock-free channels and locks. In:
  ESOP (2010). \doi{10.1007/978-3-642-11957-6\_22}

\bibitem{Liang2016LiliAPL}
Liang, H., Feng, X.: A program logic for concurrent objects under fair
  scheduling. In: POPL 2016 (2016). \doi{10.1145/2837614.2837635}

\bibitem{Liang2017LiliProgressOC}
Liang, H., Feng, X.: Progress of concurrent objects with partial methods. Proc.
  ACM Program. Lang.  \textbf{2},  20:1--20:31 (2017). \doi{10.1145/3158108}

\bibitem{MellorCrummey1991AlgorithmsFS}
Mellor-Crummey, J.M., Scott, M.L.: Algorithms for scalable synchronization on
  shared-memory multiprocessors. ACM Trans. Comput. Syst.  \textbf{9},  21--65
  (1991). \doi{10.1145/103727.103729}

\bibitem{Muehlemann1980MethodFR}
{Mühlemann}, K.: Method for reducing memory conflicts caused by busy waiting
  in multiple processor synchronisation. IEE Proceedings E - Computers and
  Digital Techniques  \textbf{127}(3),  85--87 (1980).
  \doi{10.1049/ip-e.1980.0017}

\bibitem{OHearn2001LocalRA}
O'Hearn, P.W., Reynolds, J.C., Yang, H.: Local reasoning about programs that
  alter data structures. In: CSL (2001). \doi{10.1007/3-540-44802-0\_1}

\bibitem{Owicki1976VerifyingPropiertiesGhostVariables}
Owicki, S., Gries, D.: Verifying properties of parallel programs: an axiomatic
  approach. Commun. ACM  \textbf{19},  279--285 (1976)

\bibitem{ArtifactJacobs2020VerifastIOLivenessServer}
Reinhard, T., Jacobs, B.: {VeriFast} proof of {I/O} liveness for a simple
  server with a receiver and a responder thread communicating via a shared
  buffer. (2020),
  \url{https://github.com/verifast/verifast/blob/master/examples/busywaiting/ioliveness/echo_live_mt.c}

\bibitem{ArtifactJacobs2020VerifastCLHLock}
Reinhard, T., Jacobs, B.: {VeriFast} proof of safety for {CLH} lock. (2020),
  \url{https://github.com/verifast/verifast/blob/master/examples/busywaiting/clhlock/clhlock.c}

\bibitem{ArtifactJacobs2020VerifastGhostSignalConsumerProducerBoundedFifo}
Reinhard, T., Jacobs, B.: {VeriFast} proof of termination for consumer-producer
  problem with bounded {FIFO}. (2020),
  \url{https://github.com/verifast/verifast/blob/master/examples/busywaiting/bounded_fifo.c}

\bibitem{Reinhard2020GhostSignalsTR}
Reinhard, T., Jacobs, B.: Ghost signals: Verifying termination of busy-waiting
  (technical report) (2021),
  \url{https://people.cs.kuleuven.be/~tobias.reinhard/ghostSignals--TR.pdf}

\bibitem{Reinhard2020AbruptExitPaper}
Reinhard, T., Timany, A., Jacobs, B.: A Separation Logic to Verify Termination
  of Busy-Waiting for Abrupt Program Exit, p. 26–32. Association for
  Computing Machinery, New York, NY, USA (2020),
  \url{https://doi.org/10.1145/3427761.3428345}

\bibitem{Reynolds2002SeparationLA}
Reynolds, J.C.: Separation logic: a logic for shared mutable data structures.
  Proceedings 17th Annual IEEE Symposium on Logic in Computer Science pp.
  55--74 (2002). \doi{10.1109/LICS.2002.1029817}

\bibitem{total-tada}
da~Rocha~Pinto, P., Dinsdale{-}Young, T., Gardner, P., Sutherland, J.: Modular
  termination verification for non-blocking concurrency. In: Thiemann, P. (ed.)
  Programming Languages and Systems - 25th European Symposium on Programming,
  {ESOP} 2016, Held as Part of the European Joint Conferences on Theory and
  Practice of Software, {ETAPS} 2016, Eindhoven, The Netherlands, April 2-8,
  2016, Proceedings. Lecture Notes in Computer Science, vol.~9632, pp.
  176--201. Springer (2016). \doi{10.1007/978-3-662-49498-1\_8},
  \url{https://doi.org/10.1007/978-3-662-49498-1\_8}

\bibitem{Simpson1999WKL}
Simpson, S.: Subsystems of second order arithmetic. In: Perspectives in
  mathematical logic (1999). \doi{10.1017/CBO9780511581007}

\bibitem{Tarski1955KnasterTarskiTheorem}
Tarski, A.: A lattice-theoretical fixpoint theorem and its applications.
  Pacific Journal of Mathematics  \textbf{5},  285--309 (1955).
  \doi{10.2307/2963937}

\bibitem{VafeiadisCSLSound}
Vafeiadis, V.: Concurrent separation logic and operational semantics.
  Electronic Notes in Theoretical Computer Science  \textbf{276},  335--351
  (2011). \doi{https://doi.org/10.1016/j.entcs.2011.09.029},
  \url{https://www.sciencedirect.com/science/article/pii/S1571066111001204},
  twenty-seventh Conference on the Mathematical Foundations of Programming
  Semantics (MFPS XXVII)

\end{thebibliography}
